\documentclass[%
 aip,
 amsmath,amssymb,amsfonts,reprint,
]{revtex4-1}

\usepackage{graphicx}
\usepackage{dcolumn}
\usepackage{bm}
\usepackage[utf8]{inputenc}
\usepackage{amsmath}
\usepackage{amssymb}
\usepackage[T1]{fontenc}
\usepackage{mathptmx}
\usepackage{color}
\usepackage{listings}
\usepackage{algorithm}
\usepackage[noend]{algpseudocode}
\usepackage{comment}
\usepackage{enumitem}

\usepackage[table]{xcolor}

\DeclareMathAlphabet{\mathcal}{OMS}{cmsy}{m}{n}
\DeclareMathAlphabet{\mathnormal}{OML}{cmm}{m}{n}

\def\vA{{\bm A}}
\def\vB{{\bm B}}

\def\vj{{\bm j}}

\def\vm{{\bm m}}
\def\vR{{\bm R}}

\def\gv{{\bm g}}
\def\vx{{\bm x}}
\def\vy{{\bm y}}
\def\vz{{\bm z}}

\def\vjs{{\bm j}_{\mathrm{{\mbox{$\scriptscriptstyle{\rm{s}}$}}}}}
\def\vd{{\bm{\delta}}}

\def\vv{{\bm v}}
\def\vvF{{\bm v}_{\mathrm{\mbox{$\scriptscriptstyle{\rm{F}}$}}}}
\def\uvvF{{\hat{\bm v}}_{\mathrm{\mbox{$\scriptscriptstyle{\rm{F}}$}}}}
\def\vF{\mathnormal{v}_{\mathrm{\mbox{$\scriptscriptstyle{\rm{F}}$}}}}

\def\kF{\mathnormal{k}_{\mathrm{\mbox{$\scriptscriptstyle{\rm{F}}$}}}}
\def\ps{\mathnormal{p}_{\mathrm{\mbox{$\scriptscriptstyle{\rm{s}}$}}}}
\def\vp{{\bm p}}
\def\vx{{\bm x}}
\def\vd{{\bm d}}
\def\vps{{\bm p}_{\mathrm{{\mbox{$\scriptscriptstyle{\rm{s}}$}}}}}
\def\uvps{{\hat{\bm p}}_{\mathrm{{\mbox{$\scriptscriptstyle{\rm{s}}$}}}}}

\def\vpF{{\bm{p}}_{\mathrm{\mbox{$\scriptscriptstyle{\rm{F}}$}}}}

\def\NF{N_{\mathrm{\mbox{$\scriptscriptstyle{\rm{F}}$}}}}

\def\kB{k_{\mathrm{\mbox{$\scriptscriptstyle{\rm{B}}$}}}}
\def\Tc{T_{\mathrm{\mbox{$\scriptscriptstyle{\rm{c}}$}}}}
\def\TF{T_{\mathrm{\mbox{$\scriptscriptstyle{\rm{F}}$}}}}
\def\EF{E_{\mathrm{\mbox{$\scriptscriptstyle{\rm{F}}$}}}}

\def\Ds{\Delta}
\def\Ts{\tilde \Delta}

\def\qcgrad{{i\vvF\!\cdot\!\nabla_{\!\!R}\, }}

\def\yosh{{Y_{3/2}}}
\newcommand*{\be}{\begin{equation}}
\newcommand*{\ee}{\end{equation}}
\newcommand*{\ber}{\begin{eqnarray}}
\newcommand*{\eer}{\end{eqnarray}}

\newcommand{\Neqref}[1]{Eq.~\eqref{#1}}

\def\RR{{{\bm R}}}
\def\gam{\gamma(\vpF,\vR;\varepsilon)}
\def\gamt{\tilde\gamma(\vpF,\vR;\varepsilon)}
\def\epsilonA{(\varepsilon+\tfrac{e}{c}\vvF\cdot\vA)}

\colorlet{punct}{red!60!black}
\definecolor{background}{HTML}{EEEEEE}
\definecolor{delim}{RGB}{20,105,176}
\colorlet{numb}{magenta!60!black}
\lstdefinelanguage{json}{
    basicstyle=\normalfont\ttfamily,
    numberstyle=\scriptsize,
    stepnumber=1,
    numbersep=8pt,
    showstringspaces=false,
    breaklines=true,
    frame=lines,
    backgroundcolor=\color{background},
    literate=
     *{0}{{{\color{numb}0}}}{1}
      {1}{{{\color{numb}1}}}{1}
      {2}{{{\color{numb}2}}}{1}
      {3}{{{\color{numb}3}}}{1}
      {4}{{{\color{numb}4}}}{1}
      {5}{{{\color{numb}5}}}{1}
      {6}{{{\color{numb}6}}}{1}
      {7}{{{\color{numb}7}}}{1}
      {8}{{{\color{numb}8}}}{1}
      {9}{{{\color{numb}9}}}{1}
      {:}{{{\color{punct}{:}}}}{1}
      {,}{{{\color{punct}{,}}}}{1}
      {\{}{{{\color{delim}{\{}}}}{1}
      {\}}{{{\color{delim}{\}}}}}{1}
      {[}{{{\color{delim}{[}}}}{1}
      {]}{{{\color{delim}{]}}}}{1},
}

\begin{document}

\title{SuperConga: an open-source framework for mesoscopic superconductivity}

\author{P. Holmvall}
\thanks{These authors contributed equally to this work.}
\affiliation{Department of Microtechnology and Nanoscience - MC2,
	Chalmers University of Technology, SE-412 96 G\"oteborg, Sweden}
	
\affiliation{Department of Physics and Astronomy,
	Uppsala University, Box 516, S-751 20, Uppsala, Sweden}
\author{N. Wall Wennerdal}
\thanks{These authors contributed equally to this work.}
\author{M. H\r{a}kansson}
\author{P. Stadler}
\author{O. Shevtsov}
\author{T. L\"ofwander}
\author{M. Fogelstr\"om}
\affiliation{Department of Microtechnology and Nanoscience - MC2,
	Chalmers University of Technology, SE-412 96 G\"oteborg, Sweden}

\date{\today}

\begin{abstract}
We present SuperConga, an open-source framework for simulating equilibrium properties of unconventional and ballistic singlet superconductors, confined to two-dimensional (2D) mesoscopic grains in a perpendicular external magnetic field, at arbitrary low temperature. It aims at being both fast and easy to use, enabling research without access to a computer cluster, and visualization in real-time with OpenGL. The core is written in C++ and CUDA, exploiting the embarrassingly parallel nature of the quasiclassical theory of superconductivity by utilizing the parallel computational power of modern GPUs. The framework self-consistently computes both the superconducting order-parameter and the induced vector potential, and finds the current density, free energy, induced flux density, local density of states, as well as the magnetic moment. A user-friendly Python frontend is provided, enabling simulation parameters to be defined via intuitive configuration files, or via the command-line interface, without requiring a deep understanding of implementation details. For example, complicated geometries can be created with relative ease. The framework ships with simple tools for analyzing and visualizing the results, including an interactive plotter for spectroscopy. An overview of the theory is presented, as well as examples showcasing the framework's capabilities and ease of use.
The framework is free to download from https://gitlab.com/superconga/superconga, which also links to the extensive user manual, containing even more examples, tutorials and guides.

To demonstrate and benchmark SuperConga, we study the magnetostatics, thermodynamics, and spectroscopy of various phenomena. In particular, we study flux quantization in solenoids, vortex physics, surface Andreev bound-states, and a "phase crystal''. We compare our numeric results with analytics, and present experimental observables, e.g. the magnetic moment and LDOS, measurable with for example scanning-probes, STM and magnetometry.
\end{abstract}

\maketitle

\section{\label{sec:introduction}Introduction}
Over the last decades there has been a tremendous improvement in fabrication and controllable downscaling of samples and devices.
The physical size of a sample can now be comparable to characteristic quantum length scales of the composing materials.
For a superconducting material, which is the topic of this article, the relevant length scales are the superconducting coherence length, the length scale over which the superconducting order parameter typically varies, and the magnetic penetration depth, which determines the length scale of screening.
In direct connection to the ability to make smaller samples and devices, new measurement and sensing tools have been developed. Current scanning probes
can spatially resolve minute variations in magnetic fields using scanning SQUIDS,\cite{Tsuei:2000,Kirtley:2005,Kokubo:2010,Bert:2011,Vasyukov:2013} or make detailed
spatial maps of low-lying quasiparticle states and coherences using scanning tunneling spectroscopy (STS).\cite{Hess:1989,Renner:1991,Maggio-Aprile:1995,Yazdani:1997,Pan:2000a,Pan:2000b,Hoffman:2002,Guillamon:2008,Roditchev:2015,Berthod:2017} By fabricating 
nano-scale cantilevers, the intrinsic state of a superconducting island can be monitored as changes in the cantilever-oscillation frequency.
\cite{Bolle:1999,Bleszynski-Jayich:2009,Jang:2011} Other techniques of detecting and following the superconducting state in nanoscale devices 
include Hall magnetometry,\cite{Geim:1997,Morelle:2004} scanning electron microscopy (SEM) \cite{Grigorieva:2006} and scanning Hall probe microscopy.\cite{Khotkevych:2008,Curran:2014,Ge:2017} 
Scanning single electron transistors (SET) are local probes that can map out charging, or parity, effects that become prominent for ultra-small superconductors
where the energy-level spacing and the superconducting pairing are of the same order of magnitude.\cite{Tuominen:1992,Lafarge:1993,Ralph:1995,Kubatkin:1996,Gustafsson:2013}
Experimental data gathered using the above-mentioned methods give rich and complex information of the strongly correlated material system under study. 
To fully understand this data, one needs to compare and contrast them to theoretical predictions and modelling.

Superconductivity in metals is explained by the Bardeen-Cooper-Schrieffer (BCS)
theory.\cite{bardeen_theory_1957,Schrieffer:Book,deGennes:Book,Tinkham:Book} 
Its extension and generalisation to unconventional superconductivity and superfluids\cite{Mineev:Book,Serene:1983} 
makes the theory a corner stone of condensed matter physics. 
One method to model superconducting devices, and to explore fundamental problems of superconductivity
with the BCS theory as a starting point, is the quasiclassical theory of superconductivity. It is
the extension of Landau's Fermi-liquid theory to include superconducting phenomena and was pioneered 
by Eilenberger,\cite{Eilenberger:1968} Larkin and Ovchinnikov,\cite{Larkin:1969} and
Eliashberg.\cite{Eliashberg:1971} The quasiclassical approximation relies 
on a separation between relevant energy, temperature, and length scales in the normal-metal and superconducting states.
The normal-metal state is characterised by the Fermi energy,
$\EF$, Fermi temperature, $\TF$, and inverse Fermi wave number, $1/\kF$,
while the superconducting state is characterised by the superconducting order parameter,
$\Delta$, transition temperature, $\Tc$, and coherence length, $\xi_0$.
Quasiclassical theory is then a controlled expansion in small parameters like $\Delta/\EF$,
$\Tc/\TF$, or $1/ (\xi_0  \kF)$, which are usually of order $10^{-2}-10^{-3}$ in metals. 
As a Green's function based theory,\cite{AGD:Book} it is very powerful and include in its most general form 
material effects such as impurity scattering, Fermi-liquid effects, electron-phonon interaction, and non-equilibrium situations for instance imposed by external
fields or potentials.\cite{rainer_strong-coupling_1995,Kopnin:Book} At the same time, it is versatile enough that it can be adapted to realistic device sizes and geometries.\cite{Fogelstrom:1995,Sauls:2009pw,Silaev:2015,Kevin4} In its simplest form, quasiclassical theory is equivalent to the Andreev approximation of the Bogoliubov-deGennes
equation for the ballistic case.\cite{deGennes:Book,Andreev:1964,shelankov_quasiclassical_2000}

When the mean free path, $\ell$, due to elastic impurity scattering is much smaller than the superconducting coherence length, $\xi_0$, it is possible to derive diffusion equations from the more general Eilenberger-Larkin-Ovchinnikov equations. These diffusion equations, first derived by Usadel,\cite{Usadel:1970,Alexander:1985} are more easy to handle and is widely used to describe conventional diffusive $s$-wave samples and devices.\cite{belzig_quasiclassical_1999,Bergeret:2018,Sauls:2022} Solution methods in the diffusive case include Nazarov's circuit theory\cite{Nazarov:1999} and finite-element methods.\cite{Amundsen:2016}Here, we focus instead on the more general equations, in the ballistic regime. 

A complication when solving the equations of quasiclassical theory is that one often needs to resort to numerical methods. To develop the necessary codes is technically 
demanding and time consuming. Today there is no open-source code general enough to describe ballistic devices or superconducting grains of mesoscopic size with the 
quasiclassical theory. This lack of an open-source code forces all researchers and students in the community to re-implement the theory for each individual problem.

To remedy this, in this paper we present
an open source framework for studies of two-dimensional superconducting grains. The application program interface (API) is written in C++ and CUDA, and take
advantage of the speedup made possible by running the code on Graphics Processing Units (GPUs). At the same time, the Python-based frontend is sufficiently easy 
to use that any user interested only in the physics and results never has to dwell into the technical details of the implementation. The first version of the API
that we present here can be used to study conventional and unconventional singlet superconducting grains of general geometry in two dimensions with an applied
external magnetic field. The framework is sufficiently modular that it in the future can be extended to include more aspects of quasiclassical theory.

This paper is organized in the following way. In section \ref{sec:quasiclassics} we give an introduction to  quasiclassical theory with the aim to provide a
self-contained account of the theory which the framework is based on. In this section, we also show some simple model calculations, serving as a background to the examples
worked out in later sections. In section \ref{sec:implementation} we give an overview of the main algorithm of SuperConga and give sufficient background to understand the parameters that have to be set while running the simulations. In section \ref{sec:demonstration} we show by a simple example how to run the framework using the Python frontend. Section \ref{sec:examples} contains a few more involved examples that benchmark and highlight the capabilities of SuperConga. We study the rather complicated physics and energetics of vortices in a superconducting grain. In particular we find $B_{\mathrm{c}1}$ where it becomes energetically favorable to have one vortex in a finite size disc. By comparison to formulas from literature, we are also able to extract the vortex-core energy. At high fields we demonstrate that many different vortex configurations can be stabilized with almost the same free energy. For lower fields such configurations might be more separated in energy, but can be studied as metastable states. In another example we find the vortex configurations and the field-dependent local density of states in a grain studied experimentally by Timmermans {\it et al.}\cite{Timmermans:2016} Finally, we show that in a $d$-wave superconducting annulus, the free-energy parabolas are conventional \cite{Tinkham:Book} at high temperatures, while at low temperatures zero-energy Andreev bound states at the edges modifies the energetics through their paramagnetic response to the external field. Interestingly, spontaneous circular surface currents \cite{Vorontsov:2003,Hakansson:2015,Holmvall:2020} can coexist at high external magnetic fields with superconducting phase windings $n2\pi$ ($n$ integer) around the annulus and integer flux $n\Phi_0$ threading the inner circular hole. Section \ref{sec:summary} contains a summary of the paper. The appendices contain additional information on the SuperConga external dependencies and parameter configuration file, details about self-consistency convergence acceleration, the technique to solve for the vector potential and the equations of motion, as well as tables summarizing our choice of units and order-parameter basis functions.

\section{\label{sec:quasiclassics}Quasiclassical theory of superconductivity}

The superconducting state is characterised by an order parameter
\begin{equation}
\label{eq:order_parameter}
\Delta(\vpF,\vR) = \sum_\Gamma|\Delta_\Gamma(\vR)|e^{i\chi_\Gamma(\vR)}\eta_\Gamma(\vpF),
\end{equation}
that is zero for temperatures above the superconducting transition ($T>T_{\mathrm{c}}$) and non-zero below it. All superconductors break $U(1)$-gauge symmetry and hence the minimal order parameter is described by an amplitude $|\Delta(\vR)|$ and a phase $\chi(\vR)$, both in general dependent on the spatial coordinates $\vR$. The phase is directly coupled to the superfluid momentum, $\vps(\vR)$, and the electromagnetic gauge field, $\vA(\vR)$, via the gauge invariant expression
\begin{equation}
\label{eq:phase}
\vps(\vR) = \frac{\hbar}{2}\mathbf{\nabla}\chi(\vR) - \frac{e}{c}\vA(\vR),
\end{equation}
where $e = -|e|$ is the charge of the electron, $\hbar$ is Plancks constant, and $c$ is the speed of light. In addition to breaking the $U(1)$ symmetry, some superconducting compounds also break symmetries of the crystal lattice. This possibility is encoded in the complex-valued basis function $\eta_\Gamma(\vpF)$ which depends on the Fermi momentum, $\vpF$. The index $\Gamma$ denotes the irreducible representation of the crystallographic point group that the basis function belongs to.\cite{Yip:1993} In the general case we also need to account for the spin-degree of freedom, and the order-parameter field should therefore be written as a spin-matrix $\Delta_{\alpha\beta}(\vpF,\vR)$. However, we will focus on spin-singlet superconductivity, and the form in \Neqref{eq:order_parameter} is adequate for this.

The singlet order parameter satisfies the following gap equation:
\begin{equation}
    \Ds(\vpF,\vR)=\NF T\sum_n^{|\varepsilon_n|<\varepsilon_c} \big\langle
                  V(\vpF,\vpF^\prime)\,f(\vpF^\prime,\vR;\epsilon_n)
                   \big\rangle_{\vpF^\prime} \;,
\label{eq:gapequation}
\end{equation}
where $V(\vpF,\vpF^\prime)$ is the effective superconducting pairing interaction. We use the approximation that the interaction can be separated into symmetry channels of the crystal point group
\begin{equation}
V(\vpF,\vpF^\prime)=\sum_{\Gamma} V_\Gamma \eta_\Gamma(\vpF)\eta^\dagger_\Gamma(\vpF^\prime),
\end{equation}
where $V_\Gamma$ is the pairing strength in the symmetry channel $\Gamma$. In this case, the dependencies on position and momentum direction of the order parameter separate as in Eq.~\eqref{eq:order_parameter}. The angle bracket $\langle \dots \rangle_{\vpF}$ denotes a Fermi-surface average, which in 2D is a line integral around the Fermi surface according to\cite{Graf:1993,Wennerdal:2020}
\be
\langle \dots \rangle_{\vpF} = \frac{1}{\NF} \oint_\mathrm{FS} \frac{\mathrm{d} \vpF}{(2\pi\hbar)^2|\vvF(\vpF)|} (\dots),
\ee
where the total normal-state density of states per spin at the Fermi level is
\be
\NF = \oint_\mathrm{FS} \frac{\mathrm{d} \vpF}{(2\pi\hbar)^2|\vvF(\vpF)|}.
\ee
We consider the extreme layered superconductor with conduction in stacked two-dimensional planes. In this case $\NF=\kF/2\pi\hbar \vF d$, where $\kF,\vF$ are the in-plane
Fermi-wave number and Fermi velocity respectively and $d$ is the inter plane distance.\cite{Graf:1993}
The vector potential is determined through Amp\`{e}re's law 
\begin{eqnarray}
\nabla\times\nabla\times\vA(\vR)&=&\frac{4\pi}{c} \vj(\vR) \; ,
\label{eq:ampere}
\end{eqnarray}
where the charge current density is defined as
\begin{eqnarray}
\vj(\vR) &=& 2e \NF T\sum_{n} \big\langle \vvF(\vpF)\, 
g(\vpF,\vR;\epsilon_n) \big\rangle_{\vpF} \; ,
\label{eq:current_density}
\end{eqnarray}
where the factor 2 accounts for spin degeneracy.
The Fermi velocity at $\vpF$ on the Fermi surface is $\vvF(\vpF)=\nabla_{\vp}\varepsilon(\vp)|_{\vp=\vpF}$, where $\varepsilon(\vp)$ is the dispersion.

The two functions $g(\vpF,\vR;\epsilon_n)$ and $f(\vpF,\vR;\epsilon_n)$ appearing in \Neqref{eq:gapequation} and \Neqref{eq:current_density} are components of the quasiclassical Green's function, or propagator,
\begin{equation}
\hat g(\vpF,\vR;\epsilon)= \left(\begin{array}{rc}
g(\vpF,\vR;\epsilon)& f(\vpF,\vR;\epsilon) \\ 
-\tilde f(\vpF,\vR;\epsilon)& \tilde{g}(\vpF,\vR;\epsilon)
\end{array}\right).
\label{eq:matrixg}
\end{equation}
It is a 2x2 matrix in electron-hole (Nambu) space, as indicated by the "$\hat~$"-symbol.
The propagator is determined from the quasiclassical counter part of the Gorkov-Dyson equation, the transport-like Eilenberger equation: 
\ber
0&=&\qcgrad \hat g(\vpF,\vR;\epsilon) + \nonumber\\
&&\bigg\lbrack
(\epsilon+\tfrac{e}{c}\, \vvF(\vpF)\!\cdot\!\vA(\vR))\hat \tau_3
-\hat \Delta(\vpF,\vR), \hat g(\vpF,\vR;\epsilon)\bigg\rbrack,\;\;\;
\label{eq:Eilenberger}
\eer
where $\hat\tau_3$ is the third Pauli matrix in Nambu space.
In addition to \Neqref{eq:Eilenberger}, the propagator obeys the normalization condition 
\be
\hat g^2(\vpF,\vR;\epsilon)=-\pi^2 \hat{\mathbb{I}},
\label{eq:norm}
\ee
where $\hat{\mathbb{I}}$ is the identity matrix.

In our case of weak-coupling superconductivity in the clean limit, the self energy $\hat \Delta(\vpF,\vR)$ entering in \Neqref{eq:Eilenberger} simply consists of the order parameter:
\be
\hat \Delta(\vpF,\vR)
=\left(
\begin{array}{cc} 
0 & \Ds(\vpF,\vR)\\ 
-\Ts(\vpF,\vR) & 0
\end{array}
\right).
\label{selfenergy}
\ee
There is some redundancy in the parametrisation of the propagator, and the following symmetry \cite{Serene:1983,Eschrig:2000}
\be
\tilde x(\vpF,\vR;\epsilon) =  x(-\vpF,\vR;-\epsilon^{*})^{*},
\label{symmetries}
\ee
between tilded ($\tilde x$) and un-tilded ($x$) quantities holds. The propagator may be evaluated at imaginary frequencies $\epsilon\rightarrow i\epsilon_n=i\pi T(2n+1)$ giving the Matsubara propagator or at real frequencies giving either the Retarded $(\epsilon\rightarrow \epsilon+i\delta)$, or the Advanced propagator $(\epsilon\rightarrow \epsilon-i\delta)$. Here, $\delta$ is a small and positive energy broadening. To compute the order parameter, the current, and the back-coupling through the gauge field, it is enough to work with the Matsubara propagator. For the local density of states, defined as
\begin{equation}
N(\vpF,\RR;\varepsilon) = -\frac{\NF}{\pi}\mbox{Im}\, g^\mathrm{R}(\vpF,\RR;\varepsilon),
\end{equation}
we need to also compute the retarded Green's function with the order parameter and gauge field as input.

The superconducting state has a characteristic energy scale given by the transition  temperature $T_{\mathrm{c}}$, namely $2 \pi k_\mathrm{B}T_\mathrm{c}$, and a natural length scale, the coherence length $\xi_0=\hbar \vF/ 2 \pi k_\mathrm{B}T_\mathrm{c}$. We will use the natural units $\hbar=k_\mathrm{B}=1$. The normal state density of states at the Fermi level $\NF$ is in units $[\mathrm{Energy}\times\mathrm{unit\,\, cell}\times\mathrm{spin}]^{-1}$. In addition to the coherence length, the length scale for screening of electromagnetic fields enters the theory through the penetration depth $\lambda_0$, defined as 
\begin{equation}
    \frac{1}{\lambda_0^2}=\frac{4\pi e^2}{c^2}\vF^2 \NF.
\end{equation}
We use the Ginzburg-Lanadu parameter $\kappa$, defined as $\kappa=\lambda_0/\xi_0$, to give the value of the penetration depth in relation to the coherence length. To define the magnetic units we start from the superconducting magnetic flux quantum $\Phi_0=h c/2|e|$. 
The magnetic field,
\be \vB=\nabla\times\vA, \ee
is then given in units of $B_0 = \Phi_0/\pi\xi_0^2$ and the vector potential in units of $A_0 = \Phi_0/\pi\xi_0$. Finally, current densities will be given in units of $j_0=2\pi k_\mathrm{B} T_{\mathrm{c}} |e| \vF\NF$. A summary of these units and natural scales are given in Table~\ref{table:units} in Appendix~\ref{app:units}.

The quasiclassical theory is a conserving theory in the sense that the equations for $\hat g$, $\hat \Delta$, and $\vA$ above may be derived as stationarity conditions from the Luttinger-Ward free-energy functional.\cite{Luttinger:1960,Serene:1983} For the theory to be conserving, the stationarity equations are to be solved to self consistency. At convergence, we solve \Neqref{eq:Eilenberger} with  $\Delta(\vpF,\vR)$ and $\vA(\vR)$, recalculate \Neqref{eq:gapequation} and \Neqref{eq:ampere}, and get back the same $\Delta(\vpF,\vR)$ and $\vA(\vR)$ within some given tolerance. Current conservation is then fulfilled to the same tolerance. At self-consistency we have found an extremum of the quasiclassical version of the Luttinger-Ward functional.
\cite{Serene:1983,Thuneberg:1983,Vorontsov:2003}
This functional always give a measure of the free energy with respect to the normal state,\cite{Serene:1983} $\Omega_S[T]-\Omega_N[T]$, and goes to zero as $\Delta\rightarrow 0$. To simplify notation we denote this free energy difference $\Omega[T]$, and it has the form
\begin{widetext}
\be
\Omega[T]=
\int \mathrm{d}\vR \Bigg\lbrace \frac{\left|\vB_{\mathrm{ind}}(\vR)\right|^2}{8\pi}
+\NF\sum_\Gamma|\Delta_\Gamma(\vR)|^2 \, {\rm ln} \frac{T}{T_\mathrm{c}^\Gamma}
+\pi\NF \kB  T\sum_{n} \frac{|\Delta(\vR)|^2}{|\epsilon_n|}
- {\cal{I}}(\vR)\Bigg\rbrace,
\label{qcLWFE}
\ee
where $\vB_{\mathrm{ind}}(\vR)$ is the induced magnetic field, $|\Delta(\vR)|^2=\langle |\Delta(\vpF,\vR)|^2\rangle_{\vpF}$, and
\be
{\cal I}(\vR) = \int_0^1 \mathrm{d}\lambda\, \NF \kB T \sum_n
\langle \tilde \Delta(\vpF,\vR) f_{\lambda}(\vpF,\vR;\epsilon_n)+\Delta(\vpF,\vR) \tilde f_{\lambda}(\vpF,\vR;\epsilon_n) \rangle_{\vpF}.
\label{qcLWFE_I}
\ee
The auxiliary propagator $\hat g_\lambda$ is obtained by solving the Eilenberger equation with scaled self-energy field $\lambda \hat \Delta$ so the integral over the dummy variable $\lambda$ in Eq.~(\ref{qcLWFE_I}) may be evaluated. In the framework SuperConga we also apply the original Eilenberger-form of the free-energy functional,\cite{Eilenberger:1968} in which case the term with the $\lambda$-integration is simplified to
\be
{\cal I}(\vR) = \pi \NF \kB T \sum_n \Bigg \langle
\frac{ \tilde \Delta(\vpF,\vR) f(\vpF,\vR;\epsilon_n)+\Delta(\vpF,\vR) \tilde f(\vpF,\vR;\epsilon_n)}
{\pi+i\, g(\vpF,\vR;\epsilon_n)}
\Bigg\rangle_{\vpF}.
\ee
The advantage of the Eilenberger-form is that it can be evaluated without the $\lambda$-integration at little computational cost. However, it must be used with some caution as it has been shown to not be generally valid and final results for the free energy needs to be confirmed by \Neqref{qcLWFE}.\cite{Graf:1993} 

The above form of the free energy is consistent with the following regularized gap equation
\begin{equation}
\Delta_\Gamma(\vR)\ln \frac{T}{T_{\mathrm{c}}^\Gamma}
= T \sum_n \left\langle 
\eta_\Gamma^*(\vpF) \left\lbrack f(\vpF,\vR;\epsilon_n)
-\frac{\pi\Delta(\vpF, \vR)}{|\varepsilon_n|} \right\rbrack \right\rangle_{\vpF} ,
\label{eq:gapeq_regularized}
\end{equation}
\end{widetext}
for superconductivity with the symmetry given by the representation $\Gamma$. 
The dependencies on the pairing interaction strength $V_\Gamma$ and the Matsubara sum cutoff
$\varepsilon_\mathrm{c}$ have been replaced by the measurable transition temperature
$T^\Gamma_{\mathrm{c}}$. After this regularization, the Matsubara sum is not cutoff dependent.
The connection between Eq.~\eqref{eq:gapequation} and Eq.~\eqref{eq:gapeq_regularized} can be
seen as replacing the coupling strength by
\begin{equation}
    \frac{1}{\NF\, V_\Gamma}\rightarrow \ln \frac{T}{T_{\mathrm{c}}^\Gamma}+\pi T \sum_n\frac{1}{|\varepsilon_n|}.
\end{equation}
By moving the second term on the right hand side of Eq.~\eqref{eq:gapeq_regularized} over to the left hand side, and using the fact that the basis functions are orthonormal, $\langle \eta^*_i(\vpF) \eta_j(\vpF)\rangle_{\vpF} = \delta_{ij}$, we can divide down the resulting prefactor of $\Delta_\Gamma(\vR)$, obtaining the gap equation in a fix point form:
\begin{equation}
\Delta_\Gamma(\vR)
= \frac{ T\sum_n \langle 
\eta_\Gamma^*(\vpF)f(\vpF,\vR;\epsilon_n) \rangle_{\vpF} }
{\ln \frac{T}{T_{\mathrm{c}}^\Gamma}+\pi T \sum_n\frac{1}{|\varepsilon_n|}
}.
\label{eq:gapeq_fixpoint}
\end{equation}
The currently available symmetries and the corresponding basis functions $\eta_{\Gamma}(\vpF)$ are listed in Table~\ref{table:basis_functions} in Appendix~\ref{app:units}.


\subsection{\label{sec:riccati}Riccati formalism for inhomogeneous states}

\begin{figure}
    \centering
    \includegraphics[width=1.0\columnwidth]{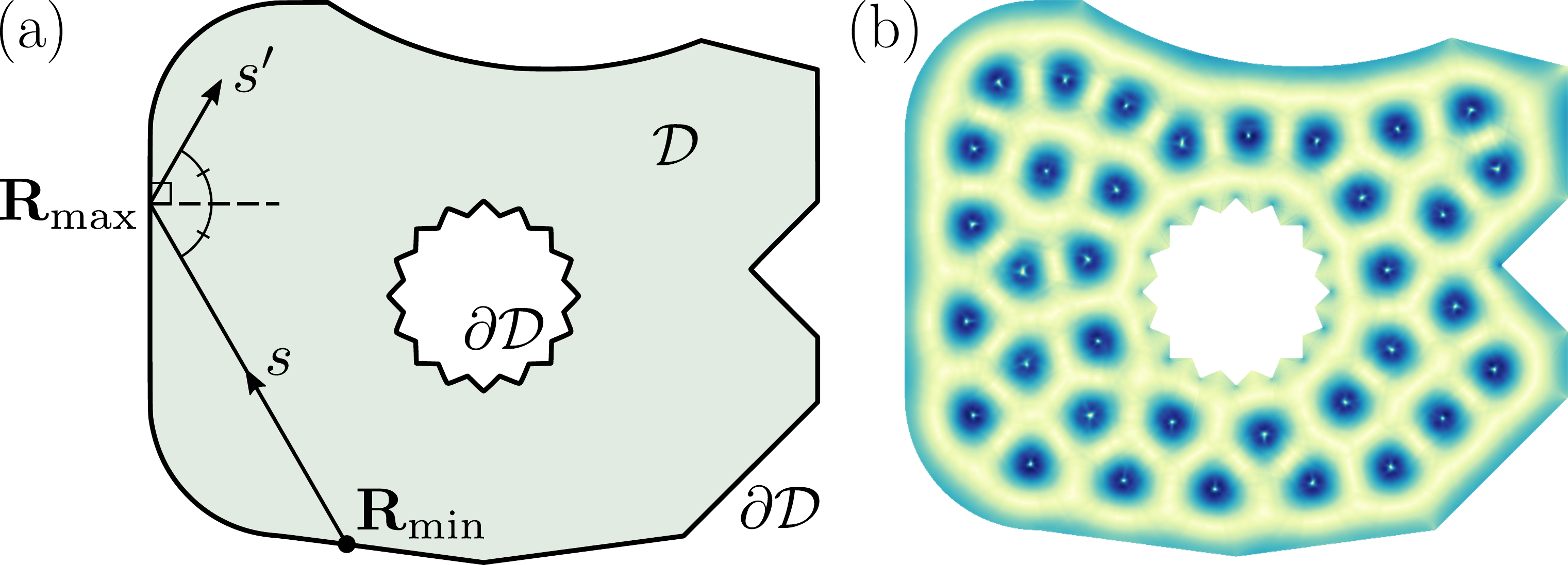}
    \caption{(a) The system domain is denoted ${\mathcal D}$, and has a boundary $\partial {\mathcal D}$. A quasiparticle trajectory parameterized by the coordinate $s$ starts at one boundary point $\vR_\mathrm{min}$ and ends at another boundary point $\vR_\mathrm{max}$. The specular boundary condition connects incoming Riccati amplitudes on trajectory $s$ to starting Riccati amplitudes on another trajectory $s'$ by simply requiring continuity of the amplitudes, i.e. $\gamma(s')=\gamma(s)$ at the boundary point.
    (b) Self-consistently computed vortex lattice in the same geometry as in panel (a), see the following sections for more details.
    }
    \label{fig:riccati_fig_A}
\end{figure}

For inhomogeneous superconducting states we need to solve the first order differential equation for the propagator $\hat g(\vpF,\vR;\varepsilon)$ in \Neqref{eq:Eilenberger}, subject to the normalization condition in \Neqref{eq:norm}. The gradient term $\qcgrad \hat g(\vpF,\vR;\varepsilon)$ couples the momentum direction and spatial coordinates. Quasiparticle trajectories are then defined by this directional derivative and we may introduce a trajectory coordinate $s$ as
\begin{equation}
\vR(s)=\vR_\mathrm{min} + s \uvvF,
\end{equation}
where $\uvvF=\vvF/\vF$ is a unit vector. Starting with an initial value, for instance at a boundary point $\vR_\mathrm{min}$, 
the propagator should be found by integrating along the trajectory until another boundary point $\vR_\mathrm{max}$ is met.
At that point, a boundary condition must be supplemented to the theory. In general, boundaries present
strong perturbations and their treatment lay outside the strict validity of the quasiclassical approximation. 
Nevertheless, one can model scattering off surfaces or at semitransparent interfaces via effective boundary conditions 
derived and carried over from a more microscopic theory.\cite{Serene:1983,Zaitsev:1984,kieselmann_self-consistent_1987,millis_quasiclassical_1988,Eschrig:2000,shelankov_quasiclassical_2000,fogelstrom_josephson_2000,zhao_nonequilibrium_2004,eschrig:2009} 
In certain cases, boundary and interface effects may arise which lay outside the standard quasiclassical approximation.
\cite{Ozana:2004,Babeev:2020a,Hainzl:2022}
In SuperConga a simple specular boundary condition is used at all boundaries, see illustration in Fig.~\ref{fig:riccati_fig_A}

In practice, the most convenient and stable formulation is based on a parametrization\cite{Nagato:1993,schopohl_quasiparticle_1995,schopohl_transformation_1998,Eschrig:2000} of the Green's function in terms of coherence functions $\gam$ and $\gamt$. They are related by the symmetry in Eq.~(\ref{symmetries}) and quantify electron-hole coherence along the trajectory in the superconducting state. The quasiclassical Green's function can be written in terms of them as
\begin{equation}
\hat g = \frac{-i\pi}{1+\gamma\tilde\gamma}\left(
\begin{array}{cc}
1-\gamma\tilde\gamma     & 2\gamma\\
2\tilde\gamma     & -1+\gamma\tilde\gamma
\end{array}
\right),
\label{eq:riccati_parameterization}
\end{equation}
where we for brevity dropped the explicit dependencies on $\vpF$, $\RR$, and $\varepsilon$.
The normalization of the Green's function in \Neqref{eq:norm} is now automatically satisfied.
The coherence functions satisfy a set of Riccati equations:
\begin{eqnarray}
i\vvF\cdot\nabla\gamma + 2\epsilonA\gamma + \tilde\Delta\gamma^2 + \Delta &=& 0,\label{eq:riccati}\\
i\vvF\cdot\nabla\tilde\gamma - 2\epsilonA\tilde\gamma + \Delta\tilde\gamma^2 + \tilde\Delta &=& 0. \label{eq:riccati_tilde}
\end{eqnarray}
The equation for $\gam$ is stable to integrate in the direction parallel to $\vvF$, while the equation for $\gamt$ is stable to be integrated in the opposite direction. After introducing the trajectory coordinate, we may let $\vvF\cdot\nabla\gamma\rightarrow \vF\partial_s\gamma$.

\section{\label{sec:implementation}Overview of the implementation}

In this section we give a brief overview of the main algorithm of SuperConga and describe a few key features of the framework. The aim of this section is to explain why certain parameters related to the implementation have to be set and how the parameter choices may influence the simulations.

\begin{figure*}

    \includegraphics[width=\textwidth]{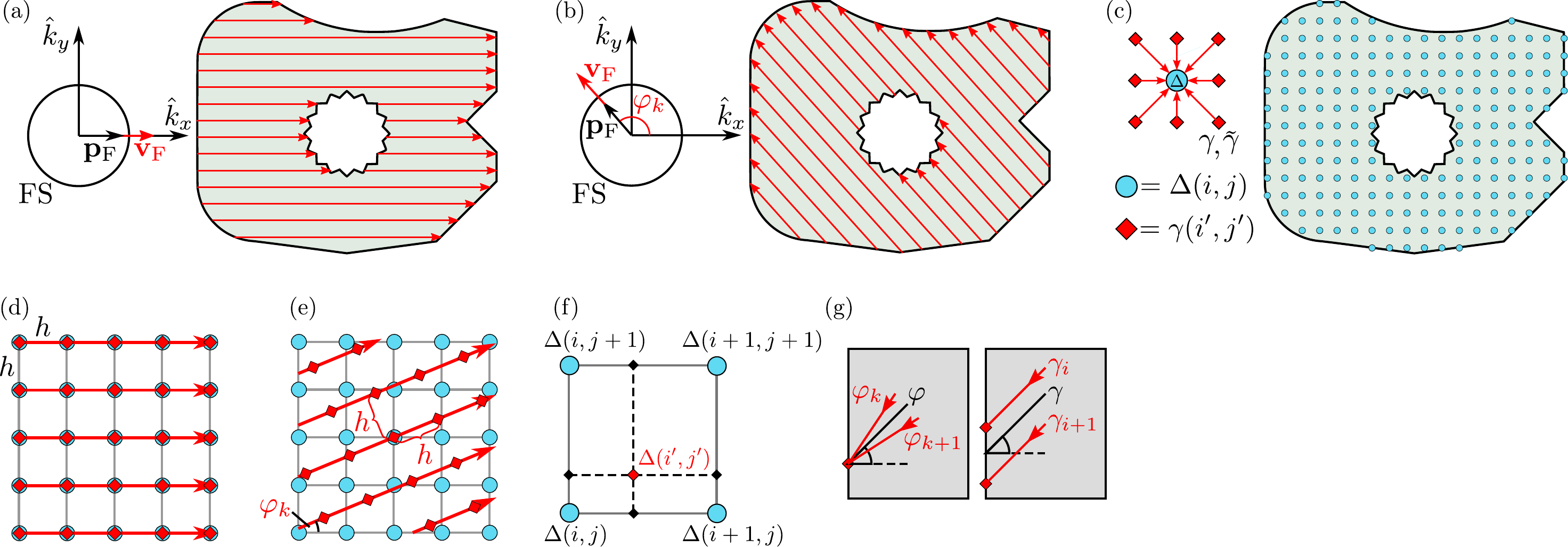}
    \caption{Vizualization of the implementation and main algorithm. (a)--(b) The Fermi surface is parametrized by the discrete angle $\varphi_k$ and the Fermi velocity $\vvF(\varphi_k)$, which together define quasiparticle trajectories in real space. As in Fig.~\ref{fig:riccati_fig_A}, each trajectory has a well-defined start point $\vR_{\mathrm{min}}$ and end point $\vR_{\mathrm{max}}$, which might either lie on an internal or external boundary.
    (c) The order parameter and vector potential are defined on a discretization of the geometry with grid spacing $h$, and are computed as the Fermi-surface average of all trajectories passing through the point.
    (d)--(f)  We let $h$ be the discrete step-size along each trajectory, as well as the separation between trajectories, defining a square lattice. For nearly all angles, the discrete geometry is incommensurate with the trajectories, and interpolation becomes necessary.
    (g) Similarly, the discrete Fermi surface makes the reflection and the matching of incoming and outgoing trajectories incommensurate, such that for a particular trajectory starting at a boundary, no incoming trajectory exists for the corresponding angle or boundary coordinate. The solution is to interpolate from neighboring incoming trajectories, in both angle and boundary coordinate.
    }
    \label{fig:algorithm_implementation_visualization}
\end{figure*}

\subsection{The main algorithm of SuperConga}

We solve the Riccati equations, Eqs.~\eqref{eq:riccati} and \eqref{eq:riccati_tilde}, numerically (see Appendix~\ref{app:riccati_stepping}) along trajectories through the grain. An example of a trajectory is shown in Fig~\ref{fig:riccati_fig_A}. Given an initial boundary value, $\gamma(\vR_{\mathrm{min}})$, we solve Eq.~\eqref{eq:riccati} until we reach $\vR_{\mathrm{max}}$. Similarly, given an initial boundary value, $\tilde{\gamma}(\vR_{\mathrm{max}})$, we solve Eq.~\eqref{eq:riccati_tilde} until we reach $\vR_{\mathrm{min}}$.

Clearly, the boundaries are special points in that we need a starting value for $\gamma(\vR_{\mathrm{min}})$ and we obtain after stepping along the trajectory an end value $\gamma(s_N)$ at another boundary, where $N$ is the number of steps taken along the trajectory $s$. All boundary values are related to each other through the boundary condition. The complication is then that the end values become, after the boundary condition has been applied, a starting value for a different trajectory. This can be solved by having an initial guess for all boundary starting values, and self-consistently compute the boundary values by repeatedly stepping along all trajectories and apply the boundary condition. Since the order parameter $\Delta(\vR)$ is also solved self-consistently by iteration, the update of the boundary values can be done in parallel. At the end of the computation, when the gap equation is satisfied, the boundary condition is also self-consistently satisfied. A caveat in SuperConga is that while the order parameter and vector potential are saved to file, the coherence functions are not, due to being to disk-space intensive. When restarting the calculation using a nominally self-consistent order parameter, the self-consistency of the boundary conditions must be reached again. This is solved via a ``burn-in'' period, where the order parameter and vector potential are kept constant for a few iterations when restarting a calculation, until the coherence functions and boundary conditions are also self-consistent.

In SuperConga the order parameter and the vector potential have been discretized in two-dimensional (2D) real space: $\Delta(\vR)=\Delta(x,y)\rightarrow\Delta(i,j)$ and $\vA(\vR)=\vA(x,y)\rightarrow \vA(i,j)$, where $i$ and $j$ are integers. The grid is a simple square grid with spacing $h$, i.e. $x=ih$ and $y=jh$, but we still take into account the exact location and shape of boundaries, such that there is no aliasing or similar artefacts. Figure~\ref{fig:algorithm_implementation_visualization} visualizes the implementation and discretization. The Fermi surface is parameterized by an azimuthal angle $\varphi\in [0,2\pi)$ and discretized according to $\varphi_k = k\Delta\varphi$ ($k$ integer). The Fermi surface averages are computed through numeric integration, via one of the following methods: Boole's, Simpson's $3/8$, Simpson's $1/3$, or the trapezoidal rule, depending on if the number of momenta is divisible by $4$, $3$, $2$, otherwise, respectively. For a particular point $\varphi_k$ on the Fermi surface, we obtain a Fermi velocity $\vvF(\varphi_k)$, which determines a direction for a set of trajectories in real space. Each trajectory coordinate has been discretized as outlined above. Considering all parallel trajectories for a certain point on the Fermi surface, the discrete grid in real space then forms a simple square grid with the same lattice spacing $h$, but rotated relative to the reference grid where the order parameter and vector potential are defined. Bilinear interpolation is then performed on the order parameter and the vector potential from the reference frame to the rotated frame. After solving the Riccati equations, the Riccati amplitudes are known in the rotated frame. The $\varphi_k$ terms in the Fermi-surface averages for the order parameter and the charge-current density are computed in the rotated frame by constructing the relevant matrix element of the Green's function, summing over energies, and then rotating back to the reference frame. In this way, the order parameter $\Delta(i,j)$, observables such as the current $\vj(i,j)$, and also the vector potential $\vA(i,j)$ can be updated through Eqs.~\eqref{eq:gapequation}-\eqref{eq:ampere}.

The boundary condition is problematic as the azimuthal angle of the specularly reflected momentum at a given surface might not exist in the set $\{ \varphi_k\}$. Furthermore, due to the rotation of the grid, the trajectories hit the boundary at different points for each Fermi velocity $\vvF(\varphi_k)$. In order to determine the initial conditions for the coherence functions, we once again perform bilinear interpolation. Here the interpolation is between spatial coordinates on the boundary $\vR \in \partial\mathcal{D}$ and azimuthal angles.

\begin{algorithm}[H]
\caption{A sketch of the SuperConga main loop}
\label{alg:suprcon}
\begin{algorithmic}
\State Define domain $\mathcal{D}$ and input parameters
\State Initialize $\Delta$ and $\vA_\mathrm{ind}$
\State Initialize $\gamma_{\partial \mathcal{D}}$ and $\tilde{\gamma}_{\partial \mathcal{D}}$ \Comment{The boundary}
\While{not converged}
\ForAll{$\vpF$,  $\epsilon_n$}
\State Rotate $\Delta$ and $\vA$ so that $\hat{\mathbf{y}} \parallel \vvF$
\State Compute $\gamma$ and $\tilde{\gamma}$ along the $y$-axis \Comment{See App.~\ref{app:riccati_stepping}}
\State Update $\gamma_{\partial \mathcal{D}}$ and $\tilde{\gamma}_{\partial \mathcal{D}}$ \Comment{Write incoming $\vpF$}
\State Update $\Delta$, $\vj$, and $\Omega$ \Comment{Eqs.~\eqref{eq:gapeq_fixpoint}, \eqref{eq:current_density}, \eqref{qcLWFE}}
\EndFor 
\State Compute $\vA_\mathrm{ind}$ \Comment{See App.~\ref{app:vector_potential}}
\State Check convergence \Comment{Eq.~\eqref{eq:global_error}}
\State $\Delta, \vA_\mathrm{ind} \gets $ Accelerator \Comment{See App.~\ref{app:accelerator}}
\State Update $\gamma_{\partial \mathcal{D}}$ and $\tilde{\gamma}_{\partial \mathcal{D}}$ \Comment{Boundary condition}
\EndWhile
\end{algorithmic}
\end{algorithm}

An overview of the algorithm of SuperConga is shown in Algorithm~\ref{alg:suprcon}. The user defines e.g. the system domain, physical parameters, the discretization $h$ in real space, the discretization $\Delta\varphi$ in momentum space, and an energy cut-off for the frequency summations [see e.g. \Neqref{eq:gapequation}].
The frequency summations are not over the usual Matsubara frequencies, but instead a more efficient summation due to Ozaki \cite{Ozaki:2007} has been implemented. In addition, a convergence criterion for the self-consistency must be set. The global error is defined as
\begin{equation}
\label{eq:global_error}
    \epsilon_\mathrm{G} = \frac{\left \| \Delta^i-\Delta^{i-1} \right \|_p}{\left \|\Delta^{i-1} \right \|_p},
\end{equation}
where $i$ here denotes the iteration number in the self-consistency loop, and $p \in \{1,2,\infty \}$ is set by the user. The loop continues until the error $\varepsilon_\mathrm{G}$ is smaller than the convergence criterion, or the given maximum number of iterations have been performed. The gap equation can be solved by direct fixed-point iteration, but in SuperConga a few methods for convergence acceleration have been implemented. In simplified terms, the accelerators may save a number of computed order parameters, $\left\{\Delta^i,\Delta^{i-1},\Delta^{i-2},...\right\}$, and make a more educated guess for $\Delta^{i+1}$. When observables, the vector potential, and free energy are also computed, the convergence criterion is the same as for the order parameter. All the details of the stepping along trajectories, interpolation, as well as the treatment of the boundaries are automatically taken care of by the API. The user must be aware of, however, that the discretization in real and momentum space can be a source of error and it is necessary to check convergence by running the same problem with finer grids. Also, depending on the problem being solved, the path to self-consistency may not be direct. For the simplest examples in this paper, this is not a problem. But for more advanced problems, it is advisable to experiment with different starting guesses for the order parameter and also carefully choose a suitable convergence acceleration strategy.

\subsection{Framework Overview}
\begin{figure}[tb!]
	\includegraphics[width=\columnwidth]{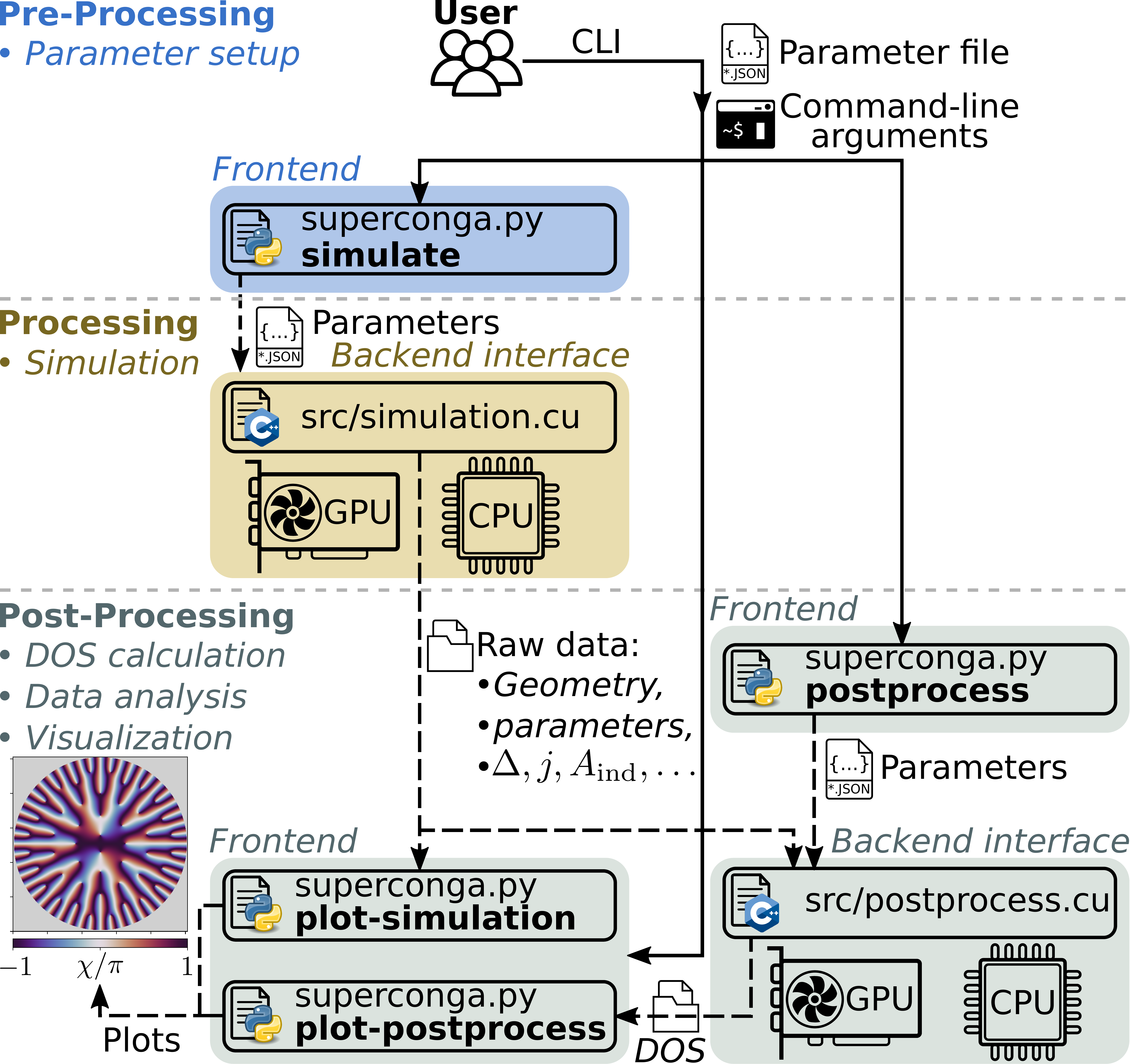}
	\caption{A usage flowchart for SuperConga. A user can use command-line interface (CLI) to pass command-line arguments and a parameter-file to the Python3 frontend \texttt{superconga.py}, which sanity checks the user input. If all input is valid, the frontend calls the binary built from the C++/CUDA backend interface \texttt{simulate.cu}. The latter includes the appropriate functionality from the backend framework (implemented as header files), and runs self-consistent simulations according to Algorithm~\ref{alg:suprcon}. The user can for example choose to enable live visualization and printing of simulation status to the terminal, and how frequently data should be saved to file. The user can perform data analysis directly on the raw data with their favorite external tool, or use the provided Python frontend, e.g. via  \lstinline{python superconga.py plot-simulation} and  \lstinline{python superconga.py plot-convergence}. Additionally, the data can be used in a postprocessing step to compute the DOS and LDOS via  \lstinline{python superconga.py postprocess} (which calls the binary built from the backend interface \texttt{postprocess.cu}). Finally, the DOS and LDOS can be illustrated with the interactive spectroscopy tool \lstinline{python superconga.py plot-postprocess}. Note that the user can also write their own backend interfaces to harness the full functionality of the SuperConga framework.
	}
	\label{fig:usage_overview}
\end{figure}

The main algorithm and functionality of SuperConga is implemented as a fairly modular framework, in the form of a \emph{backend} written in C++ and CUDA\cite{CUDA:2021,Nickolls:2008}. A user can in principle choose which modules to use and in what manner, to set up highly customized simulations, or even extend the existing functionality by exploiting the modularity. However, this requires writing C++ code that imports or adds to the desired features. While this is certainly possible and a valid approach, a main goal of SuperConga is to also enable users to perform advanced simulations without having to write any code, by instead only specifying simulation parameters, but without losing crucial functionality. To make this possible, SuperConga comes with a user friendly \emph{frontend}, and a set of \emph{backend interfaces}. The frontend manages and validates the user input, and then sends it to the appropriate backend interface. The backend interface takes care of importing and calling the necessary backend modules with the settings provided by the user. See Fig.~\ref{fig:usage_overview} for a schematic overview. SuperConga also comes with a set of ready-to-run examples (see Sec.~\ref{sec:demonstration} for a demonstration), a detailed user manual full of tutorials and guides,\cite{SuperConga:documentation} a set of helpful tools, a Singularity container which simplifies setup and improves portability, and an extensive testing framework to test the validity of the code. The latter ranges from smaller unit tests of basic functionality, to larger end-to-end physics test of e.g. current conservation and well-known analytic results. In the following, we give further insight into the backend and frontend, and list the most important modules.

\subsubsection{Backend}

The backend modules are implemented as a set of classes and functions via header files. The modularity is typically built up by an inheritance structure, from virtual base classes to derived classes. The virtual base classes describe the general signature or pattern for a module, but does not necessarily contain any functionality. This is instead taken care of by the derived classes, which inherits the signature from the corresponding virtual class, and implements the actual functionality (or offloads it to external libraries, see App.~\ref{app:dependencies} for a description of dependencies). For example, \texttt{BoundaryCondition.h} contains a virtual base class defining the general properties of a boundary condition. The derived class in \texttt{BoundaryConditionSpecular.h} inherits from \texttt{BoundaryCondition.h}, and implements a specular boundary condition. Similarly, there are virtual base classes for Fermi surfaces, geometric shapes, and even Riccati solvers, with specific implementations of e.g. a circular Fermi surface, disc- and polygon-geometries, and a Riccati solver for finite grains. The idea is that the modularity should allow for straightforward extension of anything from boundary conditions to new observables. Below is a list of the most important modules:
\begin{description}
    \item[Boundary conditions] Implementation of the boundary conditions to handle interface scattering of quasiparticles, the latter which follow the Eilenberger-Riccati trajectories. Currently only implements specular boundary conditions for superconductor-vacuum interfaces. 
    \item[Fermi surface] A class that takes care of the momentum-averaging over the Fermi surface. Can either treat circular Fermi-surfaces, or parametrize a more general Fermi-surface shape based on a tight-binding hopping model. 
    \item[Geometry] Classes and functionality for creating mesoscopic grains from basic shape primitives (i.e. discs and polygons).
    \item[Observables] Implementation of various observables and quantities, specifically the superconducting gap, charge current density, magnetic moment, LDOS, free energy, vector potential, and the magnetic induction. Also takes care of computing the residual for the self-consistency iteration. 
    \item[Accelerator] Implementation of self-consistency accelerators. The user can currently choose from basic Picard iterations, Polyak's\cite{Polyak:1964} "small heavy sphere", a variant of the Barzilai-Borwein method\cite{BarzilaiBorwein:1988}, and CongAcc which is an adaptive method developed for SuperConga. See App.~\ref{app:accelerator} for details and comparisons.
    \item[Order parameter] Class for setting up the order parameter, possibly with multiple components like $d+is$ and chiral $d$-wave. Note that the implementation is currently limited to spin-singlet with a single band and spin degenerate Fermi-surfaces.
    \item[Riccati Solver] Implements a solver for the Riccati equations, currently based on the mid-point method for confined (finite) geometries. To give an idea, this class could more generally be extended with e.g. a higher order Runge-Kutta solver, or other types of systems like bulk or semi-infinite superconductors.
    \item[Self-consistency solver] A collection of classes that take care of adding the contribution of the Riccati trajectories for each degree of freedom (energy, Fermi-surface angle, coordinate) to self-consistently solve the order parameter from the superconducting gap equation, together with the vector potential from the corresponding Maxwell equation.
\end{description}
See the API documentation for more modules and information.\cite{SuperConga:documentation}

SuperConga currently comes with two backend interfaces, that expose the functionality of the above modules to the user (via the frontend). The first backend interface is \texttt{simulation.cu}, which acts as the ``main'' program, essentially implementing Algorithm~\ref{alg:suprcon} to self-consistently solve the gap equation and Maxwell's equation. More specifically, \texttt{simulation.cu} starts by generating a superconducting grain and Fermi surface based on the user-provided geometry and model. The superconducting order parameter and its components are initialized in this grain, according to the specified basis functions, transition temperatures, and start guess. The program then continues to set up the self-consistency accelerator, Riccati solver, boundary conditions, observables and vector potential. An OpenGL instance is initialized if live visualization is enabled. Finally, the self-consistency loop is run until reaching either the convergence criterion or maximum number of iterations, upon which data is saved to disk. The second backend interface is \texttt{postprocess.cu}, which computes the LDOS based on results from a previous simulation (e.g. computed with \texttt{simulation.cu}). 

Note that both backend interfaces take as command-line input the location of a configuration file, which should contain all parameters necessary to completely specify the simulation. The simulation will not run if the configuration file is not found, or if the contents are invalid. The purpose of the frontend is to make it easier to set up all simulation parameters correctly, with validation and proper help messages to guide the user if anything goes wrong.

\subsubsection{Frontend}
The SuperConga frontend consists of a main run-file, \texttt{superconga.py}, which draws on an assortment of functionality implemented in a modular Python library, the latter located in the folder \texttt{frontend/}. The goal of the frontend is to facilitate setting up and analyzing simulations. 
For example, the frontend consists of Python modules for parsing command-line arguments, validating them and providing help for the user, reading and writing configuration files used by the backend, calling the correct backend interfaces, reading and visualizing simulation data, converting data between different formats, and modifying data by e.g. adding noise. The functionality is divided into a set of subcommands, used via \lstinline{python superconga.py <subcommand>}. A list of available frontend subcommands are obtained by calling \lstinline{python superconga.py --help}, and each subcommand also has its own help message. A short summary of each subcommand will now be given, and the next section demonstrates how to use them.

The first subcommand is \texttt{setup}, which is used to setup and generate the SuperConga build system via CMake.

The second subcommand is \texttt{compile}, which is used after the setup to compile the SuperConga framework. These two steps have to be performed before any simulations can be run.

The third subcommand is \texttt{simulate}, which calls the binary generated from \texttt{simulation.cu} to run self-consistent simulations. It takes as input the location of a configuration file, defining all simulation parameters. Optionally, each parameter can be set or overridden via the command-line interface (CLI). The parameters will then be validated for errors, attempting to provide the user with helpful messages if failing. If valid, the binary is called with the user-defined settings.

The fourth subcommand is \texttt{plot-simulation}, which takes the location of data from a self-consistent simulation, and plots computed quantities (see Fig.~\ref{fig:plotter} for an example). The user can control what to plot, in what order, and which settings to use for e.g. colormaps and fonts. If the user has specified that data should be saved when running \texttt{simulate}, then \texttt{plot-simulation} is automatically called, using the default settings, and a \texttt{PDF} is saved together with the data.

The fifth subcommand is \texttt{postprocess}, which calls the binary generated from \texttt{postprocess.cu} to calculate the LDOS from self-consistently converged data. The runner takes as command-line input the location of the data, and parameters specifying which resolution to compute the LDOS with. The latter parameters can either be set directly via the CLI, or specified in an external configuration file.

The sixth subcommand is \texttt{plot-postprocess}, which is a spectroscopy tool. It takes as input the path to a directory containing LDOS data computed with \texttt{postprocess}. If found, the data is read and visualized in an interactive plot (see Fig.~\ref{fig:ldos_plotter} for an example).

The seventh subcommand is \texttt{plot-convergence}, which plots the residuals of computed quantities versus iteration number. This tool can be run on convergence data from either \texttt{simulate} or \texttt{postprocess} subcommands.

The eight subcommand is \texttt{convert}, which is used to convert data back and forth between the file formats \texttt{h5} and \texttt{csv}.

Having given an overview of the SuperConga framework and its frontend, we will now demonstrate how to use it to set up simulations and do data analysis.

\section{Demonstration of SuperConga}
\label{sec:demonstration}

\begin{figure*}
	\includegraphics[width=\textwidth]{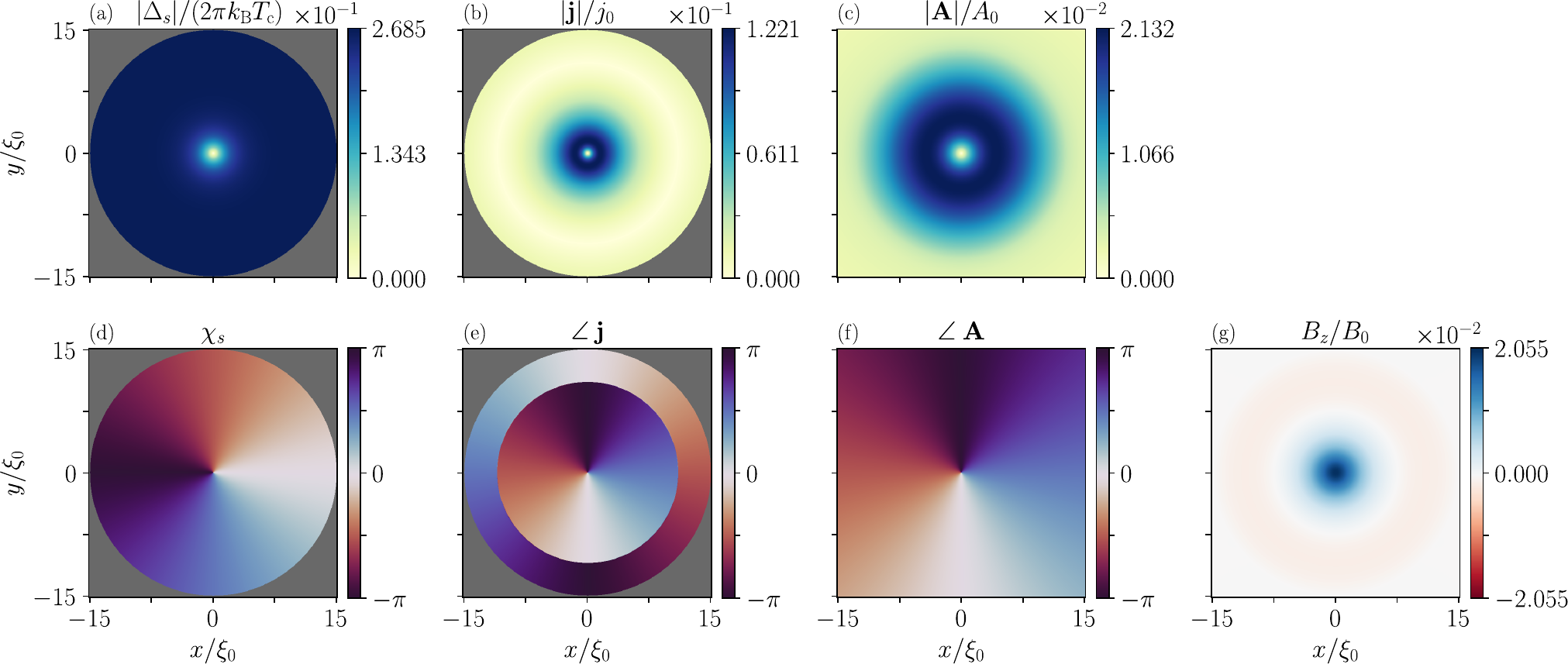}
	\caption{Plot generated with the SuperConga ``\texttt{plot-simulation}'' command, visualizing the example with a single Abrikosov vortex in a conventional superconducting disc, by following Listing~\ref{lst:running_frontend} then \ref{lst:plotter}. Here, $T=0.5\Tc$, $B_{\mathrm{ext}}=1.5\Phi_0/\mathcal{A}$, and $\kappa=5$ with full back-coupling of the magnetic gauge field. (a) The magnitude of the order parameter. (b) The magnitude of the charge-current density. (c) The magnitude of the induced vector potential. (d) The phase of the order parameter. (e) The polar angle (i.e. direction) of the charge-current density. The paramagnetic and diamagnetic regions are clearly distinguished. (f) The polar angle of the induced vector potential. (f) The induced magnetic-flux density.}
	\label{fig:plotter}
\end{figure*}
\begin{figure*}
	\includegraphics[width=\textwidth]{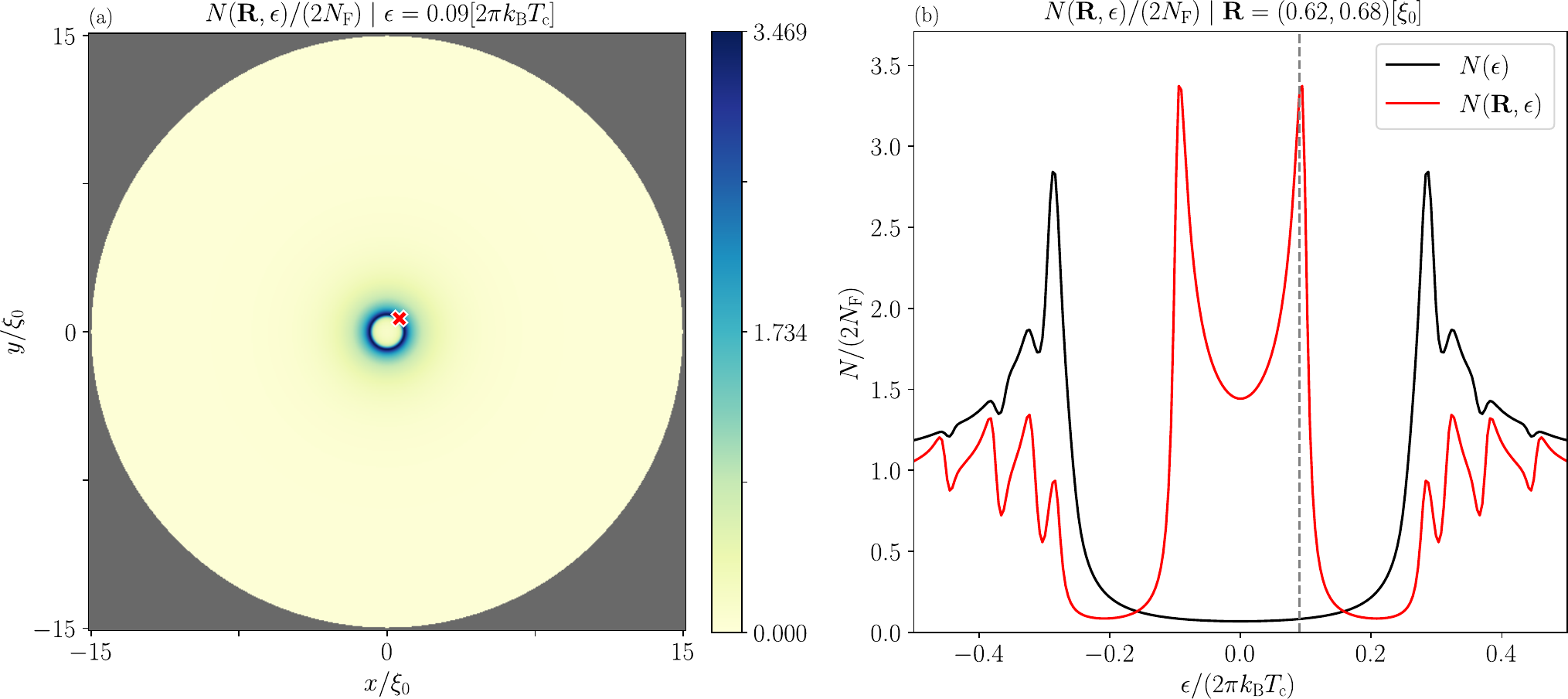}
	\caption{Screenshot of the SuperConga interactive spectroscopy command, ``\texttt{plot-postprocess}''. The plots show the local density of states (LDOS), $N(\vR, \epsilon)$, for a single vortex at the center of a conventional superconducting disc, after following Listings~\ref{lst:running_frontend}, \ref{lst:postprocessing}, and then \ref{lst:ldos_plotter} in order. Here, $T=0.5\Tc$, $B_{\mathrm{ext}}=1.5\Phi_0/\mathcal{A}$, and $\kappa=5$ with full back-coupling of the magnetic gauge field. An energy broadening of $\delta=0.008\cdot2\pi k_\mathrm{B}\Tc$ was used. (a) Heatmap of the LDOS versus spatial coordinates, at the finite energy show in the title, as well as by the vertical dashed line in panel (b). (b) LDOS (solid red curve) at the coordinates clicked by the user in the heatmap (indicated by the red cross) in panel (a), and area-averaged DOS (solid black curve). The user can change the energy shown in panel (a) by clicking in panel (b). Similarly, the user can change the spatial point being visualized in panel (b) by clicking in panel (a).
	}
	\label{fig:ldos_plotter}
\end{figure*}

The SuperConga framework ships with a few ready-to-run examples that illustrate how to set up and run simulations. These are also explained in the online documentation, together with detailed tutorials and research guides.\cite{SuperConga:documentation} One of the examples will now be shown, namely the simulation of a single Abrikosov vortex in a conventional superconducting disc. The demonstration can be split into the following generic steps:
\begin{description}
    \item[Pre-processing] setting up and defining the simulation,
    \item[Processing] running the self-consistency simulations, optionally with live visualization,
    \item[Post-processing] data analysis, visualization and spectroscopy.
\end{description}
The simulation configuration for the vortex example can be found in \texttt{examples/swave\_disc\_vortex/} in the SuperConga main folder.\cite{SuperConga:repository} Appendix~\ref{app:json-file} lists the full contents of this file, with a detailed explanation of the format. In short, each configuration file contains all necessary parameters needed to completely specify a simulation. The parameters are grouped into different sections: Listing~\ref{lst:abrikosov_example:physics} shows the group of ``physics'' parameters for the Abrikosov example, specifying the temperature $T=0.5\Tc$, the external flux $\Phi_{\mathrm{ext}} = 1.5\Phi_0$, the penetration depth $\lambda=5\xi_0$, and an $s$-wave order parameter with a phase winding of $-2\pi$ as a start guess.
\begin{lstlisting}[language=json,basicstyle=\small,caption={Part of \texttt{simulation\_config.json} in \texttt{examples/swave\_disc\_vortex/}. },label={lst:abrikosov_example:physics}]
"physics": {
  "temperature": 0.5,
  "external_flux_quanta": 1.5,
  "penetration_depth": 5.0,
  "crystal_axes_rotation": 0.0,
  "gauge": "symmetric",
  "charge_sign": -1,
  "order_parameter": {
    "s": {
      "critical_temperature": 1.0,
      "initial_phase_shift": 0.0,
      "initial_noise_stddev": 0.0,
      "vortices": [
        {
          "center_x": 0.0,
          "center_y": 0.0,
          "winding_number": -1.0
        }
      ]
    }
  }
}
\end{lstlisting}
Other parameter groups deal with for example the geometry, numerical accuracy, convergence criteria, convergence acceleration, and so on. The parameters can also be set or changed directly via command line, effectively overriding the settings in the configuration file, as shown below. Assuming that the framework has been properly installed and compiled (see the online documentation and installation guide\cite{SuperConga:repository}) the configuration file can be used to start the simulation by running the following from the root directory of SuperConga:
\begin{lstlisting}[language=bash,linewidth=\columnwidth,breaklines=true,caption={Running the example with a single Abrikosov vortex in a conventional superconducting disc.},label={lst:running_frontend}]
python superconga.py simulate -C examples/swave_disc_vortex/simulation_config.json
\end{lstlisting}
which should run a few tens of iterations until the convergence criterion in \texttt{simulation\_config.json} is fulfilled, printing simulation status for each iteration to terminal. Here, the flag \lstinline{-C} is used to specify the relative path to the configuration file. Adding the arguments ``\lstinline{-T 0.1 -B 10.0}'' will change the temperature to $0.1\Tc$, and the external flux to $10\Phi_0$. To start with an order parameter from a file, corresponding to e.g. a previous simulation or an arbitrary computed guess, the \lstinline{-L} argument can be used. Using the flag \lstinline{--help} gives further information about which parameters are available and their usage.

We note that live visualization can be turned on or off with the flags \lstinline{--visualize} and \lstinline{--no-visualize}, respectively. The main purpose of the live-visualization is to get an overview of the simulation progress. Thus it is geared towards speed rather than producing production-ready plots. For the latter, we instead recommend to visualize the fully converged results, via the data files generated by the program. The user can either use their favorite plotting tool, or use SuperConga as described in the following. Having followed the example in Listing~\ref{lst:running_frontend} the Abrikosov vortex results can be visualized by running the following from the root directory:
\begin{lstlisting}[language=bash,linewidth=\columnwidth,breaklines=true,caption={Plotting all spatially dependent quantities in the Abrikosov vortex example.},label={lst:plotter}]
python superconga.py plot-simulation -L data/examples/swave_disc_vortex
\end{lstlisting}
This plots all the computed quantities, as shown in Fig.~\ref{fig:plotter}. Various properties of the plot can be controlled via command-line arguments (as described by the help message), like fonts, colormaps, which quantites are plotted, and if saving the plot directly to file rather than drawing in a window. The latter makes it easy to automatically generate plots from a large number of simulations.

The local density of states (LDOS) can be calculated from converged data as a post-processing step, by calling:
\begin{lstlisting}[language=bash,linewidth=\columnwidth,breaklines=true,caption={Computing the LDOS for the Abrikosov vortex example.},label={lst:postprocessing}]
python superconga.py postprocess -C examples/swave_disc_vortex/postprocess_config.json
\end{lstlisting}
where the contents of \texttt{postprocess\_config.json} is shown in Listing~\ref{lst:abrikosov_example:ldos_json}.
\begin{lstlisting}[language=json,basicstyle=\small,caption={Configuration file used for computing (postprocessing) the LDOS: \texttt{postprocess\_config.json} in \texttt{examples/swave\_disc\_vortex/}.},label={lst:abrikosov_example:ldos_json}]
{
  "spectroscopy": {
    "energy_max": 0.5,
    "energy_min": 0.0,
    "energy_broadening": 0.008,
    "num_energies": 128
  },
  "numerics": {
    "convergence_criterion": 1e-4,
    "norm": "l2",
    "num_energies_per_block": 32,
    "num_fermi_momenta": 720,
    "num_iterations_burnin": -1,
    "num_iterations_max": 1000,
    "num_iterations_min": 0
  },
  "misc": {
    "data_format": "h5",
    "load_path": "data/examples/swave_disc_vortex",
    "save_path": "",
    "verbose": true
  }
}
\end{lstlisting}
This file specifies e.g. which energies and resolution to use in the LDOS calculation, and where the order-parameter data is located. The empty save path indicates that the LDOS data will be saved to the same directory as the order parameter data.

The LDOS can be particularly tricky to visualize due to the high dimensionality. To aid with this, the framework provides a spectroscopy tool in the form of an interactive LDOS visualizer. Assuming that the vortex LDOS has been calculated with the above post-processing example in Listing~\ref{lst:postprocessing}, vortex spectroscopy can be done by running the following from the root directory: 
\begin{lstlisting}[language=bash,linewidth=\columnwidth,breaklines=true,caption={Performing vortex spectroscopy.},label={lst:ldos_plotter}]
python superconga.py plot-postprocess -L data/examples/swave_disc_vortex
\end{lstlisting}
Figure~\ref{fig:ldos_plotter} shows a screenshot of the interactive LDOS plotter for this example, illustrating that the user can click in the window to choose which energy to plot the LDOS at (as a 2D-heatmap versus coordinates), and which point in space to plot the LDOS versus energy at (as a 1D curve).

\begin{figure}[th!]
	\includegraphics[width=1\columnwidth]{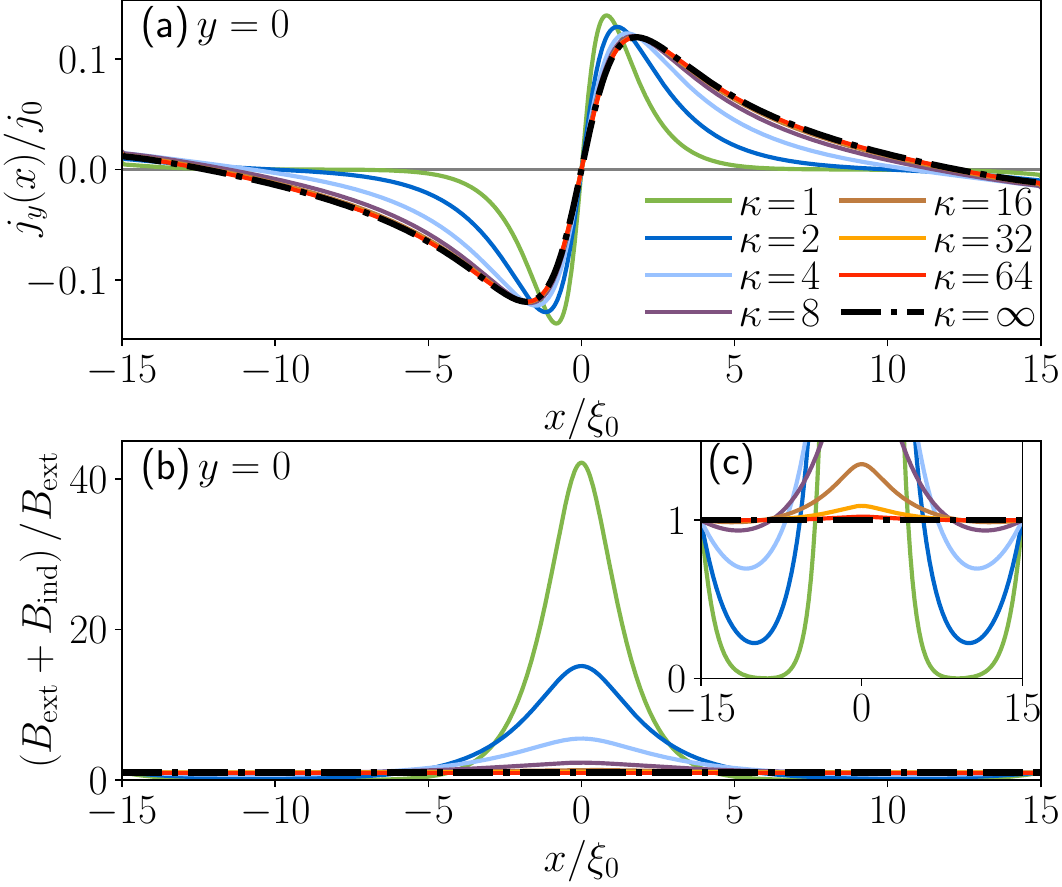}
	\caption{Spatial dependence of the (a) charge-current density, and the (b) total magnetic-flux density, as distance from a vortex core. The system is an $s$-wave disc with radius $\mathcal{R}=15\xi_0$ at temperature $T=0.5\Tc$, and external flux $\Phi_{\mathrm{ext}} = 1.5\Phi_0$ (the same as in Figs.~\ref{fig:plotter}--\ref{fig:ldos_plotter}). Different lines correspond to different values of $\kappa$, as indicated by the legend. Panel (c) is a zoom of (b), illustrating more clearly the diamagnetic and paramagnetic regions. Due to the small radius, the system is only partially screened. Full screening ($B_{\mathrm{ext}} + B_{\mathrm{ind}} = 0$) is only achieved in a narrow region for $\kappa=1$. Increasing the ratio $\mathcal{R}/\lambda$ will increase the region of full screening.}
	\label{fig:vortex_screening}
\end{figure}

To extract the temperature dependence, or the dependence on any other parameter for that matter, the most straightforward approach is to write a script which calls the program multiple times, but with unique values of the parameter. A few such examples are included in SuperConga, and the following one illustrates how to simulate an Abrikosov vortex for different values of Ginzburg-Landau coefficient $\kappa$, throughout the whole Type-II range $\kappa \in [1,\infty)$:
\begin{lstlisting}[language=bash,linewidth=\columnwidth,breaklines=true,caption={Simulating an Abrikosov vortex, for various values of the Ginzburg-Landau coefficient.},label={lst:running_sweeps}]
./examples/parameter_sweeps/swave_disc_vortex_kappa_sweep.sh
\end{lstlisting}
The results, seen in Fig.~\ref{fig:vortex_screening}, can subsequently be visualized via:
\begin{lstlisting}[language=bash,linewidth=\columnwidth,breaklines=true,caption={Plotting the vortex dependence on penetration depth.},label={lst:plotting_sweeps}]
python examples/parameter_sweeps/plot_vortex_kappa_sweep.py
\end{lstlisting}

So far, the demonstration has dealt with a superconducting disc. SuperConga allows the user to specify more general composite geometries by successively adding and removing discs and polygons. Both regular and free-form polygons are implemented, enabling quite intricate systems that e.g. has holes, are multiply connected, or even disconnected but coupled via induction. For example, Fig.~\ref{fig:riccati_fig_A} was created in this way by setting the \lstinline{"geometry"} parameter section of the configuration file. Listing~\ref{lst:geometry_example:json} shows a SQUID-like geometry, created by adding a central region shaped like an octagon, two arms in the form of a rectangle, and then removing a disc from the the central region:
\begin{lstlisting}[language=json,basicstyle=\small,caption={Example setup of a composite geometry.},label={lst:geometry_example:json}]
"geometry": [
  {
    "polygon": {
      "add": true,
      "vertices_x": [0.0, 0.0, 100.0, 100.0],
      "vertices_y": [5.0, -5.0, -5.0, 5.0]
    }
  },
  {
    "regular_polygon": {
      "add": true,
      "center_x": 50.0,
      "center_y": 0.0,
      "num_edges": 8,
      "rotation": 0,
      "side_length": 15.0
    }
  },
  {
    "disc": {
      "add": false,
      "center_x": 50.0,
      "center_y": 0.0,
      "radius": 8.0
    } 
  }
]
\end{lstlisting}
In Listing~\ref{lst:geometry_example:json}, the entry \lstinline{"geometry"} consists of a list which specifies in which order (from top to bottom) the shapes should be added or removed, the latter indicated by \lstinline{"add": true} or \lstinline{"add": false}, respectively. All coordinates and lengths are given in units of $\xi_0$, and the user can choose whichever origin they prefer, as it does not influence the physics. The disc is uniquely defined via its center coordinates and radius, a regular polygon by center coordinates, number of edges, rotation (in units of $2\pi$), and side-length. For free-range polygons, a list of $x$- and $y$-coordinates for the vertices are specified (in counter-clockwise order). Currently, the free-range polygons are limited to convex hulls. Note that an arbitrary number of each shape in principle can be listed, which was previously used to e.g. study mesoscopic roughness by generating boundaries with random removal of polygons.\cite{Holmvall:2019} Note however, that for extremely large number of added/removed shapes, the performance might eventually be reduced due to having to deal with a large set of complicated trajectories and boundary conditions.

\begin{figure*}[th!]
	\includegraphics[width=1.0\textwidth]{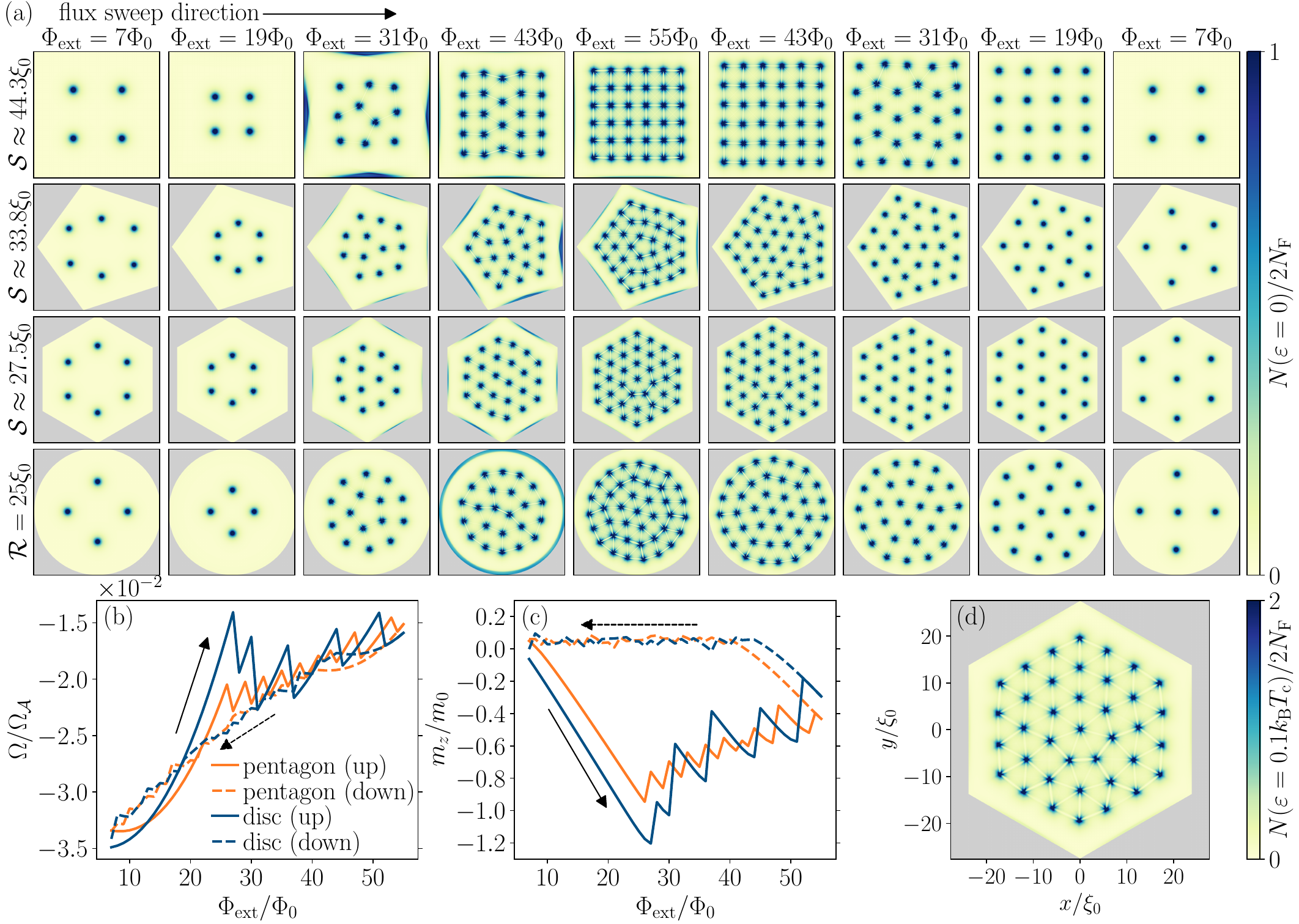}
	\caption{(a) Mesoscopic vortex lattices in conventional $s$-wave superconducting grains of different shapes (rows), exposed to different external fluxes (columns) in a directed sweep.
	We set $\kappa=10$ and $T=0.1\Tc$, and the grain side-lengths ($\mathcal{S}$) was chosen such that the area is the same as that of the disc, $\mathcal{A} = \pi\mathcal{R}^2$.
	Colors indicate the LDOS at zero energy, with localised Caroli-de Gennes-Matricon states\cite{Caroli:1964,Berthod:2017,Howon:2021} in each vortex core. To give the vortex cores a visible extent, a smearing of $\delta = 0.008 \cdot 2\pi \kB\Tc$ was used, and an arbitrary cutoff of $2\NF$ was introduced.
	The plots illustrate that vortex lattices in mesoscopic superconductors can be highly modified by finite-size effects, such as the grain shape. In particular, as the external field increases, the vortices are packed into a tighter configuration, with the introduction of vortex shells, and depending on the flux-sweep strategy, vortex-lattice "dislocations" might enter the system. In this case, the flux is varied in steps of $\Phi_0$ from low fluxes (left-most column) to high fluxes (middle column), and then from high fluxes to low fluxes again (right-most column). For each new value of the flux, the converged results from the previous simulation is used as a start guess. The different sweep directions lead to different local minima in the free energy. We note that far from the free-energy minimum, when there are few vortices compared to the external flux, there is significant pair-breaking and zero-energy states due to large screening currents at the edges of the system, see e.g. square in the third column.
	Panels (b) and (c) show how the free energy and magnetic moment evolve using this directed-sweep strategy, respectively, for the pentagon (orange) and disc (blue). Full lines are for sweeping towards higher fields and dashed for sweeps in the opposite direction. \ This asymmetry in the free energy versus applied field, between the two sweep directions, can be interpreted as a Bean-Livingston barrier that makes vortex entry energetically harder than vortex exit.\cite{Bean:1964}
	(d) LDOS at a finite energy $\varepsilon = 0.9 \kB\Tc$, at external flux $\Phi_{\mathrm{ext}} = 55\Phi_0$.
	}
	\label{fig:VortexShell}
\end{figure*}

It is straightforward to extend the above example of a single Abrikosov vortex, to vortex lattices, by increasing the external field. As an example,
Fig.~\ref{fig:VortexShell} plots the zero-energy LDOS in nanoscale grains of various geometries exposed to high external fields. Vortex cores show up as dot-like dark spots,
illustrating vortex lattices with various mesoscopic properties. In contrast to bulk lattices, the number of vortices and their separation in a finite system does not only
depend on the external field strength, but also on e.g. the boundary conditions, the system size relative to the penetration depth, and the shape of the system. While bulk
lattices are usually triangular, a mesoscopic vortex lattice can mimic the shape and symmetry of the superconducting grain.  In addition, rather than entering one by one as
the external field is increased, several vortices might suddenly enter at a time, sometimes as concentric shells (hence referred to as the vortex-shell effect). These
effects have been observed e.g. in grains shaped like discs,\cite{Geim:1997,Grigorieva:2006,Kokubo:2010} squares,\cite{Grigorieva:2006,Zhao:2008_a,Misko:2009,Zhang:2012,Zhang:2013,Timmermans:2016} triangles,\cite{Grigorieva:2006,Zhao:2008_b,Cabral:2009,Kokubo:2016,Wu:2017} and
pentagons.\cite{Huy:2013} Such scenarios are great examples of when a sufficiently sophisticated simulation tool like SuperConga is crucial, since one needs to capture the
combined effects of a confined geometry and a finite penetration depth with back-coupling of the induction. In particular, one needs to use a powerful convergence
accelerator that can properly traverse the phase space. For example, we note that with the proper choice of start guess and accelerator settings, it is in principle possible
to study giant vortices,\cite{Fink:1966,Moshchalkov:1996,Schweigert:1998,Kanda:2004,Golubovic:2005,Cren:2011} as well as vortex-antivortex
pairs.\cite{Chibotaru:2000,Zhang:2012,Zhang:2013,Iavarone:2011,Giorgio:2017,Simmendinger:2020} Careful treatment and use of the adaptive convergence accelerators show, however, that these structures are generally unstable in non-hetero-structures that are sufficiently clean, where they decay to more traditional vortex structures with a lower free energy.

\section{\label{sec:examples} Examples and results}
In this section we present several examples of studies that are quite straight forward to perform with the framework SuperConga. In a few special cases it is possible to find analytic results, which enable us to check the overall correctness of the framework. First, we study a 2D superconducting annulus. In the case of an $s$-wave superconductor we may compare with analytics, while in the $d$-wave superconducting case we demonstrate the corrections due to suppression of the order parameter at the edges and the formation of spontaneous currents at low temperature.\cite{Vorontsov:2009,Hakansson:2015,Holmvall:2017,Holmvall:2018a,Holmvall:2018b,Holmvall:2019,Holmvall:2019b,Holmvall:2020} Second, we study a superconducting disc also subjected to an external magnetic field. Here we return to the well-studied problem of vortex lattice formation, which has a long history going back to the original work of Abrikosov,\cite{Abrikosov:1957} and in particular to that of Pearl considering superconducting discs.\cite{Pearl:1964} For the cases of zero or one vortex at the center of the disc, we compare with analytic results. Then we continue with higher fields, where we have to resort to a numerical treatment. Finally, we simulate irregular superconducting islands in an external magnetic field. The study is inspired by experimental work done by Timmermans {\em et. al.} \cite{Timmermans:2016} on spectroscopy on superconducting nanostructures assembled by small squares of Mo$_{79}$Ge$_{21}$.

\subsection{Superconducting annulus in a magnetic field}
We begin by studying homogeneous superflow in a superconductor. Superflow implies that there is a finite superfluid momentum caused by phase gradients and/or a vector potential, as defined in \Neqref{eq:phase}. The fact that the phase gradient enters in the superfluid momentum becomes clear when doing a gauge transformation. In quasiclassical theory this is done as follows. We start with the order-parameter $\Delta(\vpF,\vR)=\Delta(\vR)\eta(\vpF)\exp[i\chi(\vR)]$, which in general is complex valued. The order-parameter matrix can then be decomposed as
\be 
\hat\Delta(\vpF,\vR)=\hat U(\chi) \hat\Delta_0(\vpF,\vR) \hat U^\dagger(\chi),
\label{eq:gaugetransformation}
\ee
with the transformation matrix
\be
\hat U(\chi)=\left(\begin{array}{cc}
    \mbox{e}^{i \chi(\vR)/2}& 0 \\
    0 &\mbox{e}^{-i \chi(\vR)/2}
\end{array}
\right)
= e^{i\chi(\RR)\hat\tau_3/2},
\ee
and $\Delta_0(\vR)$ a purely real amplitude (i.e. $\chi_0(\vR)\equiv 0$ while the basis function $\eta(\vpF)$ may still be
complex valued).
Applying the same transformation to the Eilenberger \Neqref{eq:Eilenberger}, suppressing the arguments of $\hat g$ and $\hat \Delta$, we see that if $\hat g$ solves
\be 
\qcgrad \hat g+[(i\epsilon_n+\tfrac{e}{c}\,\vvF\cdot\vA)\hat \tau_3-\hat\Delta,\hat g]=0,
\ee
with $\hat \Delta=\hat U \hat \Delta_0 \hat U^\dagger$, then $\hat g_0=\hat U^\dagger \hat g U$ is a solution to
\be 
\qcgrad \hat g_0+[(i\epsilon_n +\tfrac{e}{c}\,\vvF\cdot\vA - \tfrac{1}{2}\vvF\!\cdot\!\nabla \chi) \hat \tau_3-\hat\Delta_0,\hat g_0]=0.
\ee
The superfluid momentum $\vps$ in \Neqref{eq:phase} is now naturally formed and we get the Eilenberger equation in the form
\be 
\qcgrad \hat g_0+[(i\epsilon_n -\vvF\cdot\vps) \hat \tau_3-\hat\Delta_0,\hat g_0]=0,
\ee
with the purely real order-parameter matrix.

Let us look at the linear response to a small and spatially homogeneous superfluid momentum, $\ps\ll\Delta/\vF$. To do this, we write the propagator as a perturbation expansion $\hat g=\hat g_0+\delta \hat g +{\cal{O}}[\ps^2]$. Using the normalisation condition on the propagator $\hat g^2=-\pi^2 \hat{ \mathbb{I}}$ we get
\begin{equation*}
    \hat g_0^2=-\pi^2 \hat{\mathbb{I}} \quad\,\mbox{and}\quad\, \lbrace\hat g_0,\delta \hat g\rbrace=0.
\end{equation*}
Since $\hat g_0\propto i\epsilon_n\hat\tau_3-\hat \Delta_0(\vpF)$, we get via the Eilenberger equation the solution
\begin{equation}
    \delta \hat g(\vpF;\epsilon_n)=\pi\, \frac{\vvF\!\cdot\!\vps}{\Lambda_n^{3}(\vpF)}
    \left(\begin{array}{cc}
        \vert\Delta(\vpF)\vert^2 & -i\epsilon_n \Delta(\vpF) \\
        i\epsilon_n \Delta^*(\vpF) & -\vert\Delta(\vpF)\vert^2
    \end{array}\right),
\end{equation}
defining $\Lambda_n(\vpF)=\sqrt{|\Delta(\vpF)|^2+\epsilon_n^2}$. From the form of $\delta \hat g$ we can directly read out that the linear correction to the anomalous Green's function component, $f(\vpF;\epsilon_n)$, is odd in frequency and will not give a correction in leading order in $\ps$ to the order parameter $\Delta(\vpF)$. The diagonal part of $\delta \hat g$ allows us to calculate the current to linear order in $\ps$ as
\begin{equation}
    \delta \vj=2 e \NF \langle \vvF(\vpF) \yosh(\vpF)  \vvF(\vpF)\!\cdot\!\vps
     \rangle_{\vpF},
     \label{linearcurrent}
\end{equation}
with the angle-dependent Yoshida function
\begin{equation*}
\yosh(\vpF)= \pi T\sum_n \frac{|\Delta(\vpF)|^2}{\Lambda_n^3(\vpF)}.
\end{equation*}
Equation (\ref{linearcurrent}) is usually written in the well-known form 
\begin{equation}
    \vjs=e \langle \bar{\rho}_s(\vpF)\cdot \vps\rangle_{\vpF}\; , 
    \label{eq:sfdensity}
\end{equation}
defining 
\begin{equation}
\bar{\rho}_{\rm{s}}(\vpF)=2\NF \yosh(\vpF) \vvF(\vpF) \vvF^\top(\vpF),
\end{equation}
the angle-dependent superfluid-density, or superfluid stiffeness tensor.\cite{Xu:1995}

As a final step we write the change in the free energy in linear response as
\begin{equation}
\Omega = \Omega_\mathrm{bulk} + \delta \Omega_\mathrm{kin} + \delta \Omega_\mathrm{magn} \; ,
\label{FEsuperflow}
\end{equation}
where
\begin{equation}
    \Omega_{\mathrm{bulk}}={\cal{A}} \NF\Bigg\langle|\Delta(\vpF)|^2\ln \frac{T}{T_c}
    +\pi T\sum_n \frac{|\Delta(\vpF)|^4}{|\epsilon_n|\left(\Lambda_n(\vpF)+\epsilon_n\right)^2}\Bigg\rangle_{\vpF},
    \label{FEbulk}
\end{equation}
is the bulk superconducting free energy (see e.g. Ref.~[\onlinecite{Serene:1983}]), and $\cal{A}$ is the area of the superconducting sample considered. The change in free energy in \Neqref{FEsuperflow} due to superflow consists of a kinetic contribution\cite{Holmvall:2020}
\begin{equation}
    \delta \Omega_{\mathrm{kin}}=\frac{1}{2}\,\int \mathrm{d}{\vR}
    \bigg\langle\vps^\top(\vR)\bar{\rho}_s(\vpF)\vps(\vR)\bigg\rangle_{\vpF},
    \label{lFEkin}
\end{equation}
and a magnetic contribution
\begin{equation}
    \delta \Omega_{\mathrm{magn}}=\frac{1}{8\pi}\,\int \mathrm{d}{\vR}\left| \nabla\times\vA(\RR)-\vB_\mathrm{ext}
    \right|^2,
    \label{lFEmag}
\end{equation}
where $\vB_\mathrm{ext}$ is the applied external magnetic field.

\subsubsection{The solenoid gauge}

\begin{figure}[tb!]
	\includegraphics[width=\columnwidth]{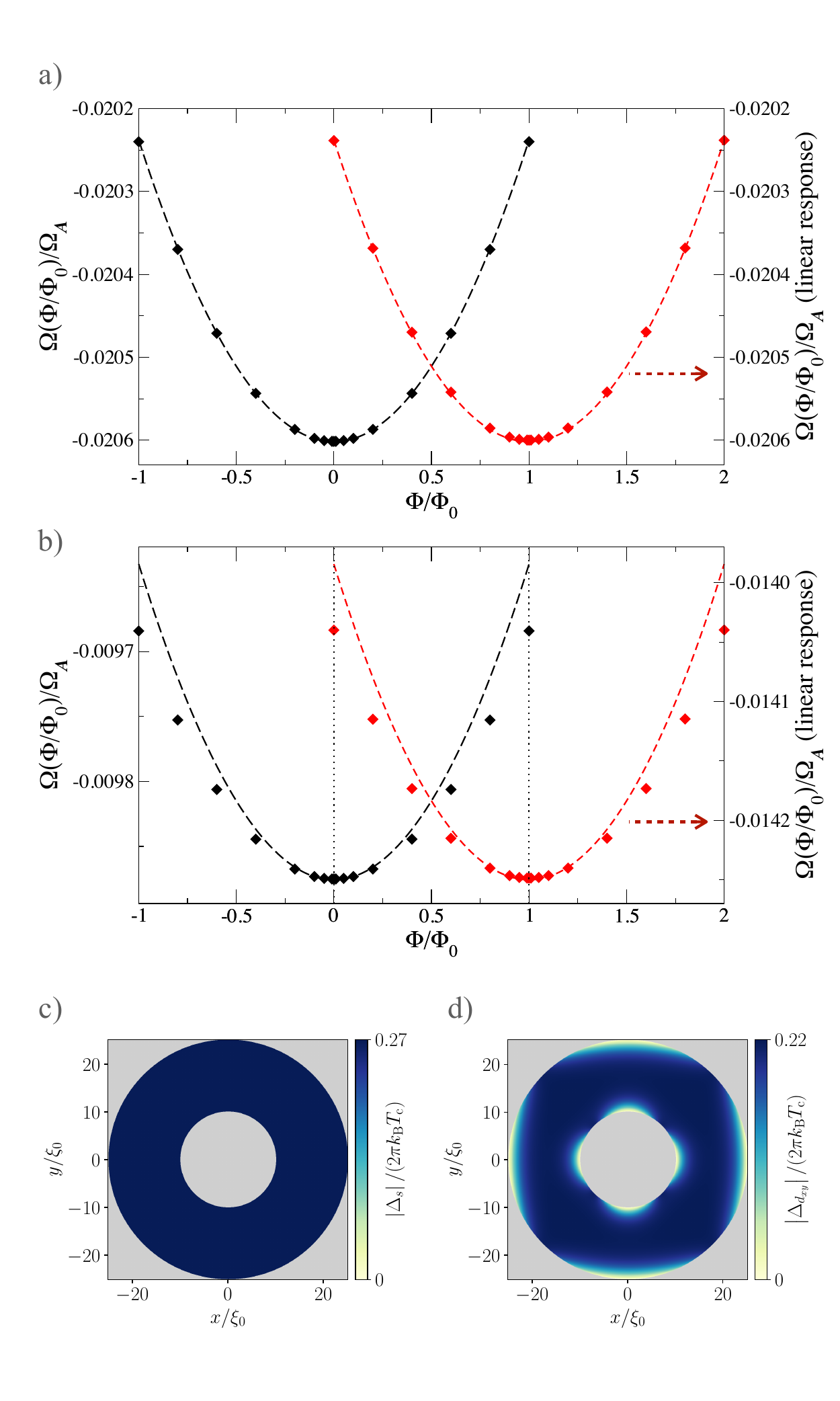}
	\caption{The magnetic flux dependence of the free energy of an $s$-wave (a) and a $d_{xy}$-wave (b) superconducting annulus using the solenoid gauge \Neqref{eq:solenoid_gauge}. The outer (inner) radius of the annulus is $\mathcal{R}_> = 25 \xi_0$ ($\mathcal{R}_< = 10\xi_0$), and the temperature is set to $T=0.5T_{\mathrm{c}}$. The order-parameter amplitude for the two cases are shown in (c), $s$-wave, and (d), $d$-wave. In (a) and (b) the diamonds are results for the free energy extracted from SuperConga with the scale on the left side $y$-axis. The dashed lines is the free energy given by \Neqref{FEbulk} and \Neqref{eq:FElinearresp} with scale on the right side $y$-axis. The black (left) and red (right) parabolas correpond to enforced superconducting phase windings of $n=0$ and $n=-1$.}
	\label{fig:solenoid_flux_sweep}
\end{figure}
With the above introduction to superflow in a superconductor, we are ready to study a superconducting annulus with outer (inner) radius $\mathcal{R}_>$ ($\mathcal{R}_<$), as illustrated in Fig.~\ref{fig:solenoid_flux_sweep}(c). The area of the annulus is $\mathcal{A}=\pi(\mathcal{R}_>^2-\mathcal{R}_<^2)$. Let us first consider the following text-book vector potential\cite{Sakurai:2011} used to illustrate the Aharonov-Bohm effect:\cite{Aharonov:1959}
\be
\vA_\mathrm{ext}(\vR)=\frac{\Phi_\mathrm{ext}}{2\pi R}\hat{\mathbf{\varphi}}\;,\; R \in [\mathcal{R}_<, \mathcal{R}_>] \; ,
\label{eq:solenoid_gauge}
\ee
where $\hat{\mathbf{\varphi}}$ is the unit vector along the azimuthal direction and $\Phi_\mathrm{ext}$ is an externally applied magnetic flux threading only the hole.
We call this particular gauge the solenoid gauge. It can be seen as a mathematical idealisation to guarantee zero external magnetic flux through the superconducting annulus, while the vector potential remains finite everywhere. One possible physical realization of this gauge would be to let the superconducting annulus encircle a strong magnet.\cite{Rybakov:2022}

Using the gauge transformation \Neqref{eq:gaugetransformation} we can impose a phase winding in the order-parameter field going around the annulus. One of the constraints on an order parameter is that it is single valued. This condition leads to that $\chi$ in \Neqref{eq:gaugetransformation} must be chosen so that the total winding going around the void is a multiple $n$ of $2\pi$, and we can write $\chi=n\varphi$ where $n$ can take any integer value. If we now look at \Neqref{eq:phase} we get, reinstating $\hbar$ and $c$,
\be
\vps
=\frac{\hbar}{2}\nabla \chi -
\frac{e}{c} \frac{\Phi_\mathrm{ext}}{2\pi R}\hat{\mathbf{\varphi}}
= \frac{\hbar}{2R}\left(n + \frac{\Phi_\mathrm{ext}}{\Phi_0} \right) \hat{\mathbf{\varphi}} \; .
\ee
We see that $\vps$ can be made to vanish for certain values of the external flux, $\Phi_\mathrm{ext}=-n \Phi_0$, i.e., the superfluid momenta vanishes throughout the superconducting annulus for every external field that matches integer multiples of the flux quantum $\Phi_0$. At those matching fields $\delta\Omega_\mathrm{kin}=0$. As a consequence of the gauge properties we see that if we have a self-consistent solution for the pair $\hat g_n,\hat \Delta_n$ at a given magnetic flux $\Phi_\mathrm{ext}$ threading the hole, we can, via an appropriate gauge transformation $\hat U$, construct a solution at any other external flux that differ from $\Phi_\mathrm{ext}$ by an integer number of flux quanta. This setup nicely demonstrates the fundamental concept of flux quantization.\cite{Deaver:1961,Doll:1961,Onsager:1961,Byers:1961,Brenig:1961} Note that the negative sign of the phase winding $n$ reflects our choice of unit, $e=-|e|$, where the particle probability current flows with the gradient of the phase, while the associated charge current is in the opposite direction, see \Neqref{eq:sfdensity}. 

\paragraph{An $s$-wave superconducting annulus in the solenoid gauge.} 
The simplest $s$-wave superconductor has an isotropic order parameter with $\Delta(\vpF)=\Delta$, where $\Delta$ is a complex scalar quantity and the basis function $\eta(\vpF)=1$. For such an order parameter, scattering off sample surfaces leaves the superconducting state unaffected. As a consequence, linear-response theory will be sufficient to describe the annulus up to quite large superfluid momenta. We can thus calculate the kinetic contribution to the free energy \Neqref{lFEkin}, using the homogeneous solution of \Neqref{eq:gapeq_regularized}, to get
\begin{align}
    \frac{\delta\Omega_{\mathrm{kin}}}{(2\pi T_c)^2\,\NF} &=& \frac{\pi}{4} \xi_0^2 \,\yosh(T)
    \ln \bigg(\frac{\mathcal{R}_>}{\mathcal{R}_<}\bigg) \left(n+\frac{\Phi_\mathrm{ext}}{\Phi_{0}}\right)^2,
    \label{eq:FElinearresp}
\end{align}
after integrating over the superconducting area and setting
\begin{equation}
    \yosh(T)=\langle \uvvF(\vpF) \yosh(\vpF;T) \uvvF(\vpF)\cdot \uvps\rangle_{\vpF}
\end{equation}
where $\uvvF(\vpF)$ and $\uvps$ are unit vectors.
For this particular gauge, we let the penetration depth $\lambda\rightarrow\infty$ and neglect screening effects. This leads to that the free-energy contribution $\delta\Omega_\mathrm{magn}$ is zero. The free energy comes in a family of parabolas as function of the external flux. Each parabola is centered around $-n\Phi_0$ where $n$ is the integer flux quantization, or phase winding, enforced on the order parameter. This is clearly shown in Fig.~\ref{fig:solenoid_flux_sweep} (a), where we also show results obtained by direct calculation using SuperConga, for phase windings $n=0$ and $n=-1$. The bottom of each parabola corresponds to the free energy $\Omega_\mathrm{bulk}(T)$ at zero external flux for the considered temperature.

We finally mention that at temperatures close to the transition temperature, $T_\mathrm{c}$, $\delta\Omega_{\mathrm{kin}}$ can made to exceed the energy gain $\Omega_{\mathrm{bulk}}$ when the external flux is close to $(n+\frac{1}{2})\Phi_0$. 
This is a manifestation of the Little-Parks effect,\cite{Little:1962,Groff:1968,Buisson:1990,Tinkham:Book} where one can induce a field-modulation of the critical temperature of a superconducting film enclosing a hole.

\begin{figure}[th!]
	\includegraphics[width=0.95\columnwidth]{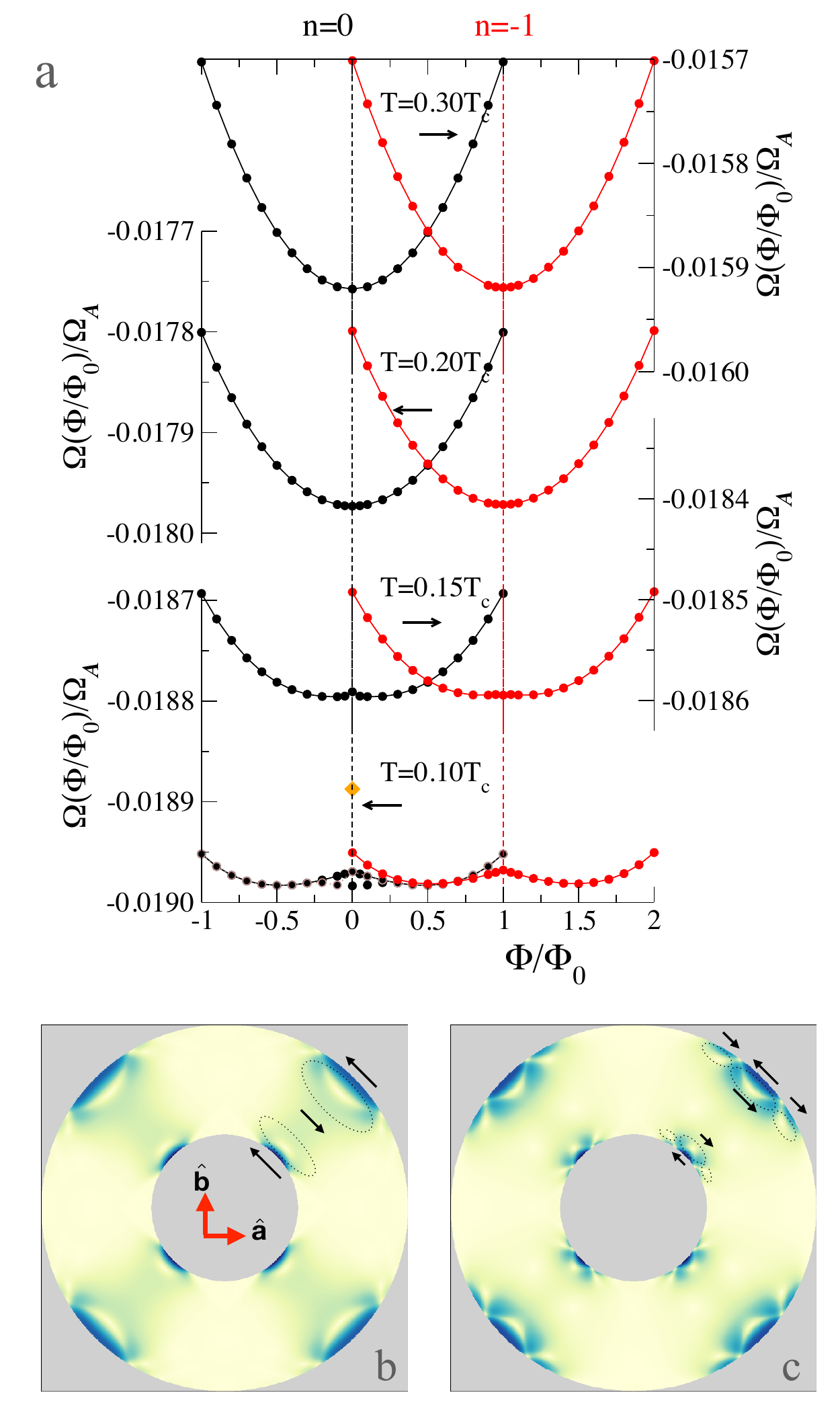}
	\caption{(a) The free energy of a $d_{x^2-y^2}$ superconducting annulus as function of external flux using the solenoid gauge. The outer (inner) radius of the annulus is $\mathcal{R}_> = 25 \xi_0$ ($\mathcal{R}_< = 10\xi_0$). The penetration depth is assumed to be much larger than the coherence length so that screening can be neglected. We look at two cases with and without a phase winding $2\pi$ around the annulus. In panel (a), from top to bottom, we lower the temperature from $T=0.3T_\mathrm{c}$ down to $T=0.1 T_\mathrm{c}$. In panels (b)--(c) we show the configuration of spontaneous currents that signifies the two low-energy states at zero magnetic field. It is the current configuration showed in (c) that has the lowest free energy. The magnitudes of the current density are at maximum $\lesssim 0.1 j_0$. The orange diamond in panel (a) is a special meta-stable state with no spontaneous currents.}
    \label{fig:d-wave_solenoid}
\end{figure}

\paragraph{A $d$-wave superconducting annulus in the solenoid gauge.}
Next, we change the superconductor to be a $d$-wave superconductor. The order-parameter field for the $d$-wave depends on position on the Fermi surface, $\vpF$, via the basis function, either the $d_{x^2-y^2}$ or $d_{xy}$ as listed in Table~\ref{table:basis_functions}.
This  momentum dependence leads to that scattering off a surface may be pair breaking and one must solve for the spatial dependence of the order parameter using \Neqref{eq:gapeq_regularized}.

As for the $s$-wave superconductor, the free energy in an external magnetic field come in a family of parabolas $\sim(n+\Phi_\mathrm{ext}/\Phi_0)^2$. However, as shown in Fig.~\ref{fig:solenoid_flux_sweep}(b), neither $\Omega_{\mathrm{bulk}}(T)$ nor $\delta \Omega_{\mathrm{kin}}(T,n,\Phi_\mathrm{ext})$, are captured correctly by a calculation with a homogeneous order parameter magntitude and superflow treated in linear response. The order-parameter magnitude of a $d_{xy}$-wave superconducting annulus is shown in Fig.~\ref{fig:solenoid_flux_sweep}(d). As seen there are regions near the surfaces where the magnitude is severely reduced, which leads to correction to $\Omega_\mathrm{bulk}(T)$.

The pair-breaking at interfaces that can be found in unconventional superconductors is due to low-energy Andreev states that are localised in a region of a few coherence lengths width at the surface.\cite{Hu:1994} These states modify low-temperature properties of the superconductor. A disc-shaped unconventional superconductor was studied by Suzuki and Asano finding a paramagnetic instability at low temperatures as a response to a small magnetic field.\cite{Suzuki:2014} We find similar paramagnetic behavior for the annular superconductor at low temperatures and relate the effects to spontaneous generation of phase gradients,\cite{Vorontsov:2009,Hakansson:2015} in a state known as a ``phase crystal''.\cite{Holmvall:2020,Wennerdal:2020,Chakraborty:2022} 

In Fig.~\ref{fig:d-wave_solenoid}, the free-energy is shown for a set of temperatures for the two cases where the initial guess for the order-parameter field is homogeneous, and where we lock a phase winding of $2\pi$ around the annulus. Below $T\sim0.2 T_\mathrm{c}$, the existence of zero-energy Andreev states start to drive the superconductor to generate a finite $\vps$ by spontaneous phase gradients.\cite{Vorontsov:2009,Hakansson:2015} This finite superfluid momentum gives rise to a Doppler shift, $\vvF\cdot\vps$, that lifts the zero-energy states away from the Fermi level and thus lowers the free energy of the system. The free-energy minimum below $T^*\approx 0.18 T_\mathrm{c}$ is shifted to occur at finite magnetic flux. At the lowest temperature we show here, $T=0.1T_\mathrm{c}$, we can  find several meta-stable states at small fluxes. The orange rhombus in Fig. \ref{fig:d-wave_solenoid} has a purely real order parameter and the zero-energy states are not shifted away from the Fermi level. This state is only stable at $\Phi_{\mathrm{ext}}\equiv 0$ and any small seed of a phase gradient in the order parameter, or a minute magnetic perturbation, will generate a configuration with spontaneous currents. The two low-energy states we find are combinations of two possible states as showed in Fig.~\ref{fig:d-wave_solenoid}.

\subsubsection{\label{sec:radial_gauge}Solving London-Maxwell equation in a symmetric gauge}

In contrast to the case with the solenoid gauge above, a more common situation experimentally is when the external magnetic field $\vB_\mathrm{ext}=B_\mathrm{ext}\hat \vz$ is applied in a homogeneous fashion and penetrates also the superconductor. We incorporate this case through a symmetrical gauge $\vA_\mathrm{ext}(\vR)=\tfrac{1}{2}B_\mathrm{ext}(-y\,\hat \vx+x\,\hat \vy)$ subject to the condition $\nabla\cdot\vA=0$.

Let us study the London-Maxwell equation for the vector potential $\vA$ in the symmetrical gauge
\cite{deGennes:Book,Pearl:1964,Fetter:1980,Tinkham:Book,Kogan:2004}
\begin{equation}
-\nabla^2\vA=\frac{4\pi e}{c}\langle\bar\rho_s(\vpF)\vps\rangle_{\vpF} = -\frac{1}{\lambda^2(T)} 
\left( \frac{\Phi_0}{2\pi R}\,n\,\hat{\mathbf{\varphi}} + \vA \right), \label{LMeqRing}
\end{equation}
for the same annulus geometry as in the previous section. We allow for a phase winding $2\pi n$ of the order parameter going around the annulus,
and include screening of the externally applied field. The temperature-dependent penetration depth $\lambda(T)$ entering \Neqref{LMeqRing} is defined as
\begin{equation}
    \frac{1}{\lambda^2(T)}=\frac{1}{\lambda^2_0}\yosh(T).
\end{equation}
As long as the inner radius is much larger than the coherence length and the external magnetic field is moderate in strength, the superfluid momentum will 
be small, and we can disregard any reduction of the order-parameter amplitude for an $s$-wave annulus.
In this case linear-response theory remains valid and gives a platform to verify more involved calculations produced by SuperConga. Below, we introduce the dimensionless applied magnetic flux 
\begin{equation}
    \alpha = \frac{\Phi_{\mathrm{ext}}}{\Phi_0}.
\end{equation}

\paragraph{Analytic solution of the London-Maxwell equation.}
Equation (\ref{LMeqRing}) is a diffusion equation and can be solved analytically. We give the solution here for
completeness. In a cylindrically symmetric case, the solution to \Neqref{LMeqRing} is given by the modified Bessel
functions of first and second kind, $\mathrm{I}_\mu(x)$ and $\mathrm{K}_\mu(x)$ respectively. Writing
$\vA=A(R)\hat{\mathbf{\phi}}$, the general solution, in the external magnetic field $\vB_\mathrm{ext}\,=\alpha
\Phi_0 \hat{\vz}/\mathcal{A}$, may be written as 
\begin{eqnarray}
 A(R)&=& \Phi_0 \lambda \left[ a\, \mathrm{I}_1 \left(\frac{R}{\lambda} \right) + 
 b\, \mathrm{K}_1 \left(\frac{R}{\lambda} \right) \right]  \nonumber \\ 
 && - \frac{n \Phi_0}{2 \pi R} \;,\; R \in [\mathcal{R}_<, \mathcal{R}_>], \\ 
 A(R)&=& \beta \frac{\Phi_0 }{\mathcal{A}_\mathrm{h}}\frac{R}{2} \;,\; R \in [0, \mathcal{R}_<),
\label{Aring}
\end{eqnarray}
where $\mathcal{A}= \pi (\mathcal{R}_>^2 - \mathcal{R}_<^2)$ and $\mathcal{A}_\mathrm{h} = \pi \mathcal{R}_<^2$ are the areas of the annulus and of the inner hole. The term $\sim 1/R$ is the particular solution needed when $n\ne0$. The unknown constants $(a,b,\beta)$ are determined so that $\vA$ is bounded for $R<\mathcal{R}_<$, and continuous at $R=\mathcal{R}_<$ and that the magnetic field $\vB=\nabla\times \vA$ is continuous at $R=\mathcal{R}_>$. We state them for completeness 
\begin{eqnarray}
a &=& \frac{1}{M}\bigg(\frac{\mathrm{K}_2^<}{\mathcal{A}}\alpha +
\frac{\mathrm{K}_0^>}{\mathcal{A}_\mathrm{h}}n\bigg),\\
b &=& \frac{1}{M}\bigg(\frac{\mathrm{I}_2^<}{\mathcal{A}}\alpha +
\frac{\mathrm{I}_0^>}{\mathcal{A}_\mathrm{h}}n\bigg),\\
\beta &=& \frac{1}{ M}\bigg(\frac{2\pi  \lambda^2}{\mathcal{A} }\alpha 
- \left[\mathrm{I}_0^> \mathrm{K}_0^< - \mathrm{I}_0^< \mathrm{K}_0^>\right] n \bigg),
\end{eqnarray}
with $M=\mathrm{I}_0^> \mathrm{K}_2^<-\mathrm{I}_2^< \mathrm{K}_0^>$ and $\mathrm{I}_\mu^{>,<}=\mathrm{I}_\mu(\frac{R_{>,<}}{\lambda})$ and the same for $\mathrm{K}_\mu^{>,<}$.
The corresponding magnetic field, $\vB=B(R)\hat{\mathbf{z}}$ in and around the annulus is
\begin{eqnarray}
B(R) &=& \Phi_0 \left[ a\,\mathrm{I}_0 \left(\frac{R}{\lambda} \right)- b\, \mathrm{K}_0 \left( \frac{R}{\lambda} \right)\right] \;,\; R \in [\mathcal{R}_<, \mathcal{R}_>],\nonumber\\ 
B(R) &=& \beta\frac{\Phi_0}{\mathcal{A}_\mathrm{h}}\;,\; R \in [0, \mathcal{R}_<) \; .
\label{Bring}
\end{eqnarray}
The constant $\beta$ gives the flux strength in the inner hole in a similar way as $\alpha$ does for the annulus. 

\begin{figure}[t]
	\includegraphics[width=\columnwidth]{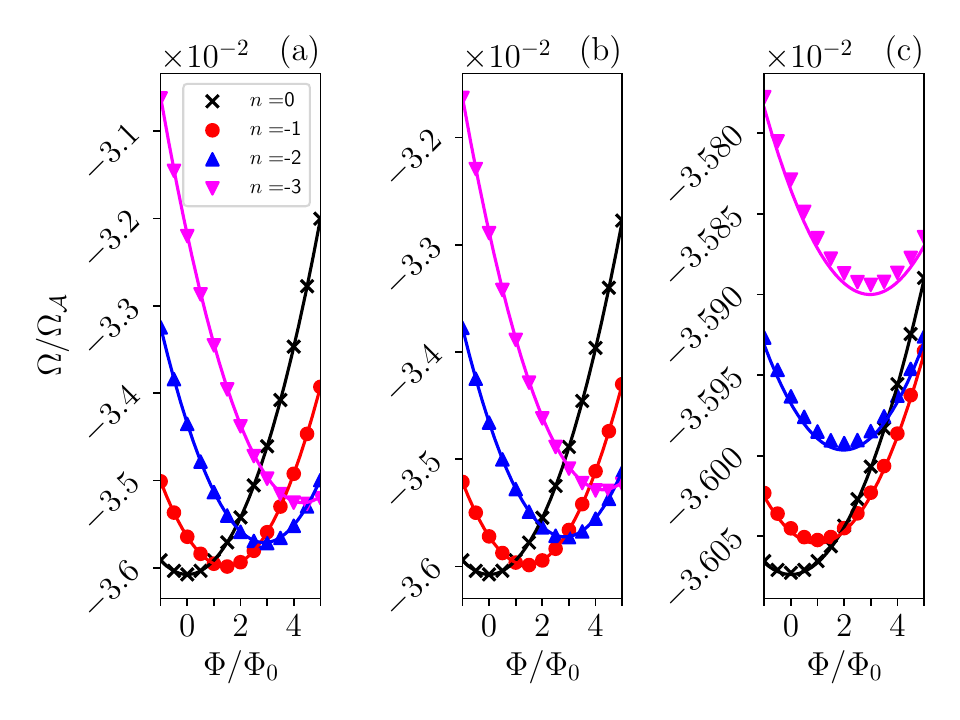}
	\caption{The free energy for an $s$-wave superconductor in an annulus, with an outer (inner) radius $\mathcal{R}_> = 25\xi_0$ ($\mathcal{R}_< = 10\xi_0$), in an external magnetic field in symmetric gauge, at $T=0.2T_\mathrm{c}$. The magnetic flux-density is given by $B_z = \Phi_{\mathrm{ext}} / \mathcal{A}$, where $\mathcal{A} = \pi (\mathcal{R}^2_> - \mathcal{R}^2_<)$. Four different values of the order-parameter phase-winding, $n$, are shown in each plot. The points are numerical results from SuperConga, and the solid lines are analytical results. The only parameter that differs between the plots is the Ginzburg-Landau parameter $\kappa = \lambda/\xi_0$, which is (a) $\kappa = 100$, (b) $\kappa = 20$, and (c) $\kappa = 2$.}
	\label{fig:annulus_radial_gauge}
\end{figure}

\paragraph{An $s$-wave superconducting annulus:}
For the $s$-wave annulus, the equations (\ref{Aring}) and (\ref{Bring}) give a good account for the response to an external field. The free energy of the superconducting annulus can thus be evaluated using Eqs. (\ref{lFEkin})--(\ref{lFEmag}), and the results compare well with results obtained by SuperConga. By inspection, one sees that the functional form of the free-energy correction due to an external field ($\alpha$) and a possible phase winding ($n$) will be\cite{Kogan:2004} 
\begin{equation}
\delta \Omega_{\mathrm{kin}}+ \delta \Omega_{\mathrm{magn}}=
\delta\Omega_{0}(\omega_{\chi} n^2 +(\omega_{B} n -\alpha)^2).
\label{eq:dfe}
\end{equation}
We determine the factors $\delta \Omega_0,\omega_{\chi}$ and $\omega_B$ numerically. They depend both on geometry and on temperature via the temperature dependence of the penetration depth. 

The results for the free energy are shown in Fig. \ref{fig:annulus_radial_gauge} for $T=0.2T_\mathrm{c}$ and for a set of different penetration depths. The main purpose here is to verify SuperConga by comparison to the results obtained by linear response. The agreement is excellent for small phase windings and large penetration depths. However, the agreement worsens with decreasing $\lambda$ and increasing $n$. This is a numerical issue that is remedied by a finer spatial resolution of the simulation. This highlights that one needs verify results from SuperConga by varying the parameters that are related to numerical accuracy. These are typically spatial resolution of the geometry, angular resolution of the Fermi-surface averages, and the cut-off in the frequency sums.
\begin{figure}[t]
	\includegraphics[width=\columnwidth]{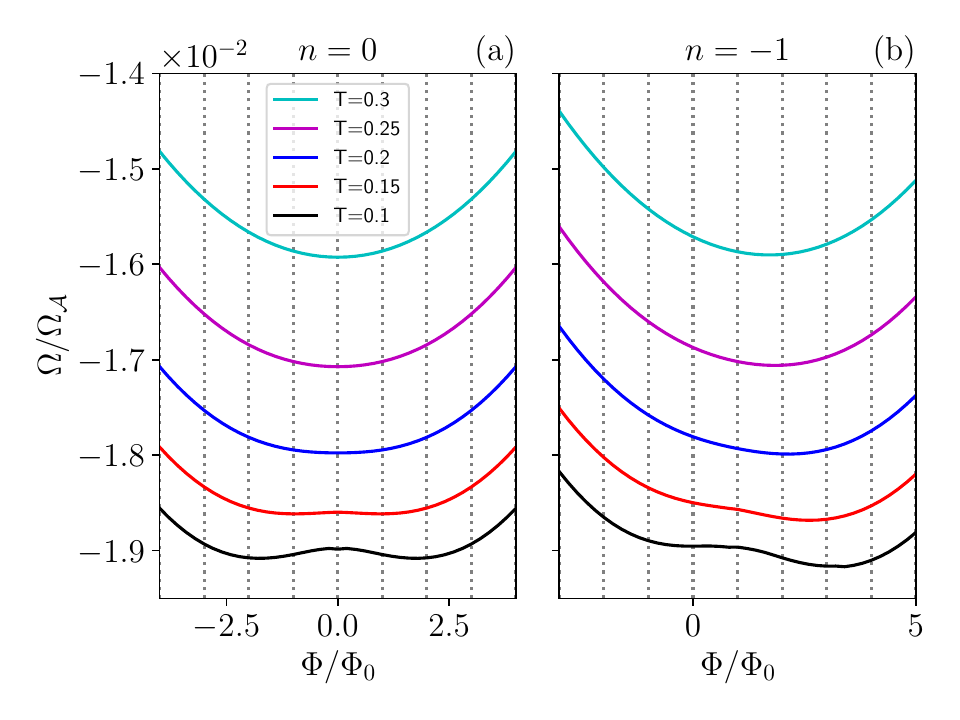}
	\caption{The free energy as function of external magnetic field for a $d$-wave superconducting annulus, with an outer (inner) radius $\mathcal{R}_> = 25\xi_0$ ($\mathcal{R}_< = 10\xi_0$), obtained using SuperConga. The penetration depth is much larger than the coherence length, $\lambda=50\xi_0$. We look at two cases; without ($n=0$) and with ($n=-1$) a phase winding of $2\pi$ around the annulus. From top to bottom we lower the temperature from $T=0.3T_\mathrm{c}$ down to $T=0.1 T_\mathrm{c}$ as in Fig.~\ref{fig:d-wave_solenoid}. The dotted vertical lines are a guide to the eye.}
	\label{fig:d-wave_radial}
\end{figure}

\paragraph{A $d$-wave superconducting annulus:}
For the $d$-wave superconductor we solve Eqs.~(\ref{eq:ampere})-(\ref{eq:current_density}) and Eq.~(\ref{eq:Eilenberger}) self-consistently. We set the zero-temperature penetration depth to $50\xi_0$ as most $d$-wave superconductors are extreme type-II superconductors. The results are shown in Fig. \ref{fig:d-wave_radial} for zero and one phase winding locked into the annulus. Well above $T\approx 0.2T_\mathrm{c}$ we find that the free energy has the parabolic functional dependence on applied field as for the $s$-wave superconductor. At lower temperatures, the low-energy surface Andreev states dominate, and their paramagnetic response modifies the free energy to host two minima at finite external flux, signalling that it is energetically favorable to form spontaneous currents as discussed above using the solenoid gauge.

\paragraph{Magnetic moment.}
The presence of a circulating current in an annulus can be detected as a magnetic moment 
\begin{equation}
    \vm = \frac{1}{2} \int_\mathcal{V} \mathrm{d}\vR\, \vR \times \vj_{3D}(\vR).
\end{equation}
Since the supercurrent, $\vj(\vR)$, we compute with SuperConga is confined to flow in a two-dimensional plane, the current density in a stack of layers is given as 
$$\vj_{3D}(\vR)=d \sum_{layers}  \delta(z-z_{l}) \vj(\vR),$$
i.e. the magnetic moment is given as 
\begin{equation}
    \vm= \frac{N_{l} d}{2} \int_\mathcal{A} \mathrm{d}\vR\, \vR \times \vj(\vR) \; ,
\label{eq:magnetic_moment_def}
\end{equation}
where $N_{l} d $ is the thickness of the grain along the $z$-axis assuming $N_l$ layers separated by the c-axis distance $d$. The unit for
the magnetic moment is $m_0=j_0\xi_0\mathcal{V}$ where
\begin{equation}
    j_0\xi_0=\frac{c^2}{4\pi}\frac{B_0}{\kappa^2}\mbox{\,\,(cgs)}=\frac{1}{\mu_0}\frac{B_0}{\kappa^2}\mbox{\,\,(SI)},
\end{equation}
with $B_0=\Phi_0/(\pi\xi_0^2)$ and $\mathcal{V}=N_l d \mathcal{A}$. The parameters that determine $m_0$ are thus the geometrical size (and shape) of the object and
the magnetic penetration depth specific of the superconducting material. $m_0$ can also be expressed in the perhaps more familiar form 
$m_0=2 \mu_\mathrm{B} n \mathcal{V}$, where $\mu_B=\hbar|e|/2m_e$ is the Bohr 
magneton and $n=v_F p_F \NF$ the number density of charge carriers. Inserting numerical values for $\Phi_0$ and $\mu_0$ and setting the length scale to $nm$:s for $\lambda_0$ and $\mathcal{V}$ gives 
\begin{equation}
    \frac{m_0}{(\mathcal{V}/\lambda_0^2)}\approx  0.5\times 10^{-18} \mbox{Am}^2= 0.5\times 10^{-21} \mbox{emu}.
\end{equation}  

\begin{figure}[tb!]
	\includegraphics[width=1.05\columnwidth]{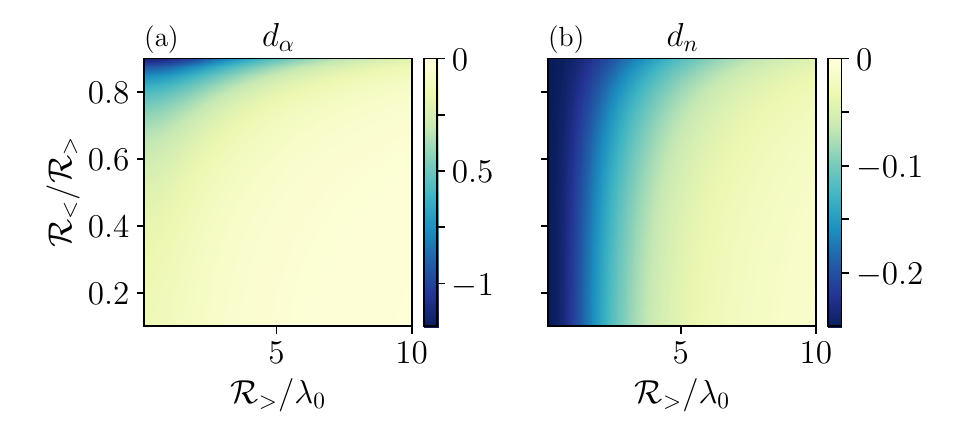}
	\caption{(a) The dependence of $d_\alpha$ and (b) $d_n$, on the relative sizes of the inner (outer) radius $\mathcal{R}_<$ ($\mathcal{R}_>$) and the penetration depth $\lambda_0$. Here, $d_\alpha$ and $d_n$ are factors entering the expression for the magnetic moment of an $s$-wave superconducting annulus, Eq.~\eqref{magneticannulus}. Hence, together with this equation, the figures show how a larger magnetic moment, and therefore experimental signature, can be obtained by creating annulus samples with different inner and outer radii.}
	\label{fig:mm_factors}
\end{figure}

\begin{figure}[t]
	\includegraphics[width=\columnwidth]{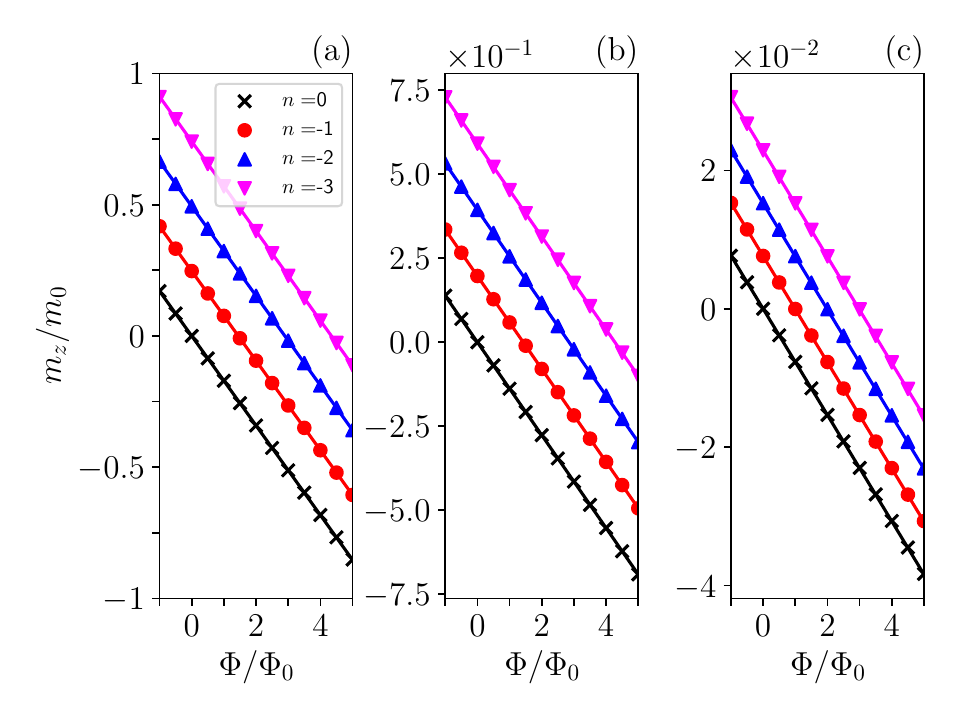}
	\caption{The magnetic moment, Eq.~\eqref{eq:magnetic_moment_def}, for an $s$-wave superconducting annulus, with an outer (inner) radius $\mathcal{R}_> = 25\xi_0$ ($\mathcal{R}_< = 10\xi_0$), in an external magnetic field in symmetric gauge, at $T=0.2T_\mathrm{c}$. The external flux density is given by $\vB_\mathrm{ext} = \Phi_{\mathrm{ext}} {\hat{\vz}} / \mathcal{A}$, where $\mathcal{A} = \pi (\mathcal{R}^2_> - \mathcal{R}^2_<)$. Four different values of the order-parameter phase-winding, $n$, are shown in each plot. The points are numerical results from SuperConga, and the solid lines are analytical results, Eq.~\eqref{magneticannulus}. The only parameter that differs between the plots is the Ginzburg-Landau parameter $\kappa = \lambda_0/\xi_0$, which is (a) $\kappa = 100$, (b) $\kappa = 20$, and (c) $\kappa = 2$.}
	\label{fig:M_annulus_radial_gauge}
\end{figure}

Inserting the analytical form of the vector potential from Eq.~\eqref{Aring} into Eq.~\eqref{linearcurrent}, we can evaluate
the magnetic field ($\alpha$) and phase-winding ($n$) dependence of the magnetic moment of a stack of 2D $s$-wave annuli to be 
\begin{eqnarray}
\vm(\alpha,n) &=&
m_0\left(d_\alpha \alpha + d_n n\right) \hat{\vz}\; ,\label{magneticannulus}
\end{eqnarray}
where 
\begin{eqnarray}
d_\alpha &=& \frac{\pi \lambda^2_0}{\mathcal{A}} \frac{\mathcal{A} + \mathcal{A}_\mathrm{h}}{M \mathcal{A}} \left( I^<_2 K^>_2 - I^>_2 K^<_2 \right) \; , \\
d_n &=& \frac{\pi \lambda^2_0}{\mathcal{A}} \left( \frac{2\pi \lambda^2}{M \mathcal{A}_\mathrm{h}} - 1 \right) \; .
\end{eqnarray}
How $d_\alpha$ and $d_n$ depend on the annulus radii and the penetration depth is shown inf Fig.~\ref{fig:mm_factors}. In Fig.~\ref{fig:M_annulus_radial_gauge} 
we show how the magnetic moment depends on the penetration depth and the forced phase winding as function of applied field. The computed values of the magnetic
moment using SuperConga fall right on the analytical ones of Eq.~(\ref{magneticannulus}). 

\begin{figure}[t]
	\includegraphics[width=\columnwidth]{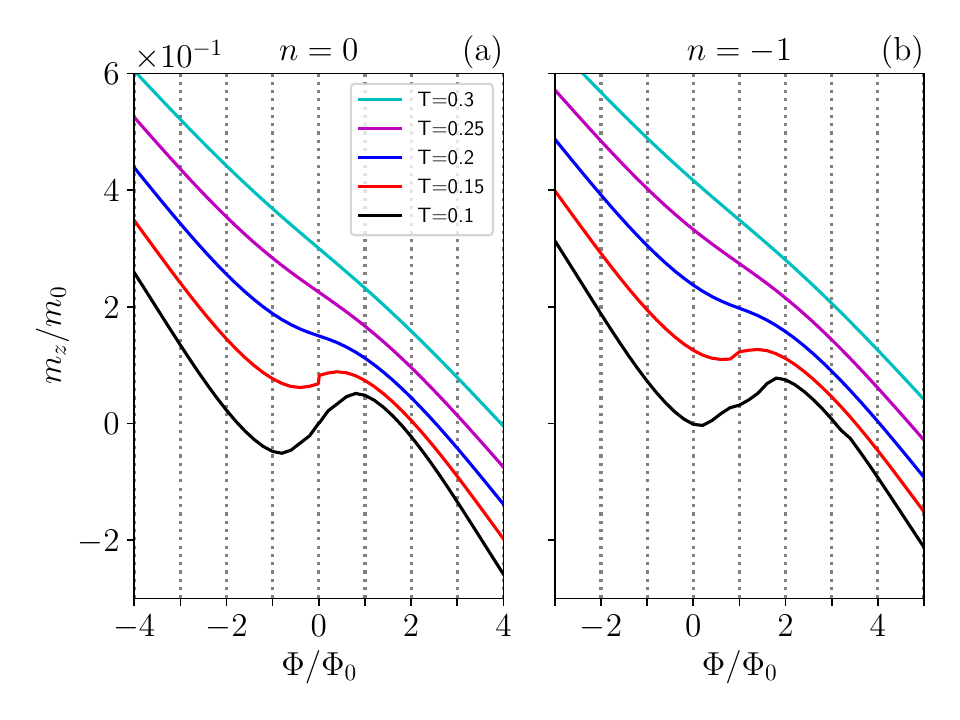}
	\caption{
	The figure shows how the flux dependence of the magnetic moment $m_z(\Phi)$ in a $d$-wave superconducting annulus changes with temperature. The configuration is the same as in Fig.~\ref{fig:d-wave_radial}. 
	The two panels differ by (a) having no order-parameter phase-winding $(n=0)$ and (b) having one $(n=-1)$. 
	In both panels the magnetic moment goes from being linear in applied flux at higher temperatures $T\gtrsim 0.25T_c$ to highly non-linear, and with a complete reversal in the sign of the slope close to $\Phi/\Phi_0=n$ at 
	low temperatures $T\lesssim 0.15T_c$.
	This  temperature dependence of the magnetic moment is a direct consequence of 
	surface-scattering induced Andreev states in a d-wave superconductor. The non-linear flux dependence at low temperatures is a signature that a "phase crystal'' is 
	established. 
	The dotted vertical lines in both panels are a guide for the eye. Each line is shifted $0.075$ units upward from the line below.}
	\label{fig:d-wave_radial_magnetic_moment}
\end{figure}

While the magnetic moment of the $s$-wave annulus holds little surprise, the magnetic moment of the $d$-wave annulus shows trace of a low-temperature 
transition where screening currents are modified by the paramagnetic nature of the zero-energy Andreev states. 
\cite{Higashitani:1997,Fogelstrom:1997,Fogelstrom:2004,Vorontsov:2009,Hakansson:2015} The free energy dependence on temperature and magnetic field 
may be hard to directly measure whereas the magnetic moment of a $d$-wave annulus could be a possible route to detecting this low-temperature 
transition. In Fig.~\ref{fig:d-wave_radial_magnetic_moment} we show the magnetic moment of a $d$-wave superconducting annulus as function of external flux at several temperatures. The dimensions are the same as for the s-wave case, i.e. $\mathcal{R}_>=25\xi_0$ and
$\mathcal{R}_<=10\xi_0$, and the penetration depth is $\lambda_0=50\xi_0$. For low temperatures and small external fields, the signature
of the Andreev states is that the magnetic moment of the annulus is reversed compared to that at high fields.

\subsection{\label{sec:swave_disc}A mesoscopic $s$-wave superconducting disc}

As another example we consider an $s$-wave superconductor in the form of a 2D-disc, with radius $\mathcal{R}$, in an external magnetic field. We will allow for arbitrary strength of the applied magnetic field so that multiple vortices can enter the disc. The general solution can and will break the continuous rotational symmetry of the disc around its central axis. Investigating this problem requires that we solve the Eilenberger equation, \Neqref{eq:Eilenberger}, with a self-consistently determined order-parameter field, \Neqref{eq:gapequation}, and vector potential \Neqref{eq:ampere}. 
The magnetic field where the first vortex enters is the first critical field of a superconductor and is referred to as $B_\mathrm{c1}$. At this 
field it becomes energetically favourable for the superconducting material to host one vortex located in its interior well away from its physical
edges. This compared to generating large screening currents flowing at the sample edges. The first critical field will depend on several parameters.
Parameters intrinsic to the superconducting material are the penetration depth, $\lambda(T)$, and the coherence length, $\xi(T)$. Both depend on
temperature via the temperature dependence of the superconducting energy gap $\Delta(T)$. Extrinsic parameters are e.g. the physical size and shape of 
the superconducting sample, which in our example here will simply be the radius $\mathcal{R}$ of the disc. 
Also temperature and magnetic-field strength 
are externally controlled. We will assume that our superconductor is a layered 2D-material with negligible inter-layer coupling. As the field is increased above $B_\mathrm{c1}$, a cascade of mesoscopic vortex configurations are possible, which are separated by relatively small energy
barriers. At high fields a vortex lattice may be established.\cite{Abrikosov:1957}. Reaching the second critical field, $B_\mathrm{c2}$,
bulk superconductivity ultimately vanishes. With our framework SuperConga we can treat the three relevant length scales of this problem, i.e.
$\xi_0,\lambda_0$, and $\mathcal{R}$, on the same footing at general temperatures and at general magnetic fields up to $B_\mathrm{c2}$ (and in principle $B_\mathrm{c3}$). Throughout this
section we focus on a disc with a radius fixed to 25$\xi_0$ at a temperature set to $0.2T_c$ if not otherwise stated. The analytic results we give
serve as an important tool to verify the results extracted from SuperConga.

Let us first focus on $B_\mathrm{c1}$ which was nicely treated by A. Fetter in Ref.~[\onlinecite{Fetter:1980}] for a thin-film superconducting disc. 
In the case that the penetration depth is the largest scale in the system the back-coupling of the current can be omitted.  
The corresponding superfluid momentum is
\begin{equation*}
    \vps(R)=\frac{\hbar}{2}\bigg(\frac{n}{R} + \alpha \frac{R}{\mathcal{R}^2}\bigg) \hat{\varphi}.
\end{equation*}
Calculating the free-energy density correction due to the superfluid momentum above with $(n=-1)$ and without $(n=0)$ a single vortex gives:
\begin{eqnarray}
    n=0&:&\,(\Omega(\alpha)-\Omega_{\mathrm{bulk}})/{\cal{A}}=\delta \omega_{\mathrm{kin}}\alpha^2,
    \label{0vor}\\
    n=-1&:&\,(\Omega(\alpha)-\Omega_{\mathrm{bulk}})/{\cal{A}}=\delta \omega_{\mathrm{kin}}(\epsilon_{\rm core}-4 \alpha+\alpha^2). \;\;\;
    \label{1vor}
\end{eqnarray}
where $\Omega_{\mathrm{bulk}}$ is the bulk free-energy contribution Eq.~(\ref{FEbulk}) and 
\begin{equation}
    \delta\omega_{\mathrm{kin}}=\frac{1}{16}\bigg(\frac{\xi_0}{\mathcal{R}}\bigg)^2 \yosh(T),
    \label{lFKinfty}
\end{equation}
gives the free-energy contribution from the induced supercurrent due to the combined effect of the external field and to the phase-winding of $2\pi$ 
around the vortex core. The term $\epsilon_{\rm core}$ in principle consists of two terms. The first is proportional to 
$\ln(\mathcal{R}/\mathcal{R}_\mathrm{c})$, with the vortex-core radius $\mathcal{R}_\mathrm{c}$, due to the $1/R$ dependence of $\vps$. The second, 
$\epsilon_0$, is due to the gradient-energy terms that come form the suppression of the order-parameter amplitude needed to accommodate the singularity 
in the vortex-core center. This second term is not captured in the simple London-Maxwell theory as it requires a theory that includes also the 
coherence-length scale $\xi_0$ in the calculations. 
\begin{figure}[!t]
\includegraphics[width=\columnwidth]{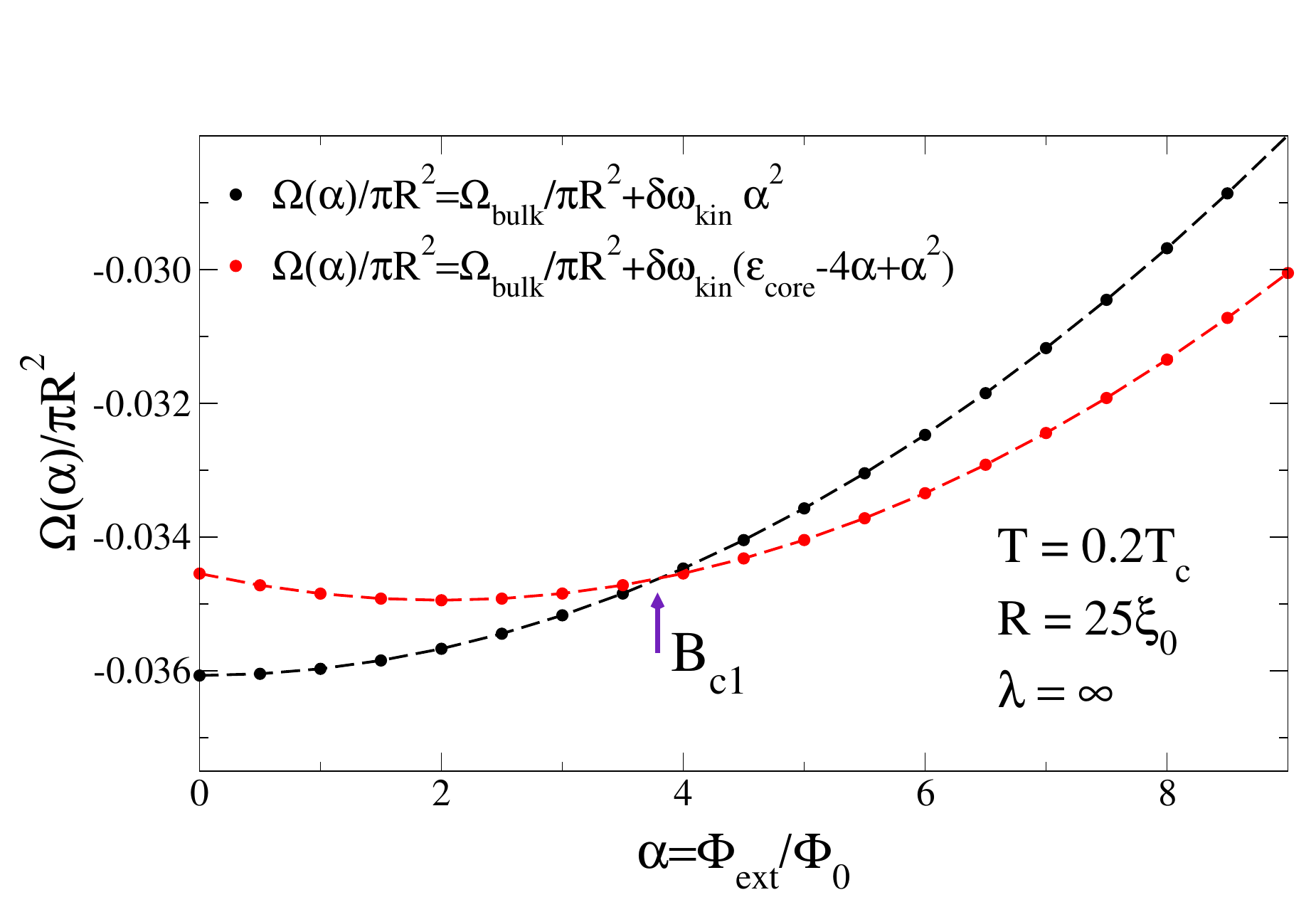}
\caption{
 The figure shows how the first critical field ($B_\mathrm{c1}$) can be extracted from the intersection of the solutions for zero ($n=0$) and one ($n=1$) Abrikosov vortex.
 Here, the free energy per unit area is plotted as a function of externally applied field for an $s$-wave superconducting disc of radius 25$\xi_0$, computed analytically (points) and numerically with SuperConga (dashed lines). Here we show the low field free-energy in the case $\lambda=\infty$ and at a temperature set to $0.2T_c$.
 The first critical field ($B_\mathrm{c1}$) can be extracted from the intersection of the solutions for zero ($n=0$, black curves ) and one ($n=1$, red curves) Abrikosov vortex.}
\label{fig:lambda_inf}
\end{figure}
In Fig. \ref{fig:lambda_inf} we show calculations made with SuperConga for $\lambda_0=\infty$. As seen, numerical results fall straight on the theoretical prediction for $(\Omega(\alpha)-\Omega_{\mathrm{bulk}})/{\cal{A}}$ given by Eqs. (\ref{0vor})-(\ref{1vor}). The main physical observable to extract is $B_\mathrm{c1}$, defined as the field where having a vortex becomes energetically favorable compared to the vortex-free case. For this disc at this temperature $B_\mathrm{c1}$ occurs at $\alpha\approx3.8$.

\begin{figure}[!t]
	\includegraphics[width=\columnwidth]{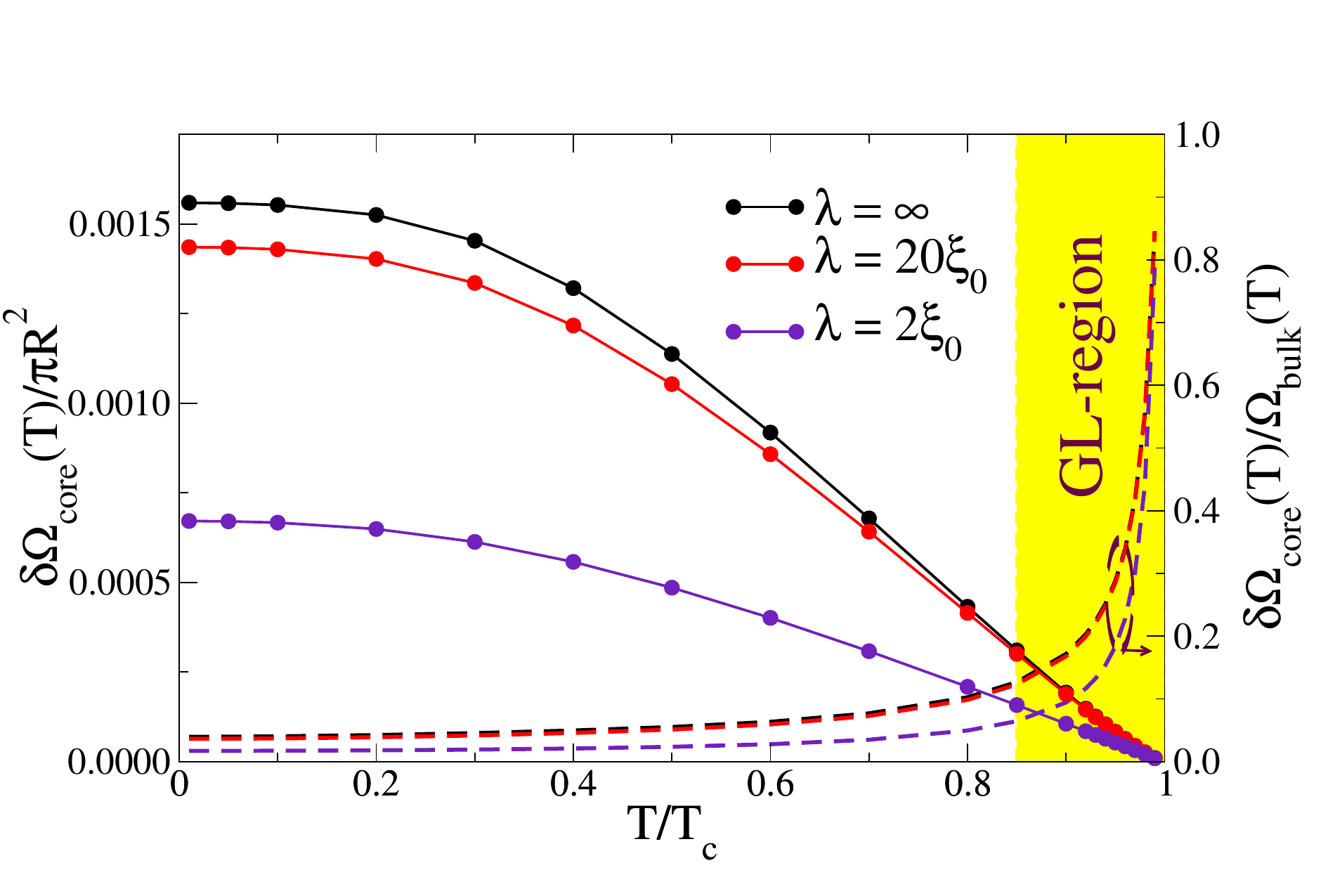}
 	\caption{
	The vortex-core energy $\delta\Omega_{\rm core}=\delta\omega_{\mathrm {kin}}\epsilon_{\rm core}$,
 	extracted as the difference in free energy between the one-vortex 
 	and the zero-vortex case, at zero-field $(\alpha=0)$. 
 	Here, $\delta\Omega_{\rm core}(T)$ is computed with SuperConga, and shown for the full temperature range from $T\approx0.0$ to $T\lesssim T_c$.
 	Solid and dashed line correspond to the left and right axes, respectively, and line colors denote the penetration
 	depth as indicated by the label.
 	}
	\label{fig:relative_core_energy_vs_T}
\end{figure}

\begin{figure*}[th!]
	\includegraphics[width=1.02\textwidth]{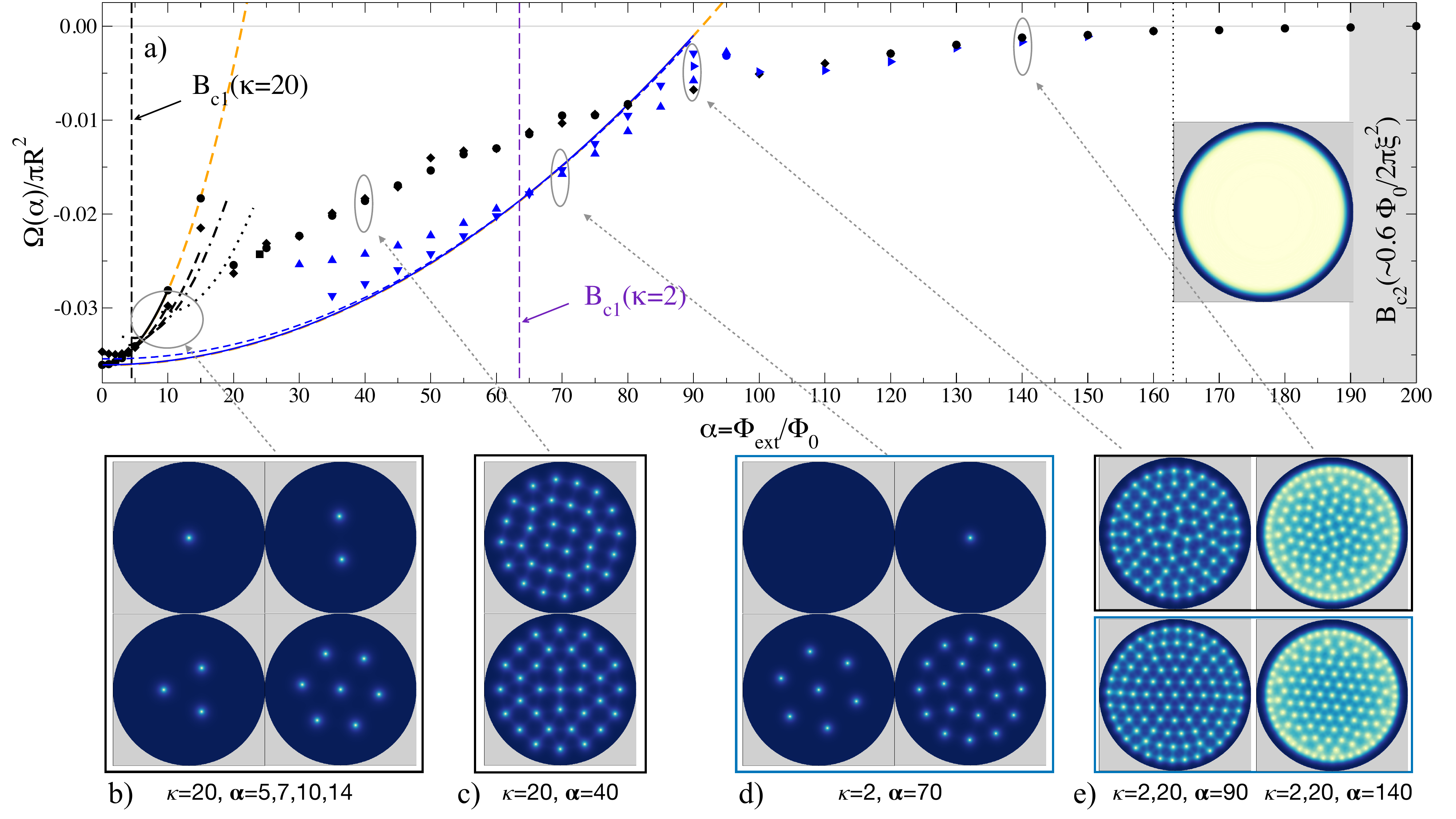}
	\caption{(a) The free energy as a function of externally applied field for an $s$-wave superconducting disc of radius 25$\xi_0$. The temperature is set to $0.2T_c$. We consider two values for the penetration depth 
	$\lambda_0=20\xi_0\sim \mathcal{R}$ and $\lambda_0=2\xi_0\ll \mathcal{R}$ and follow the route from the zero-field case via the first critical field 
	$B_\mathrm{c1}$, where the first vortex enters the disc, to the critical field $B_\mathrm{c2}$ where the (bulk) superconductivity is killed by the 
	magnetic field. Black lines and symbols refer to $\kappa=20$ while the same in indigo refer to $\kappa=2$. The two orange dashed 
	curves are given using the analytical expression Eq.~(\ref{lFKfinite}) for the two respective values of $\kappa$. 
	The lower panels (b)--(e)
	show possible vortex configurations at different fields. In panel (b) we show the four vortex states that live on the parabolas
	at low fields. As the field is increased above $B_\mathrm{c1}(\kappa=20)$ the stable configuration contain more and more vortices. Panel (c) shows that at higher fields,
	different vortex configurations are possible at the same field with very little difference in free energy between the configurations. 
	In panel (d) we show 4 vortex configurations just above $B_\mathrm{c1}(\kappa=2)$. This indicates that for marginal type-II superconductors the vortex
	lattice is established just above $B_\mathrm{c1}$. In panel (e) we show configurations for both values of $\kappa$. The inset in panel (a) shows the extreme
	high-field configuration with only a surface layer having a sizable order-parameter amplitude. It sustains a large circulating supercurrent by a
	phase that winds
	rapidly around the disc perimeter. This surface superconductivity vanishes at an even higher field $B_\mathrm{c3}$.}
	\label{fig:swave_disc_in_field}
\end{figure*}

When the penetration depth is comparable to or smaller than the size of the disc we need to analyze Eqs. (\ref{Aring})-(\ref{Bring}) in the limit that $\mathcal{R}_<\rightarrow 0$. The result is 
\begin{eqnarray}
 A_{\phi}(R)&=&
 \lambda B_\mathrm{eff}(\alpha,n) \mathrm{I}_1 \left( \frac{R}{\lambda} \right) 
 + n \frac{\Phi_0}{2\pi} \left[ \frac{1}{R} - \frac{1}{\lambda} \mathrm{K}_1 \left( \frac{R}{\lambda} \right) \right], \;\;\;
\label{Alfinite}\\
 B_z(R)&=&
 B_\mathrm{eff} (\alpha,n) \mathrm{I}_0 \left( \frac{R}{\lambda} \right) 
 + n \frac{\Phi_0}{2\pi\lambda^2} \mathrm{K}_0 \left( \frac{R}{\lambda} \right),
 \label{Blfinite}
\end{eqnarray}
with
\be
B_\mathrm{eff}(\alpha,n) =
\frac{\Phi_0}{\pi \mathcal{R}^2} \frac{1}{\mathrm{I}_0 \left(\frac{\mathcal{R}}{\lambda}\right)} \left[ \alpha - \frac{n}{2} \left(\frac{\mathcal{R}}{\lambda}\right)^2 \mathrm{K}_0 \left(\frac{\mathcal{R}}{\lambda}\right) \right]
\; .
\ee

The functional form of Eqs.~(\ref{Alfinite})-(\ref{Blfinite}) allows an explicit evaluation of the free-energy
contributions $\delta\Omega_{\mathrm{kin}}$ and  $\delta\Omega_{\mathrm{magn}}$ in the case of no vortictity $(n=0)$ and 
$\delta \omega_{\mathrm{kin}}$ is modified for finite $\lambda$ in the following way
\begin{equation}
    \delta \omega_{\mathrm{kin}}(\lambda)=\bigg(k(\frac{\mathcal{R}}{\lambda})+b(\frac{\mathcal{R}}{\lambda})\bigg)\bigg(\frac{\xi_0}{\mathcal{R}}\bigg)^2\yosh(T),
     \label{lFKfinite}
    \end{equation}
with
\begin{eqnarray}
k(x)&=&\frac{1}{x^2}\frac{\mathrm{I}^2_1(x)+\frac{2}{x}\mathrm{I}_0(x)\mathrm{I}_1(x)-\mathrm{I}^2_0(x)}{2  \mathrm{I}^2_0(x)},\\
b(x)&=&\frac{2\mathrm{I}^2_0(x)-\frac{4}{x}\mathrm{I}_0(x)\mathrm{I}_1(x)-\mathrm{I}^2_1(x)}{2 \mathrm{I}^2_0(x)}.
\end{eqnarray}
In the limit that $x\rightarrow 0$, corresponding to $\lambda\rightarrow \infty$, $k(x)\rightarrow 1/16$ and 
$b(x)\rightarrow 0$ and we recover the prefactor $\delta\omega_{\mathrm{kin}}$ (Eq. (\ref{lFKinfty})) as we should. 

The vector potential, \Neqref{Alfinite}, is finite in the origin. This is not the case for the magnetic field, \Neqref{Blfinite}, 
which diverges as $\ln(R)$ when $R \rightarrow 0$ in the origin\cite{deGennes:Book,Tinkham:Book}. 
This divergence leads to that superconductivity must vanish locally in the vortex core (see e.g. Fig. \ref{fig:plotter}). 
The suppression of superconductivity happens over a couple of coherence lengths, and we determine the order-parameter profile using SuperConga. The functional form of the free energy, Eqs. (\ref{0vor})-(\ref{1vor}), holds also for finite screening but 
with $\delta \omega_{\mathrm{kin}}(\lambda)$ and with a modified coefficient for linear-in-field term in \Neqref{1vor}
(see also \Neqref{eq:dfe}). 
In Fig. \ref{fig:relative_core_energy_vs_T} we show how the free-energy of a single 
vortex core, i.e. $\delta\omega_{\mathrm {kin}}(\lambda)\epsilon_{\rm core}$, depends on
temperature. The penetration depth is considered for the values $\kappa=(2,20,\infty)$ and the external field is zero.
The temperature dependence of $\xi$ and $\lambda$ gives that close to $\Tc$, in the regime of validity of Ginzburg-Landau theory, the
physical size of the disc will be the smallest length-scale of the system. In this limit the relative contribution of
the vortex core to the free energy dominates, as seen in Fig. \ref{fig:relative_core_energy_vs_T}. Moving to lower 
temperatures Ginzburg-Landau theory is strictly speaking no longer applicable and we need to resort to the quasiclassical 
theory. Below $T=0.8T_c$, the vortex-core contribution for this geometry becomes less dominant and settles to be less than 5\% of the total free energy.
Instead, the free energy mainly depends on how well the supercurrents originating from the phase-winding around the vortex core
are screened, i.e. focus is primarily on the ratio $\lambda_0/\mathcal{R}$. We consider two cases of screening: a typical 
type-II superconductor with $\kappa=20.0$ for which the penetration depth compare to the size
of the disc, and a marginal type II superconductor with $\kappa=2.0$ with $\lambda_0 \ll \mathcal{R}$. 
In Fig.~\ref{fig:relative_core_energy_vs_T} we see that strong screening $(\kappa=2)$ 
reduces the energy cost of a single vortex compared to the case of moderate screening $(\kappa=20)$ so that we can extract  
$\delta\omega_{\rm kin}\epsilon_{\rm core}(\kappa=2)\approx0.5\delta\omega_{\rm kin}\epsilon_{\rm core}(\kappa=20)$. 

In Fig.~\ref{fig:swave_disc_in_field} we show a large set of computed free energies as a function of applied external magnetic field. The magnetic
field $B=\alpha \Phi_0/\mathcal{A}$ ranges from $\alpha=0$ to $200$ so that we pass $B_\mathrm{c1}$, where the first vortex enters, to $B_\mathrm{c2}$ where the
superconducting state is no longer stable and the disc becomes normal. 
Starting with the first critical field, $B_\mathrm{c1}$, we lift results from literature.\cite{Pearl:1964,Fetter:1980}  
For a superconducting disc of radius $\mathcal{R}$ where $\lambda_0/\mathcal{R}\sim 1$ the first critical field is given as
\begin{equation}
    B_\mathrm{c1}^\mathrm{disc}=\frac{\Phi_0}{\mathcal{A}}\bigg(\ln\frac{\mathcal{R}}{\mathcal{R}_\mathrm{c}}+\epsilon_0\bigg),
    \label{eq:Hc1disc}
\end{equation}
with $\epsilon_0$ due to the gradient terms. 
For a {\em pancake vortex} in the extreme 2D-case with $\lambda_0/\mathcal{R}\ll 1$, the first critical field for a vortex in the 2D superconducting 
sheet is given as\cite{Clem:1991,Graf:1993}
\begin{equation}
    B_\mathrm{c1}^\mathrm{sheet}=\frac{\Phi_0}{4\pi \lambda^2}\bigg(\ln \kappa+\epsilon_0\bigg).
    \label{eq:Hc1bulk}
\end{equation}
The lower critical field, $B_\mathrm{c1}$, that we extract correspond well to earlier theory if we 
associate the case $\kappa=20$ to \Neqref{eq:Hc1disc} and $\kappa=2$ to \Neqref{eq:Hc1bulk}. 
Extracting from the data in Fig.~\ref{fig:swave_disc_in_field} we find
\begin{eqnarray}
 &B_\mathrm{c1}&(\kappa=20)=4.4 \frac{\Phi_0}{\mathcal{A}},\\
 &B_\mathrm{c1}&(\kappa=2)=63.3 \frac{\Phi_0}{\mathcal{A}} = 1.62 \frac{\Phi_0}{4\pi \lambda_0^2}.
\end{eqnarray}
The numerical prefactor $4.4=\ln (\mathcal{R}/\mathcal{R}_c)+\epsilon_0$ contains two unknown parameters $\mathcal{R}_c$ and $\epsilon_{0}$ for
$\kappa=20$. On the other hand the prefactor in the case $\kappa=2$, $1.62=\ln\kappa +\epsilon_0$ contains only one unknown and we extract
$\epsilon_0=1.62-\ln \kappa \approx 0.93$. Taking the results from Fig.~\ref{fig:relative_core_energy_vs_T} we can determine the nominal 
vortex-core size $\mathcal{R}_c$ as $\ln\, \mathcal{R}_c=\ln \mathcal{R}+2 \epsilon_0(\kappa=2)-4.4\approx 0.68$ or $\mathcal{R}_c=2.7 \xi_0$. 
This value is quite reasonable estimate for a vortex-core size if we compare with the order-parameter field displayed in Fig. \ref{fig:plotter} 
(or in the panels b-e in Fig. \ref{fig:swave_disc_in_field}).

For large external fields, the upper critical field, $B_\mathrm{c2}$, is given by the field scale when vortex cores start to 
overlap,\cite{Tinkham:Book} i.e.
\begin{equation}
    B_\mathrm{c2}=\eta\frac{\Phi_0}{2\pi \xi_0^2}.
\end{equation}
We add a prefactor, $\eta$, as in our finite geometry there is always a vortex-free surface layer around the perimeter of the sample 
that reduces the effective vortex-lattice area. The vortex-free ring at the perimeter of our disc has a width $\sim 5\xi_0$ and thus 
we can estimate $\eta=(20/25)^2=0.64$. The upper critical field for our 
disc would be at $\alpha=\frac{1}{2} (\mathcal{R}/\xi_0)^2=312.5$ but we find that superconductivity vanishes at $\alpha\approx200$. 
This reduction of $B_\mathrm{c2}$
captured (by construction) by the smaller effective area of the vortex as we described above. 

For intermediate fields, $B_\mathrm{c1}< B < B_\mathrm{c2}$, a vortex lattice emerges. In Fig. \ref{fig:swave_disc_in_field} we 
follow how the vortex lattice is formed for the two cases,
$\kappa=2,20$. It is a very rich system with several local free-energy minima corresponding to different meta-stable configurations of vortices in
the disc. For a stronger type-II superconductor, $\kappa=20$, the lattice structure is less pronounced and only at high fields can we see a square
lattice evolving in the center of the disk. On the route towards higher fields one can find transitions between states comprising of distinct number of
vortices. In the following example below we come back to this and connect to recent experiments.

For the marginal type-II superconductor the vortex lattice is established quite rapidly in increasing field above $B_\mathrm{c1}$. The vortex lattice is
triangular with the outer most layer of vortices creating an ever denser necklace of vortices following the perimeter. 

\begin{figure}[t]
	\includegraphics[width=\columnwidth]{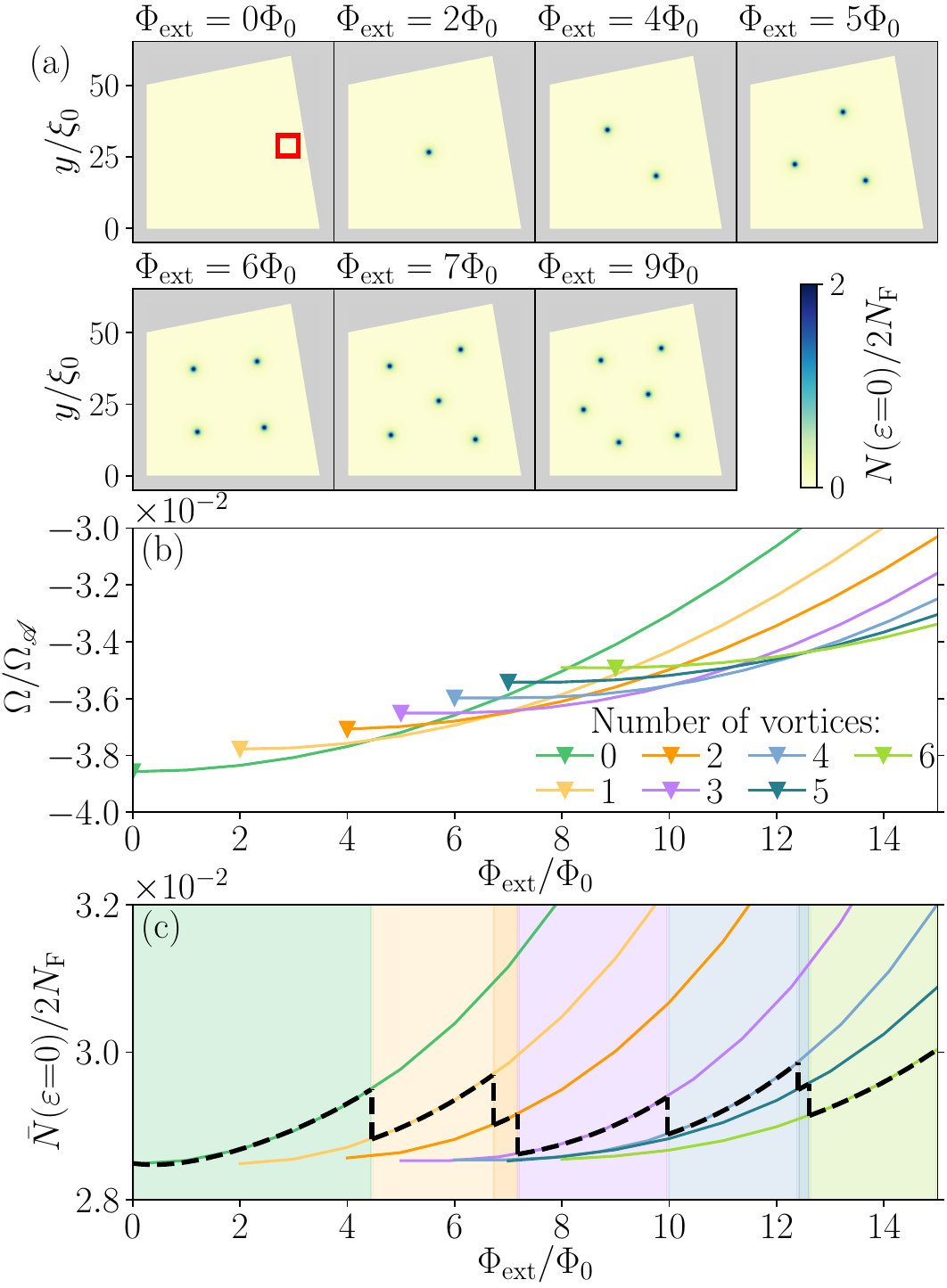}
	\caption{Simulations of conventional superconductors shaped like an irregular square, inspired by experiments in 
 Ref.~[\onlinecite{Timmermans:2016}], with similar values of the side length $\mathcal{S} \sim 50\xi_0$, Ginzburg-Landau
 coefficient $\kappa=83$, and temperature $T=0.1\Tc$. (a) Zero-energy local density of states at the minimum of the free energy
 with the specific number of vortices shown.
 (b) Magnetic field dependence of the free energy with the minimal value for each number of vortex marked with a
 triangle. (c) Average density of states as function of magnetic field. The average is over the surface region marked as red
 square in the first panel of (a). The smearing is $\delta=0.008 \cdot 2\pi \kB\Tc$.}
	\label{fig:vortex_irregular_grain}
\end{figure}

\subsection{\label{sec:Timmermans}Spectroscopy on superconducting nanostructures}
As we have seen above, the vortex physics in a finite geometry might be highly modified compared to a bulk system. 
In particular, the number of vortices and their separation do not only depend on the external field strength, but also on the
geometry, the nature of boundaries, and the relation between system size and superconducting length scales, i.e., the coherence
length and the penetration depth. This problem has attracted considerable interest both theoretically and experimentally during
the last years 
\cite{Geim:1997,Morelle:2004,Grigorieva:2006,Kokubo:2010,Zhao:2008_a,Misko:2009,Zhang:2012,Zhang:2013,Timmermans:2016,Zhao:2008_b,Cabral:2009,Kokubo:2016,Wu:2017,Huy:2013}.

Figure~\ref{fig:vortex_irregular_grain} shows the arrangement of 1--6 vortices in a finite and irregular-shaped superconducting grain, inspired by the experiment of Timmermans {\em et al.}~\cite{Timmermans:2016}
Finding the arrangement and number of vortices for a given magnetic field is a difficult task due to the complicated magnetic field dependence of the free energy. For this reason, Fig.~\ref{fig:vortex_irregular_grain}(a) shows the local density of states at the minimum of the free energy. The magnetic field dependence of the free energy is shown in Fig.~\ref{fig:vortex_irregular_grain}(b) for every number of a vortex. Each vortex follows a parabolic dependence on the magnetic field with a minimum marked as a triangle. The theoretically obtained vortex positions are in qualitative agreement with the experiment results in Ref.~[\onlinecite{Timmermans:2016}]. 

The transition between different vortices entering the grain can be visualized by measuring the zero-bias conductance 
close to the edge of the grain (Fig. 2(b) in Ref.~[\onlinecite{Timmermans:2016}]). Increasing the magnetic field 
starting from zero, screening currents prevent the magnetic field to penetrate the superconductor. The screening 
currents are responsible for a reduction of the superconducting gap and an increase of the density of states. 
The magnetic-field dependence of the density of states close to the edge is shown in 
Fig.~\ref{fig:vortex_irregular_grain}(c) for each number of a vortex. The entrance of a vortex in the grain is 
accompanied by a reduction of the density of states at zero energy. These abrupt reduction of the local density of states are
related to the reduction in the zero-bias conductance measurements in Ref.~[\onlinecite{Timmermans:2016}].

\section{\label{sec:summary}Summary}

SuperConga is an open-source framework for simulating mesoscopic superconducting grains within the quasiclassical theory of superconductivity. Its main strengths are ease-of-use, speed, and modularity. A Python frontend enables the user to quickly set up simulations of multi-component singlet superconducting grains, in general 2D-geometries and in an applied magnetic field. The framework solves for the order parameters self-consistently, and includes the back-coupling of the vector potential due to induced supercurrents. Real-time visualization is provided during simulations, and the Python frontend includes functionality for additional data analysis and visualization, as well as interactive spectroscopy of the local density of states.

The SuperConga framework is free to download under the GNU LGPL v3 license or higher, from its GitLab repository\cite{SuperConga:repository} https://gitlab.com/superconga/superconga. An extensive user manual has been published online,\cite{SuperConga:documentation} containing numerous pedagogical examples, tutorials and guides. The framework has been developed and tested for Unix-based environments, and generally runs on most modern laptops, desktops, as well as in cluster environments. The framework relies on the high-performance capabilities offered by GPU acceleration and CUDA\cite{CUDA:2021,Nickolls:2008}, and therefore requires a CUDA-capable device (i.e. NVIDIA GPU)\cite{NVIDIA_device:2021} to run. To the best of our knowledge, SuperConga is the only open-source code that uses quasiclassical theory to describe mesoscopic superconducting grains and ballistic devices. It therefore adds the missing ``mesoscopic link'' between more phenomenological methods, such as London-Maxwell\cite{Horn:2022} and Ginzburg-Landau,\cite{Sadovskyy:2015,Rybakov:2022,Rybakov:2022:b} and fully microscopic methods, such as  density-functional theory\cite{Kresse:1996:a,Kresse:1996:b,Giannozzi:2009,Giannozzi:2017,Nakhaee:2020} and tight-binding approaches.\cite{groth_kwant_2014,Bjornson:2019,Lothman:2020,Nagai:2020}

This paper outlines the functionality of version 1.0 of SuperConga. We foresee that the project website will evolve continuously, and that a community of developers can be formed to enable further improvements that increase the functionality and scope of the framework to include, e.g., impurity scattering, spin degrees of freedom, multiband physics, more general boundary conditions, and non-equlibrium phenomena.

\begin{acknowledgments}
N.W-W. and P.H. have contributed equally to this work.
We thank Anton B. Vorontsov and Kevin Marc Seja for valuable discussions.
We thank the Swedish Research Council (VR) for financial support.
The computations were enabled by resources provided by the Swedish National Infrastructure for Computing (SNIC) at the computing centers C3SE, HPC2N and NSC, partially funded by the Swedish Research Council through grant agreement no. 2018-05973.
\end{acknowledgments}

\clearpage
\appendix

\section{Dependencies}
\label{app:dependencies}

This appendix lists external dependencies and packages that SuperConga offloads functionality to. For a more detailed description, please refer to the installation guide in the main README of the repository.\cite{SuperConga:repository} SuperConga uses git as a version control system, with the central repository being hosted on \href{https://gitlab.com/suprcongpu/suprcongpu}{GitLab}.\cite{SuperConga:repository} The SuperConga backend is written in C++, with most of the heavy lifting being done by CUDA\cite{CUDA:2021,Nickolls:2008} kernels and Thrust~\cite{Bell:2012,Kaczmarski:2013,Wynters:2013} algorithms running on the GPU. This functionality is provided by the CUDA toolkit, which also provides the NVCC compiler for compiling the CUDA code. Alternatively, Clang can be used for compilation. To run the compiled code, a CUDA-enabled device (an NVIDIA GPU) is required. Note that we use CMake\cite{Cedilnik:2021} as the build-system generator, and Ninja\cite{Martin:2020} as the build system, but that it is possible to replace either of these with another choice. Linear algebra on the CPU uses Armadillo\cite{Sanderson:2016,Sanderson:2018}, and all tests are implemented in the doctest\cite{Kirilov:2021} framework. The real-time plotting is done using OpenGL\cite{OpenGL:2021, Woo:1999} and ArrayFire-Forge\cite{Yalamanchili:2015}. The package JsonCpp\cite{Dunn:2020} is used to handle JSON\cite{Pezoa:2016} objects on the backend side, and HighFive\cite{HighFive:2021} for HDF5\cite{hdf5:1997} support. The SuperConga frontend requires a Python3 environment, in particular with the modules NumPy\cite{Harris:2020}, Matplotlib\cite{Hunter:2007}, h5py\cite{Collette:2014} and pandas\cite{McKinney:2010}.

\section{The simulation-parameter file}\label{app:json-file}

The simulation configuration is specified in a \texttt{simulation\_config.json} file. It is a JSON\cite{Pezoa:2016} file, which consists of a number of key-value pairs on the form \lstinline[language=json]{"key": value}. The \texttt{value} of a key can be more key-value pairs, numbers, strings, etc. Below is the JSON file with the parameters used in the demonstration example in Section~\ref{sec:demonstration}.
\begin{lstlisting}[language=json,basicstyle=\small,escapeinside={(*}{*)}, caption={examples/swave\_disc\_vortex/simulation\_config.json.},label={lst:abrikosov_example:full_config}]
{
  "physics": {
    "temperature": 0.5,
    "external_flux_quanta": 1.5,
    "penetration_depth": 5.0,
    "crystal_axes_rotation": 0.0,
    "gauge": "symmetric",
    "charge_sign": -1,
    "order_parameter": {
      "s": {
        "critical_temperature": 1.0,
        "initial_phase_shift": 0.0,
        "initial_noise_stddev": 0.0,
        "vortices": [
          {
            "center_x": 0.0,
            "center_y": 0.0,
            "winding_number": -1.0
          }
        ]
      }
    }
  },
  "geometry": [
    {
      "disc": {
        "add": true,
        "center_x": 0.0,
        "center_y": 0.0,
        "radius": 15.0
      }
    }
  ],
  "numerics": {
    "convergence_criterion": 1e-05,
    "energy_cutoff": 16.0,
    "norm": "l2",
    "num_fermi_momenta": 32,
    "num_iterations_burnin": 0,
    "num_iterations_max": 10000,
    "num_iterations_min": 0,
    "points_per_coherence_length": 20.0,
    "vector_potential_error": 1e-06
  },
  "accelerator": {
    "name": "congacc"
  },
  "misc": {
    "data_format": "h5",
    "load_path": "",
    "save_frequency": -1,
    "save_path": "data/examples/swave_disc_vortex",
    "verbose": true,
    "visualize": true
  }
}
\end{lstlisting}
Under \texttt{physics} are all parameters directly related to the physics of the simulation, except the geometry which is specified separately. The \texttt{temperature} is given in units of the highest critical temperature of the order-parameter components, $T_\mathrm{c} \equiv \max \{ T_{\mathrm{c}, \Gamma} \}$. The \texttt{external\_flux\_quanta} is given in units of $\Phi_0$. The \texttt{penetration\_depth} is given in units of $\xi_0$. If the penetration depth is zero or negative it is considered to be infinite and the induced vector potential will be zero. The \texttt{crystal\_axes\_rotation}, i.e. the rotation of the crystal $ab$-axes relative to the simulation coordinate system, is given in units of a full turn, i.e. $2\pi$. The choices for the \texttt{gauge} is either \texttt{symmetric}, \texttt{landau} or \texttt{solenoid}. The sign of the carrier charge is controlled by \texttt{charge\_sign}.

The field \texttt{order\_parameter} is a list of components. The supported order-parameter components are $d_{x^2 - y^2}$, $d_{xy}$, $g_{xy(x^2 - y^2)}$, and $s$. The \texttt{critical\_temperature} is only meaningful with multiple components as it is the \emph{relative} $ T_{\mathrm{c}, \Gamma}$ that is relevant. Each component is initialized as
\begin{eqnarray}
\Delta_\Gamma(\vR) &=& \Delta^{(\mathrm{bulk})}_\Gamma e^{i \chi_\Gamma(\vR)} + Z \; ,\\
\chi_\Gamma(\vR) &=&2\pi \delta_\Gamma + \sum_v w^{(v)}_\Gamma \phi \left(\vR - {\bm C}^{(v)}_\Gamma \right) \; , \\
Z &\sim& \mathcal{N}(0, \sigma^2_\Gamma) \; ,
\end{eqnarray}
where $\delta_\Gamma$ is controlled by \texttt{initial\_phase\_shift}, and $Z$ is complex additive Gaussian noise with a standard deviation $\sigma_\Gamma$ controlled by \texttt{initial\_noise\_stddev} in units of $2\pi k_\mathrm{B} T_\mathrm{c}$. The field \texttt{vortices} is list of vortices, with each vortex being specified by its center position, ${\bm C}_\Gamma = (\mathtt{center\_x},\mathtt{center\_y})$, and phase winding $w_\Gamma = \mathtt{phase\_winding}$. Furthermore, $\phi(\vR) = \mathrm{atan2}(y,x)$ is the polar angle of the coordinate $\vR$.  We may add additional components by adding entries to the \texttt{"order\_parameter"} field in the above JSON file. For example, we may add a subdominant $d_{xy}$-wave component initialized with a phase shift of $\pi/2$, some Gaussian noise, and no vortices:
\begin{lstlisting}[language=json,firstnumber=1,basicstyle=\small,caption={Example of a multi-component order parameter.},label={lst:multi_orderparameter}]
"order_parameter": {
  "s": {
    "critical_temperature": 1.0,
    "initial_phase_shift": 0.0,
    "initial_noise_stddev": 0.0,
    "vortices": [
      {
        "center_x": 0.0,
        "center_y": 0.0,
        "winding_number": -1.0
      }
    ]
  },
  "dxy": {
    "critical_temperature": 0.1,
    "initial_phase_shift": 0.25,
    "initial_noise_stddev": 0.01,
    "vortices": [
      {
        "center_x": 0.0,
        "center_y": 0.0,
        "winding_number": -1.0
      }
    ]
  }
}
\end{lstlisting}

The \texttt{geometry} is built of three different kinds of components; \texttt{disc}, \texttt{regular\_polygon}, and \texttt{polygon}. A \texttt{disc} is specified by its center which is set by \texttt{center\_x} and \texttt{center\_y}, and its \texttt{radius}. All parameters are in units of $\xi_0$. A \texttt{regular\_polygon} is specified by its center coordinates \texttt{center\_x} and \texttt{center\_y}, the number of edges \texttt{num\_edges}, its \texttt{rotation}, and the \texttt{side\_length}. The side length and center is given in units of $\xi_0$, and the rotation in units of a full turn, i.e. $2\pi$. A \texttt{polygon} is specified by the $x$- and $y$-coordinates of its vertices, \texttt{vertices\_x} and \texttt{vertices\_y} respectively. The vertices should be given in counter-clockwise order. All components has a parameter \texttt{add}, which determines if the area of the component should be added (\texttt{true}), or subtracted (\texttt{false}), from the geometry. For more general geometries it is possible to add and subtract several objects by adding to the \texttt{geometry} field. For example, we may put a square hole of side length $5\xi_0$ in the disk at $10\xi_0$ to the right of the center:
\begin{lstlisting}[language=json,firstnumber=1,basicstyle=\small,caption={Example of a compound geometry.},label={lst:compound_geometry}]
"geometry": [
  {
    "disc": {
      "add": true,
      "center_x": 0.0,
      "center_y": 0.0,
      "radius": 15.0
    }
  },
  {
    "regular_polygon": {
      "add": false,
      "center_x": 10.0,
      "center_y": 0.0,
      "num_edges": 4,
      "rotation": 0,
      "side_length": 5.0
  }
  }
],
\end{lstlisting}
Note that the order of the components is important if any component is subtracted.

Under \texttt{numerics} are parameters controlling the numerics of the simulation. The \texttt{convergence\_criterion} determines when the simulation is considered converged. When all computed quantities; each order-parameter component, the charge current density, the induced vector potential, and the free energy, has a residual (as given by Eq.~\eqref{eq:global_error}) smaller than \texttt{convergence\_criterion} the simulation should stop. The \texttt{energy\_cutoff} is the largest energy, in units of $2\pi k_\mathrm{B} T_\mathrm{c}$, to include in the energy sums. The $p$-norm to use when computing the residuals is controlled by \texttt{norm}, with $p \in \{1,2,\infty \}$. The \texttt{num\_fermi\_momenta} is the number of discrete momenta to use when approximating the Fermi-surface averages $\langle \ldots \rangle_{\vpF}$. Note that the terms in the Fermi-surface averages are computed sequentially in contrast to the terms in the energy sum which are computed in parallel on the GPU.

The field \texttt{num\_iterations\_burnin} controls the number of iterations during which the order-parameter and vector potential are kept constant, while the coherence functions are converged and updated self-consistently, in particular on the boundary, i.e. $\gamma_{\partial \mathcal{D}}$ and $\tilde{\gamma}_{\partial \mathcal{D}}$. Once the burn-in period is over, all quantities are updated self-consistently again. This is particularly useful when resuming a previous simulation from file, since the order parameter and vector potentials are stored, but the coherence functions are not (due to occupying too much disk space). If the number of burn-in iterations is negative, it will run until convergence. The maximum and minimum of iterations to run is controlled by \texttt{num\_iterations\_max} and \texttt{num\_iterations\_min}, respectively. When \texttt{num\_iterations\_max} iterations has been run the simulation stops regardless of whether convergence has been achieved. Similarily, the simulation does \emph{not} stop until \texttt{num\_iterations\_min} iterations has been run even if the simulation is considered converged. This is useful for avoiding endless simulations, or prematurely stopping a simulation that is moving slowly on a plateau, respectively. The spatial resolution of the simulation is set by \texttt{points\_per\_coherence\_length}. The \texttt{vector\_potential\_error} is explained in App.~\ref{app:vector_potential}.

All parameters of the accelerator are explained in App.~\ref{app:accelerator}.

Lastly, under \texttt{misc} are miscellaneous parameters not affecting the simulation \emph{per se}. \texttt{data\_format} specifies which data format to use when saving files. The options are \texttt{h5} and \texttt{csv} yielding compressed HDF5\cite{hdf5:1997} files or plain-text CSV files, respectively. Both data formats are standard and have excellent support in e.g. MATLAB\cite{MATLAB:2021} and Python (via h5py\cite{Collette:2014} and pandas\cite{McKinney:2010}). By setting \texttt{load\_path} SuperConga will read the files located there and use them as the initial values for the order parameter and the vector potential. How often to save the results are controlled by \texttt{save\_frequency}. If \texttt{save\_frequency} is negative the results will only be saved at the end of the simulation, and if it is zero the results will not be saved at all, otherwise it will save every $n$th iteration. The results are saved to \texttt{save\_path}. If \texttt{verbose} is \texttt{true} the progress of the simulation will be written to the terminal, and if \texttt{visualize} is \texttt{true} the simulation will be visualized live. 

\section{Solving the Riccati equations}
\label{app:riccati_stepping}

Using the trajectory coordinate, $s$, Eq.~\eqref{eq:riccati} in standard form is
\begin{eqnarray}
\label{eq:riccati_standard_form}
\frac{\partial \gamma}{\partial s} &=& f(s, \gamma(s)) \; , \\
f(s, \gamma(s)) &=& \frac{i}{\vF} \left[ \Delta^*(s) \gamma^2(s) + 2 z(s) \gamma(s) + \Delta(s) \right] \; ,
\end{eqnarray}
with the dependence on $\vpF$ and $\varepsilon$ dropped, and
\begin{equation}
    z(s) \equiv \varepsilon + \tfrac{e}{c} \vvF \cdot \vA(s) \; ,
\end{equation}
for brevity. Instead of using the stepping method,\cite{eschrig:2009,grein_inverse_2013} i.e. using an analytical formula for $\gamma(s)$ but approximating $\Delta(s)$ and $z(s)$ as piecewise constant, we solve Eq.~\eqref{eq:riccati_standard_form} along a trajectory using the implicit mid-point method, which we find to be both faster and more accurate. The implicit mid-point method has the following update rule,
\begin{eqnarray}
\gamma_{j+1} &=& \gamma_j + h f \left(s_{j + \frac{1}{2}}, \frac{1}{2} \left(\gamma_j + \gamma_{j+1} \right) \right) \; ,
\end{eqnarray}
with the discretized trajectory coordinate $s_j=jh$, where $j$ is an integer and $h$ is the length between points. This yields a quadratic equation in $\gamma_{j+1}$ to solve for each step along the trajectory;
\begin{eqnarray}
0 &=& c_2 \gamma^2_{j+1} + c_1 \gamma_{j+1} + c_0 \; , \\
c_2 &=& \frac{i h}{4 \vF} \Delta^*_{j + \frac{1}{2}} \; , \\
c_1 &=& 2 c_2 \gamma_j +  \frac{i h}{\vF} z_{j + \frac{1}{2}} - 1 \; , \\
c_0 &=& \left(c_1 - c_2 \gamma_j + 2 \right) \gamma_j - 4 c^*_2 \; ,
\end{eqnarray}
with the solution
\begin{eqnarray}
\label{eq:riccati_midpoint}
\gamma_{j+1} &=& \frac{2 c_0 }{-c_1 + \sqrt{c^2_1 - 4 c_2 c_0}} \; .
\end{eqnarray}
Half-step values, $\Delta_{j + \frac{1}{2}}$ and $z_{j + \frac{1}{2}}$, are calculated by linear interpolation. Nearest-neighbour extrapolation is used at the boundary. If we have $\gamma(s_j)$ at one point we obtain $\gamma(s_{j+1})$ via Eq.~\eqref{eq:riccati_midpoint}. The solution along the whole trajectory is obtained by starting with the initial boundary value $\gamma(s_0=\vR_{\mathrm{min}})$ and stepping along the trajectory to obtain $\gamma(s_j)$ with $j>0$. The stepping proceeds until a second boundary is reached (denoted $\vR_{\mathrm{max}}$, see Fig~\ref{fig:riccati_fig_A}). We compute $\tilde{\gamma}(s)$ by solving Eq.~\eqref{eq:riccati_tilde} in the same manner, but stepping in the opposite direction from $\vR_{\mathrm{max}}$ to $\vR_{\mathrm{min}}$.

\section{\label{app:vector_potential}Computing the vector potential}
The induced vector potential is obtained by solving Eq.~\eqref{eq:ampere}, in the Coloumb gauge, using the Green's function of the 2D Laplacian operator. The current is computed on a $N\times N$ grid, with every grid cell having the area $h^2$. We treat the current as being piecewise constant within each grid cell. With our choice of units, this yields
\begin{widetext}
\begin{eqnarray}
\label{eq:app:vector_potential:solution}
\vA^\mathrm{ind}_{m,n} &=& -\frac{1}{\kappa^2} \sum^{N - 1}_{m^\prime=0} \sum^{N - 1}_{n^\prime=0} G_{|m^\prime-m|, |n^\prime-n|} \vj_{m^\prime n^\prime}  \; , \\
G_{m, n} 
&=& \frac{1}{4\pi} \int^{+\frac{1}{2}h}_{-\frac{1}{2}h} \int^{+\frac{1}{2}h}_{-\frac{1}{2}h} \ln\left ( \frac{\left (x^\prime + h m \right )^2 + \left (y^\prime + h n \right )^2}{\xi^2_0} \right ) \mathrm{d}x^\prime \mathrm{d}y^\prime \\
&=& \frac{1}{4\pi} \left(\frac{h}{\xi_0}\right)^2 \left( 2\ln\left(\frac{h}{\xi_0}\right) - \ln\left(2\right)- 3 + \frac{1}{4} \sum_{a=\pm} \sum_{b=\pm} \Gamma (am,bn) \right) \; , \\
\Gamma(m,n) &=& \alpha(m,n) + \beta(m,n) + \beta(n,m) \; , \\
\alpha(m,n) &=& \left(1+2m\right)\left(1+2n\right)  \ln\left(\frac{1}{2}\left[ 1+2m \right]^2 + \frac{1}{2}\left[ 1+2n \right]^2\right) \; , \\
\beta(m,n) &=& \left( -2m^2 + 2n^2 + 4n + 1\right) \tan ^{-1}\left(\frac{1 + 2 m}{1 + 2 n}\right)  \; .
\end{eqnarray}
\end{widetext}

$G$ is a normal matrix, so it can be eigendecomposed as $G = V \Lambda V^\top$, where $V$ is a unitary matrix with (normalized) eigenvectors as its columns, and $\Lambda$ is a diagonal matrix with the eigenvalues as diagonal elements. Equivalently,
\begin{eqnarray}
G_{m,n} &=& \sum^{N - 1}_{k=0}  \lambda_k V^{(k)}_{m} V^{(k)}_{n}  \; ,
\end{eqnarray}
where $V^{(k)}_m$ is the eigenvector corresponding to the eigenvalue $\lambda_k$. Thus
\begin{eqnarray}
\vA^\mathrm{ind}_{m,n} &=&-\frac{1}{\kappa^2} \sum^{N - 1}_{k=0} \sum^{N - 1}_{m^\prime=0} \sum^{N - 1}_{n^\prime=0} \lambda_k V^{(k)}_{|m^\prime-m|} V^{(k)}_{|n^\prime-n|} \vj_{m^\prime n^\prime}  \; ,
\end{eqnarray}
and the cross-correlation with the eigenvectors can be done separately. The eigendecomposition is done with Armadillo\cite{Sanderson:2016,Sanderson:2018}. In order to increase performance only $N^*$ eigenvalues are kept. It is computed as the smallest integer obeying
\begin{eqnarray}
\sqrt{1 - \frac{\sum^{N^*-1}_{n=0} \lambda^2_n}{\sum^{N-1}_{n=0} \lambda^2_n}} &<& \epsilon \; ,
\end{eqnarray}
with the eigenvalues sorted $|\lambda_0| \geq |\lambda_1| \geq ... \geq |\lambda_{N-1}|$, and $\epsilon > 0$ being a tolerance chosen by the user via the parameter \texttt{vector\_potential\_error} in the configuration file. If $\epsilon = 0$ no eigendecomposition is done, and the current density is cross-correlated with the full 2D Green's function.

Lastly, the mean of the charge-current density is in general not zero due to numerical errors or bad initialization. This has the undesired effect of introducing a constant term in the induced vector potential, which gives rise to an overall phase gradient through the grain and an associated current. For small penetration depths, this can prevent the simulation from converging. In order to remedy this, the mean of the charge-current density is subtracted before solving Eq.~\eqref{eq:ampere}.

\section{Convergence accelerators}
\label{app:accelerator}

SuperConga solves Eqs.~\eqref{eq:gapeq_regularized} and \eqref{eq:ampere} self-consistently. With our choice of units, the solution to Eq.~\eqref{eq:ampere} is given by
\begin{eqnarray}
\label{eq:app:accelerator:ampere}
\kappa^2 \vA_\mathrm{ind}(\vR) &=& - \int_\mathcal{A} \mathrm{d}\vR^\prime \, G(\vR, \vR^\prime) \vj(\vR^\prime) \; ,
\end{eqnarray}
in the Coloumb gauge, where $G(\vR, \vR^\prime)$ is the Green's function of the 2D Laplacian. In order to simplify the notation we concatenate the left- and right-hand sides of Eqs.~\eqref{eq:gapeq_regularized} and \eqref{eq:app:accelerator:ampere}, yielding
\begin{eqnarray}
    \label{eq:app:accelerator:selfcons_concat}
    \vx &=& \gv (\vx) \; , \\
    \label{eq:app:accelerator:x_concat}
    \vx &\equiv& \left( \mathrm{Re} \left[ \Delta_{\Gamma_1} \right], \mathrm{Im} \left[ \Delta_{\Gamma_1} \right],\ldots, \kappa^2 A^\mathrm{ind}_x, \kappa^2 A^\mathrm{ind}_y \right)^\top \; .
\end{eqnarray}
Note that the vector potential is scaled by $\kappa^2$ in order to have similar magnitudes of all elements in $\vx$. SuperConga provides several different methods of solving Eq.~\eqref{eq:app:accelerator:selfcons_concat}. Namely, basic Picard iterations, Polyak's\cite{Polyak:1964} "small heavy sphere", a variant of the Barzilai-Borwein method\cite{BarzilaiBorwein:1988}, and a custom method. The user controls which method to use by setting the \texttt{name} in the \texttt{accelerator} in Listing.~\ref{lst:abrikosov_example:full_config}. By only setting the name, all internal parameters of the accelerator will be set to default values. The user can change the internal parameters by specifying them in the configuration file, or via the CLI. 

In the following discussion of the different accelerators, it is convenient to define the difference between the right- and left-hand sides of Eq.~\eqref{eq:app:accelerator:selfcons_concat},
\begin{equation}
    \label{eq:app:accelerator:difference}
    \vd \equiv \gv(\vx) - \vx \; ,
\end{equation}
which enters in all accelerator methods.

\subsection{Picard}

The simplest way of solving Eq.~\eqref{eq:app:accelerator:selfcons_concat} is by using Picard iterations,
\begin{eqnarray}
    \label{app:accelearator:picard}
    \vx_{i+1} &=& \vx_i + \alpha \vd_i  \; ,
\end{eqnarray}
where $\alpha > 0$ is the the step size. Picard iterations can converge very slowly, or not at all. Hence, the need for a method accelerating the iterations, and preferably stabilizing them. Calling the Picard method an accelerator is a misnomer as it is the baseline. The other methods should improve on this method. How to use this accelerator is shown in Listing~\ref{lst:app:accelerator:picard}.
\begin{lstlisting}[language=json,firstnumber=1,basicstyle=\small,caption={Picard accelerator with default values.},label={lst:app:accelerator:picard}]
"accelerator": {
  "name": "picard",
  "step_size": 1.0
}
\end{lstlisting}

\subsection{Polyak}

In Polyak's method, $\vx$ behaves similarly to the position of a particle moving through a viscous fluid, with a low Reynold's number, and a potential field (Eq.~\eqref{eq:app:accelerator:difference} taking the role of the negative potential gradient),
\begin{eqnarray}
    \label{app:accelearator:polyak}
     \vv_{i+1} &=& (1 - \beta) \vv_{i} + \alpha \vd_i \; , \\
    \vx_{i+1} &=& \vx_{i} + \vv_{i+1} \; ,
\end{eqnarray}
where $\beta \in (0, 1)$ is the drag, and $\alpha > 0$ is the step size. The velocity is initialized to zero, $\vv_0 = \mathbf{0}$. How to use this accelerator is shown in Listing~\ref{lst:app:accelerator:polyak}.
\begin{lstlisting}[language=json,firstnumber=1,basicstyle=\small,caption={Polyak accelerator with default values.},label={lst:app:accelerator:polyak}]
"accelerator": {
  "name": "polyak",
  "step_size": 2.0,
  "drag": 0.5
}
\end{lstlisting}

\subsection{Barzilai-Borwein}

The Barzilai-Borwein (BB) method\cite{BarzilaiBorwein:1988} is similar to the Picard method but with an adaptive step size,
\begin{eqnarray}
    \label{app:accelearator:barzilai_borwein}
    s^\mathrm{BB1}_i &=& \frac{\| \vd_{i - 1} \|^2_2 } {\left| \vd_{i - 1} \cdot \left( \vd_i - \vd_{i - 1} \right) \right|} \; , \\
    s^\mathrm{BB2}_i &=& \frac{\left| \vd_{i - 1} \cdot \left( \vd_i - \vd_{i - 1} \right) \right| } {\| \vd_i - \vd_{i - 1} \|^2_2} \; , \\
    \alpha_{i+1} &=& \alpha_{i} \cdot \left \{ \begin{matrix} s^\mathrm{BB1}_i & \mathrm{if}\; i \; \mathrm{odd} \\ s^\mathrm{BB2}_i & \mathrm{if}\; i \; \mathrm{even} \end{matrix} \right . \; , \\
    \vx_{i+1} &=& \vx_{i} + \alpha_{i+1} \vd_i \; ,
\end{eqnarray}
where $s^\mathrm{BB1}$ and $s^\mathrm{BB2}$ are two different variants of how to scale the step size. Note that we force the step size to be positive. Alternating between the variants gives better performance on the examples provided by SuperConga than using either one of them on its own. What step size to use during the first iteration is a free parameter and can be set by the user. How to use this accelerator is shown in Listing~\ref{lst:app:accelerator:barzilai_borwein}.
\begin{lstlisting}[language=json,firstnumber=1,basicstyle=\small,caption={Barzilai-Borwein accelerator with default values.},label={lst:app:accelerator:barzilai_borwein}]
"accelerator": {
  "name": "barzilai-borwein",
  "step_size": 1.0,
  "step_size_max": 100.0,
  "step_size_min": 0.001
}
\end{lstlisting}

\subsection{CongAcc}

CongAcc, short for (Super)Conga Accelerator, is our custom made accelerator. Introducing the \emph{component} index, $c$, where a component is either the real or imaginary part of a order-parameter component, or the $x$- or $y$-component of the induced vector potential, in Eq.~\eqref{eq:app:accelerator:x_concat}, the update rule is as follows,
\begin{eqnarray}
s_{i,c} &=& S_C(\vm_{i,c}, \vd_{i,c}) \; , \\
\alpha_{i+1,c} &=& \alpha_{i,c} k^{s_{i,c}} \; , \\
w_{i+1,c} &=& \left \{ \begin{matrix} w_{i,c} & \mathrm{if}\; s_{i,c} \geq s_\mathrm{min} \\
w^* & \mathrm{otherwise} \;  \end{matrix} \right . \; , \\
\vm_{i+1,c} &=& w_{i+1,c} \vm_{i,c} + (1 - w_{i+1,c}) \vd_{i,c} \; , \\
\vx_{i+1,c} &=& \vx_{i,c} + \alpha_{i+1,c} \vm_{i+1,c} \; ,
\end{eqnarray}
where $S_C$ is the cosine similarity, $\vm_{i,c}$ is an exponential moving average of the component difference $\vd_{i,c}$, and $k$ is a constant determining how much the step size should maximally increase (decrease) which can be set by the user. If $s_{i,c} < s_\mathrm{min}$, where $s_\mathrm{min}$ is the minimum tolerated similarity, the weight $w^*$ is obtained by solving
\begin{eqnarray}
\label{app:accelerator:eq:similarity_weight}
S_C(w^* \vm_{i,c} + (1 - w^*) \vd_{i,c}, \vd_{i,c}) &=& s_\mathrm{min} \; 
\end{eqnarray}
ensuring that the step will be similar to $\vd_{i,c}$. Equation~\eqref{app:accelerator:eq:similarity_weight} is solved approximately using the bisection method. Note that we set $w_{0,c}=0$ during the first iteration. 

The reasoning is that if $\vm_{i,c}$ and $\vd_{i,c}$ roughly point in the same direction, then we should have taken a larger step the previous iteration, and thus the step size is increased. Analogously, if they roughly point in opposite directions, the step size should decrease.

How to use this accelerator is shown in Listing~\ref{lst:app:accelerator:cognac}.
\begin{lstlisting}[language=json,firstnumber=1,basicstyle=\small,caption={CongAcc accelerator with default values.},label={lst:app:accelerator:cognac}]
"accelerator": {
  "name": "congacc",
  "step_size": 0.5,
  "step_size_factor": 1.234,
  "cos_similarity_min": 0.7,
  "step_size_max": 100.0,
  "step_size_min": 0.001
}
\end{lstlisting}

This accelerator should be regarded as experimental as no thorough analysis of it has been done. 

\subsection{Comparing the accelerators}

\begin{table*}
\begin{center}
\begin{tabular}{|c|c|c|c|c|c|c|c||c|c|c|c|}
\hline
\rowcolor{lightgray}\multicolumn{12}{|c|}{SuperConga examples and accelerator comparison}  \\
\hline
& Name & $T/\Tc$ & $\Phi / \Phi_0$ &  $\kappa$ & Symmetry & Initialization & Geometry & Picard & Polyak & BB & CongAcc \\
\hline
\hline
(a) & \texttt{dwave\_chiral} & $0.5$ & $0$ & $\infty$  & $d_{xy} + i d_{x^2-y^2}$ & Bulk & Disc: $\mathcal{R}=12.5\xi_0$ & $80$ & $34$ & $25$ & $37$ \\
\hline
(b) & \texttt{dwave\_octagon} & $0.5$ & $0$ & $\infty$ & $d_{x^2-y^2}$ & Bulk & Octagon: $\mathcal{S}=10\xi_0$ & $71$ & $33$ & $25$ & $32$ \\
\hline
(c) & \texttt{dwave\_plus\_swave} & $0.5$ & $0$ & $\infty$ & $d_{xy} + s$ & Bulk & Disc: $\mathcal{R}=12.5\xi_0$ & $69$ & $36$ & $30$ & $33$ \\
\hline
(d) & \texttt{dwave\_phase\_crystal} & $0.1$\footnote{The temperature is low, $T < T^*$, so the solution is a phase crystal.\cite{Hakansson:2015,Holmvall:2020}} & $0$ & $100$ & $d_{x^2-y^2}$ & Bulk + vortex\footnote{The vortex has a phase winding $n=-1$ and is positioned outside of the grain, yielding a soft phase-gradient through the grain.} & Irregular polygon\footnote{The shape of the grain is a square, $\mathcal{S} = 25\xi_0$, with one corner removed, yielding four $[100]$-interfaces, and one pair-breaking $[110]$-interface.\cite{Wennerdal:2020}} & $3027$ & $809$ & $1054$ & $439$ \\
\hline
(e) & \texttt{swave\_abrikosov\_lattice} & $0.2$ & $20$ & $10$ & $s$ & Bulk + giant vortex\footnote{The giant vortex in the middle has a phase winding $n=-13$. It is unstable and decays into a lattice of singly-quantized Abrikosov vortices.} & Square: $\mathcal{S}=25\xi_0$ & $1001$ & $313$ & $211$ & $133$ \\
\hline
(f) & \texttt{swave\_disc\_meissner} & $0.5$ & $0.5$ & $5$ & $s$ & Bulk & Disc: $\mathcal{R}=15\xi_0$ & F & F &  $20$ & $43$ \\
\hline
(g) & \texttt{swave\_disc\_vortex} & $0.5$ & $1.5$ & $5$ & $s$ & Bulk + vortex\footnote{The vortex in the middle has a phase winding $n=-1$} & Disc: $\mathcal{R}=15\xi_0$ & F & F & $45$ & $42$\\
\hline
\end{tabular}
\end{center}
\caption{How many iterations each accelerator, with default parameters, require in order to reach convergence on the examples provided with SuperConga. An F means that it did not converge during the $10^4$ iterations the simulation was run. The convergence criterion is set to $10^{-5}$ for all examples. In the geometry column, a radius is denoted $\mathcal{R}$, and a polygon side length $\mathcal{S}$.}
\label{app:accelerator:table:comparison}
\end{table*}

Table~\ref{app:accelerator:table:comparison} shows the number of iterations needed to reach convergence in all the examples, (a)--(g), for each of the different accelerators available in SuperConga (using default parameters). The examples (a)--(c) are fairly easy as even Picard iterations converge quickly. It is examples (d)--(g) that really show the strength of the more advanced methods. In (d) Polyak, BB and CongAcc provide roughly a speedup of $3\sim 7$ compared to Picard. Both Picard and Polyak fail to converge in example (f) and (g). This is due to the small penetration depth in those examples, making the default step sizes too large. They can be made to converge by manually changing their parameters, however. The adaptive methods, BB and CongAcc, have no problem with bad initial step sizes; they will quickly adjust the step size to something appropriate.

Listing~\ref{app:accelerator:lst:running_example} shows how to run one of the available examples with a specific accelerator, where \texttt{<example>} and \texttt{<accelerator>} should be replaced with the example and accelerator names, respectively.
\begin{lstlisting}[language=bash,linewidth=\columnwidth,breaklines=true,caption={Running an examples with one of the accelerators.},label={app:accelerator:lst:running_example}]
python superconga.py simulate -C examples/<example> -A <accelerator>
\end{lstlisting}

\section{Units and basis functions}
\label{app:units}

We summarize in Table~\ref{table:units} our choice of units and the resulting natural scales for observables and other quantities. The available order parameter basis functions are listed in Table~\ref{table:basis_functions}

\renewcommand{\arraystretch}{1.5}
\begin{table*}
\begin{center}
\begin{tabular}{c|l|l}
\hline
\rowcolor{lightgray}\multicolumn{3}{|c|}{Input to quasiclassical theory} \\
\hline
$\vF$ && Fermi velocity\\
\hline
$N_\mathrm{F}$ && normal-state density of states per spin at the Fermi level\\
\hline
$T_\mathrm{c}$ && superconducting transition temperature (critical temperature)\\
\hline
\rowcolor{lightgray}\multicolumn{3}{|c|}{Geometry} \\
\hline
$\mathcal{A}$ && total superconducting computational area\footnote{We use calligraphic capital letters for quantities related to geometry, such as area ${\cal A}$ or disc radius ${\cal R}$.}\\
\hline
$\mathcal{V}$ & $=N_ld\mathcal{A}$ & volume of a stack of $N_l$ layers with interlayer distance $d$\\
\hline
\rowcolor{lightgray}\multicolumn{3}{|c|}{Natural units} \\
\hline
$T_\mathrm{c}$ && {\it temperature} in units of critical temperature\\
\hline
$2\pi k_\mathrm{B} T_\mathrm{c}$ && {\it energy} in units of critical temperature\\
\hline
$\xi_0$ & $=\frac{\hbar \vF}{2\pi k_\mathrm{B}T_\mathrm{c}}$ & {\it length} in units of superconducting coherence length\\
\hline
$\Phi_0$ & $=\frac{hc}{2|e|}$ & {\it magnetic flux} in units of superconducting flux quantum\\
\hline
\rowcolor{lightgray}\multicolumn{3}{|c|}{Derived units and observables} \\
\hline
$\lambda_0$ & $\frac{1}{\lambda_0^2}=\frac{4\pi e^2}{c^2} N_\mathrm{F}\vF^2$ & penetration depth. Here we use Gaussian units. In SI units $4\pi/c^2 \rightarrow \mu_0$ with $\mu_0 \approx 4\pi\times 10^{-7}J/A^2m$ \\
\hline
$\kappa$ & $=\frac{\lambda_0}{\xi_0}$ & penetration depth in units of coherence length\\
\hline
$\alpha$ & $=\frac{B_\mathrm{ext}{\cal A}}{\Phi_0}$ & symmetric gauge: external magnetic flux penetrating the superconducting computational area in units of the flux quantum\\
 && solenoid gauge: external magnetic flux penetrating a circular hole centered at origo in units of the flux quantum\\
\hline
$A_0$ & $=\frac{\Phi_0}{\pi\xi_0}$ & vector potential\\
\hline
$B_0$ & $=\frac{\Phi_0}{\pi\xi_0^2}$ & magnetic flux density\\
\hline
$j_0$ & $=2\pi k_\mathrm{B}T_\mathrm{c}|e| \vF N_\mathrm{F}$ & charge-current density\\
\hline
$m_0$ & $=j_0\xi_0\mathcal{V}$ & magnetic moment\\
\hline
$\Omega_\mathcal{A}$ & $=(2\pi k_\mathrm{B}T_\mathrm{c})^2 N_\mathrm{F} {\cal A}$ & free energy\\
\hline
\end{tabular}
\end{center}
\caption{Natural units and scales for observables used in this paper and in the framework SuperConga. We use the convention $e=-|e|$ for the charge of the electron, but allow for simulations with either positive or negative charge carriers. Often, we let $\hbar=k_\mathrm{B}=c=1$.}
\label{table:units}
\end{table*}

\begin{table*}
\centering
\begin{tabular}{|c|c|c|}
\hline
\rowcolor{lightgray}\multicolumn{3}{|c|}{Order-parameter basis functions} \\
\hline
Symmetry class for $D_{4h}$ & Order parameter $\Delta({\bf k})$ & Basis function $\eta(\varphi)$ for circular Fermi surface\\
\hline
$A_{1g}$ ($s$) & $1$ & $1$\\
\hline
$A_{2g}$ ($g_{xy(x^2-y^2)}$) & $k_x k_y (k_x^2-k_y^2)$ &$\sqrt{2}\sin(4\varphi)$\\
\hline
$B_{1g}$ ($d_{x^2-y^2}$) & $k_x^2-k_y^2$ & $\sqrt{2}\cos(2\varphi)$\\
\hline
$B_{2g}$ ($d_{xy}$) & $k_x k_y$ & $\sqrt{2}\sin(2\varphi)$\\
\hline
\end{tabular}
\caption{The available symmetries of the order parameter that can be chosen in SuperConga. Note that any linear combination forming a multi-component order parameter is also allowed.}
\label{table:basis_functions}
\end{table*}

\clearpage

\bibliography{ordered.bib}

\end{document}